\documentclass[12pt,a4paper,twoside]{scrartcl}
						
					\usepackage[utf8]{inputenc} 
					\usepackage[english]{babel} 
					\usepackage[T1]{fontenc}  	
						\usepackage{float}
					\usepackage{framed}
					\usepackage{ulem}
					\usepackage{multirow}
					\usepackage{subfig}
					\usepackage{amsmath}
					\usepackage{amsfonts}
					\usepackage[font=small,labelfont=bf]{caption}
					\usepackage{amssymb}
					\usepackage{graphicx}
					\usepackage{placeins}
					\usepackage{a4wide}
					\usepackage{gensymb}
					\usepackage{stmaryrd}
					\usepackage{pgfplots}
					\usepackage{bbm}
					\usepackage{amsthm}
					\usepackage{pdflscape}
					\usepackage{url}
					\usepackage{tabularx,booktabs}
					\usepackage{fix-cm}
					\usepackage{xcolor}
					\usepackage{titlesec}
					\definecolor{gray75}{gray}{0.75}
					\newcommand{\hsp}{\hspace{0pt}}
					
					\usepackage[breaklinks,pdfborder={0 0 0},bookmarksopen=true]{hyperref}
					\numberwithin{figure}{section}
\begin{document}
					\title{}
					\begin{titlepage}
					\center
					\hrule\hrule \vspace{0.2cm}
					{\Huge COMPUTER SIMULATIONS OF GEL FORMATION IN COLLOIDAL SYSTEMS OF STICKY RODS }\\

					\vspace{0.2cm}
					\hrule\hrule \vspace{1.5cm}
					\text{\Huge Master thesis}\\[0.5cm]
					\text{\LARGE by}\\
					\text{\huge Johannes Krotz}\\[1cm]
					
					\text{\large submitted on the}\\[0.5cm]
					\text{\Large 3.06.2019}\\[0.5cm]
					\text{\large to}
					\begin{figure}[h!]
					\centering
					\includegraphics[width=0.8\linewidth]{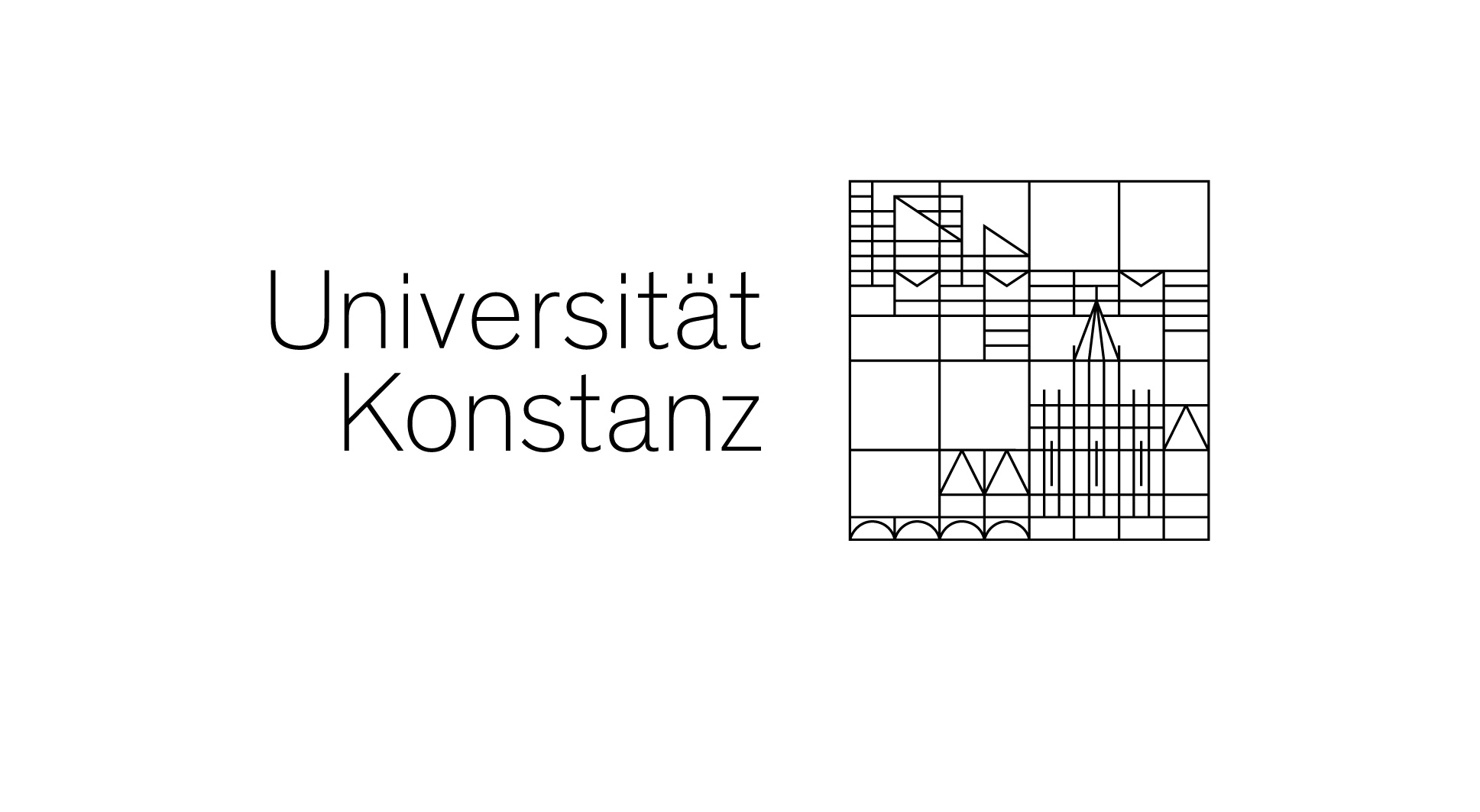}
					
					\label{fig:rulogo}
					\end{figure}
					\hrule\hrule
					\vspace{0.2cm}
					\text{\LARGE Faculty of Science}\\
					\text{\LARGE Department of Physics }\\
					\begin{flushleft}
					\text{\Large \hspace{0.5cm} evaluated by Professor Dr. P. Nielaba (University of Konstanz)}  \\ \text{\Large \hspace{0.5cm} and Professor Dr. C. Bechinger (University of Konstanz)}\\
					
					\end{flushleft}
					\hrule\hrule
					\thispagestyle{empty}
					\end{titlepage}
					\pagenumbering{gobble}

			

					%
					%
					\renewcommand{\contentsname}{Table of Contents}
			\titleformat{\section}[hang]{\flushleft
				\fontseries{b}\fontsize{80}{100}\selectfont}{\fontseries{b}\fontsize{100}{130}\selectfont \textcolor{gray75}\thesection\hsp}{0pt}{ \Huge\bfseries }[]
				
				\titleformat{\section}[hang]
				{\flushleft\fontseries{b}\fontsize{80}{100}\selectfont}
				{\fontseries{b}\fontsize{100}{130}\selectfont \textcolor{gray75}{\thesection}\hsp}
				{0pt}
				{\Huge\bfseries}
				[]
				
				\titleformat{\subsection}[hang]
				{\normalfont\large\bfseries}
				{\thesubsection}
				{1em}
				{}
				
				\titleformat{\subsubsection}[hang]
				{\normalfont\normalsize\bfseries}
				{\thesubsubsection}
				{1em}
				{}
				
				\titlespacing*{\section}      {0pt}{3.5ex plus 1ex minus .2ex}{2.3ex plus .2ex}
				\titlespacing*{\subsection}   {0pt}{2.3ex plus .2ex}{1.5ex plus .2ex}
				\titlespacing*{\subsubsection}{0pt}{1.5ex plus .2ex}{1.0ex plus .2ex}
				
					\section*{Acknowledgments:}
					I would like to express my gratitude and appreciation to everyone, who supported my work on this thesis and thus made its completion possible.\\
					In particular I would like to mention
					\begin{itemize}
						\setlength{\itemsep}{0.cm}								
					\item \textbf{Ullrich Siems}, who had the idea for this thesis, introduced me to the topic, lay the groundwork for all simulations included and found even the best-hidden bugs in any code I wrote, all of which I, of course, only implemented to test him. He passed all my test brilliantly.
					\item \textbf{Professor Peter Nielaba}, whose lectures awakened my interest in computer simulations and  who freely invited me into his group to work on this thesis. Apart from evaluating this thesis he always generously supported me in many ways.
					\item \textbf{Professor Clemens Bechinger}, who volunteered to evaluate this thesis and whose group contributed experimental data to compare to my simulations.
					\item \textbf{Bastian Trepka} and \textbf{Jacob Steindl}, who contributed experimental data to compare to my simulations.
					\item The "colloid-office" \textbf{Tobias Vater}, \textbf{Anton Lüders} and  \textbf{Jacob Holder} for fruitful discussions and way to long coffee-breaks.
					\item All other members of the \textbf{AG Nielaba}, whose ideas and advice roughly saved as much time as I wasted discussing everything but work with them.
					\item My \text{friends}, who prevented me from finishing this thesis way earlier. The \textbf{Bierslot-Crew} might be the most faulty of wasting my time, which I am very thankful for.
					\item My \textbf{family}, who always supported me in everything I do.
					\item \textbf{Everyone else}, who  influenced my personal and professional development and made my time at the university of Konstanz memorable.						
					\end{itemize}

					\begin{figure}[H]
	\centering
	
	\includegraphics[width=0.7\linewidth]{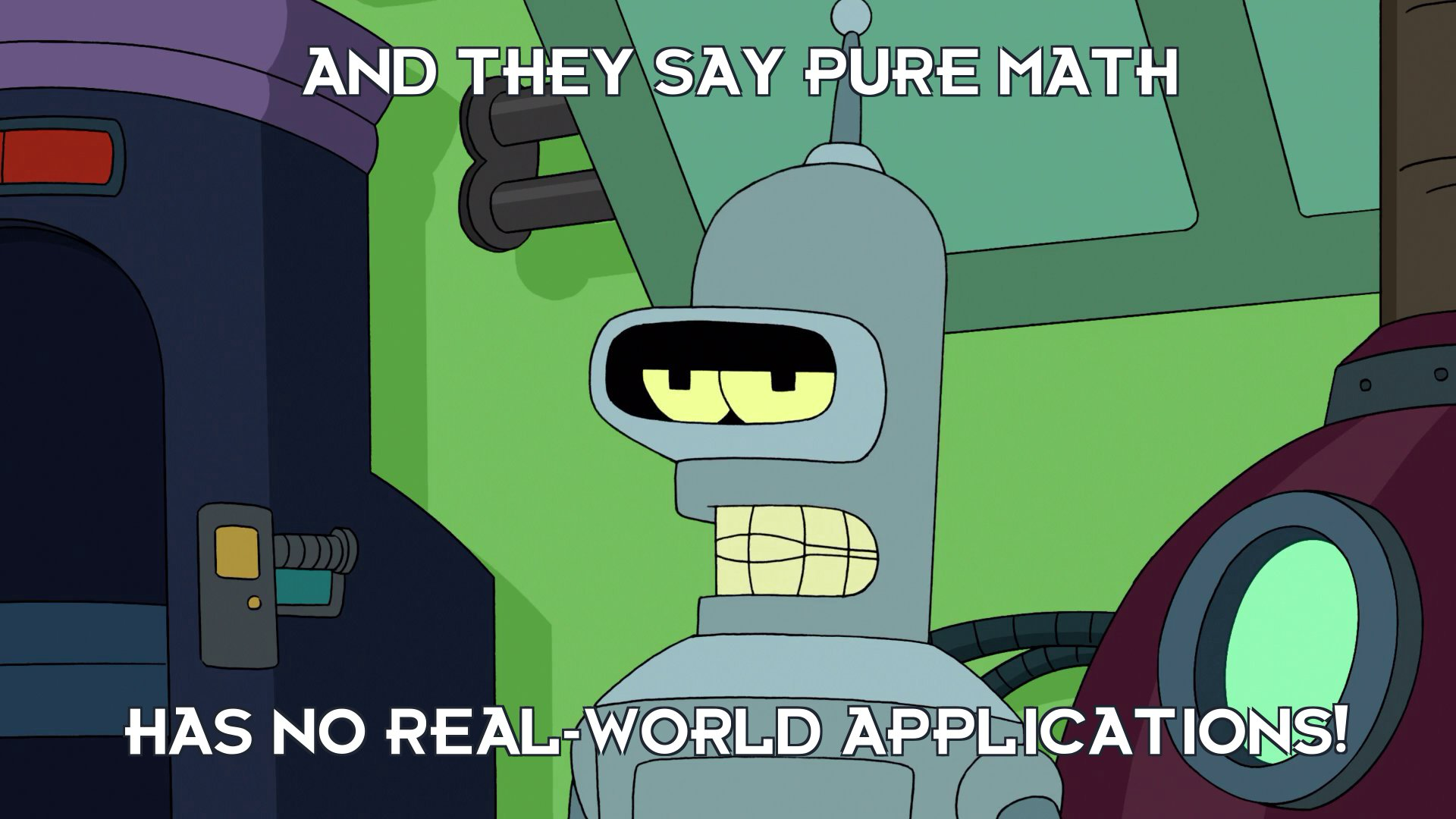}
	\end{figure}
	
\newpage
					\tableofcontents
					\makeatletter \def\l@subsection{\@tocline{2}{0pt}{1pc}{5pc}{}} \def\l@subsection{\@tocline{2}{0pt}{2pc}{6pc}{}} \makeatother
					
					
					\renewcommand{\c}{\cdot}
					\renewcommand{\proofname}{Beweis:}
					\renewcommand{\theta}{\vartheta}
					\newcommand{\R}{\mathbb{R}}
					\renewcommand{\phi}{\varphi}
					\renewcommand{\d}{\partial}
					\renewcommand{\(}{\left(}
					\renewcommand{\)}{\right)}
					\renewcommand{\subset}{\subseteq}
					\renewcommand{\epsilon}{\varepsilon}

					 \newtheorem{theorem}{Theorem}[section]
					 \newtheorem{lemma}[theorem]{Lemma} \newtheorem{satz}[theorem]{Satz}
					 \newtheorem{proposition}[theorem]{Proposition}
					 \newtheorem{corollary}[theorem]{Korollar}
					
					 \theoremstyle{definition}
					  \newtheorem{bemerkung}[theorem]{Bemerkung} \newtheorem{definition}[theorem]{Definition}
					 \newtheorem{example}[theorem]{Beispiel}
					 \newtheorem{beispiel}[theorem]{Beispiel}
					 \newtheorem{beispiele}[theorem]{Beispiele}
					 \newtheorem{notation}[theorem]{Notation}
					 
					\newpage
					
					 	 \pagenumbering{arabic}
					 \setcounter{page}{1}
				
					 \section{Introduction and Conclusion\label{sec:1}}
				
					In this work we studied computer simulations of colloidal dispersions. We particularly simulated sticky spheres and rods with a Brownian dynamics algorithm and analyzed their capability to form percolating porous networks with the rheological properties of a gel.\\
					The motivation to study these systems with respect to their gelation-properties originally came from Ullrich Siems, who was working on computer simulations of spherocylinders, i.e. rods \cite{Ullidiss} for quite some time, and Bastian Trepka, who grew $EuO$-based nanorods in order to produce aerogels. 
					\begin{figure}[H]
	\centering
	\includegraphics[width=0.5\linewidth]{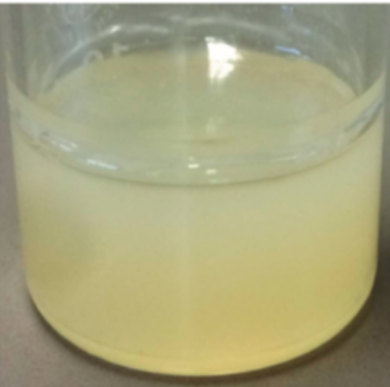}
	\caption{Photographic image of a gelated dispersion of $Eu_2O_3$-benzoate
		nanorods \cite{Euro}}
	\label{fig:gel}
	\end{figure}

					 When growing these nanorods Bastian Trepka observed, that they and their solvent formed a gel once the aspect ratio (ratio between length and width of a rod) grew larger than $\approx 20$. A picture of one of these gels can be seen in figure \ref{fig:gel}. The primary idea of this thesis was to extend Ullrich Siems simulation program so far, that we can gain some insight into Bastian Trepka's gelated systems and hopefully reproduce them.   \\
					 During the course of this work we will first give a short reminder of the physical and computational basics necessary to interpret our results in the chapters 2 and 3. Chapter 4 will give a detailed  overview of all the simulations we conducted, including all necessary parameters to eventually reproduce our results. Finally chapter 5 is a comprehensive presentation and evaluation of our results. There, in section 5.1, we will take a look at our simulations of spherical particles. With these simulations we reproduced the results on gel-forming particles by Santos, Campanella and Carignano \cite{kugelgel} in order to test our program and methods. Chapter 5.2 then holds all the results concerning our simulations of sticky rods including an analysis of their rheological properties and the topology of the generated networks. We use chapter 6 to compare these results to the experimental data collected by Jacob Steindl with Bastian Trepka's gels. Ultimately chapter 7 contains a short list of ideas and suggestions on how further research could be done using the data, tools and knowledge accumulated during the fabrication of this thesis and during the studies preceding it. \\
					 
					 Even though first comparisons to experimental results show little indication that the model used by us is suitable to properly describe the gelated systems of Bastian Trepka, which was our original intention, we deem our work to be an overall success
					 and consider our results very promising at least. \\
					 We were able to reproduce the paper of Santos, Campanella and Carignano \cite{kugelgel} in every aspect we intended and extended their results by a large margin from spheres to even more complicated systems of sticky rods.Furthermore, we further were able to clearly observe an connection between the aspect ratio and the density of rods to their capability to form porous, percolating networks, a result, which in general is well established \cite{Stabpercol}, but to our knowledge was never before observed in systems of spherocylindrical colloids interacting via a Kihara-like potential, that were used by us.
					 Our analysis of the topological structure of these percolating networks shows interesting results and might give useful insights in comparable structures or be used to automatically compare pictures of real gels among each other and to results of simulations as well. \\
					 Finally the rheological analysis shows us that we were actually able to simulate several systems of sticky rods that exhibit the characteristic properties of a gel.

					All in all  this work proofs that our tools and methods are functional and can be used to simulate gelated systems on the basis of sticky rods. Starting from our work here, there a plenty of paths wide open to be followed in future projects.

					 \section{Physical Basics}
					 In this chapter we will discuss the physical basics necessary to understand the further course of this work. First an overview of colloids and colloidal systems will be given, followed by a summary of Brownian motion. In order to define gels in section \ref{sec:gel} we will introduce the storage and loss module in section \ref{sec:mod}. In the last section of this chapter then we will discuss the trajectories of spherocylinders in a shear flow, called Jeffery-orbits. 
					 \subsection{Colloids}
					 A \textbf{colloid} is a mixture  of two substances one of which  is microscopically dispersed in the other. A colloid constitutes two phases, the dispersed phase and the continuous phase. This is an essential difference to a solution, where solute and solvent only constitute one phase. It is very common  to refer to the dispersed substance alone as colloid, or if that substance consists of distinguishable entities as colloids. In this case the continuous phase is called the dispersion medium. From here on we will also use this terminology and speak of colloidal system, if we want to refer to the mixture of dispersed substance and dispersion medium. \cite{einfuehrungkoll}\\
					  Historically the use of colloids dates back as far as the earliest records of human civilization. Cave painting in Lussac-les-Ch$\hat{\text{a}}$teaux (France) who are around 15000 years old and writings about the Egyptian pharaohs, dating back as early as 3000 B.C. were created using colloidal pigments. In fact many of mankind's earliest inventions like paper, pottery, soap or other cosmetics require even today the manipulation of colloidal systems.\\
					  Colloidal science dates back as early as the middle of the 19th century, when Francesco Selmi(1877-1881) first described the characteristics of the "pseudosolutions". In 1861 Thomas Graham(1805-1869) then coined the term "colloid", which is Greek for glue, emphasizing the low rate of diffusion and lack of crystalline order in these pseudo solutions Due to this low diffusion rate Graham determined the lower bound for the size of colloids to be around $1nm$. Also considering the failure of colloids to sediment under the influence of gravity (at least for reasonably long times) he was further able to determine an upper bound for the colloid size of around $1\mu m$. These size limitations are still often seen as additional condition in the definition of colloids.\cite{colloidaldomain}\\
					  With the discovery of Brownian motion (compare section \ref{sec:BM}) colloidal science further established a relation between theoretical physics and the descriptive chemistry of that time creating a gateway to physical chemistry and statistical mechanics. Due to their already frequent use through many industries (paints, ceramics, gels, food, cosmetics,...) and their abundance in living organisms (polymers, DNA, blood,...) the importance of colloids in a modern technological society is almost self-evident.\cite{colloidaldomain} (Since coffee also is a colloid suspension the role of colloids in sustaining our civilization can hardly be overestimated.)\\
					  To get a more hands-on impression of the characteristics of colloids some examples are listed in table \ref{tab:excol}. They are sorted by the state of matter (solid, liquid, gas) the colloid and the dispersion medium are in.\\
					  \begin{table}[h]
					  	\begin{centering}	  		
					  \begin{tabular}{lcccc}
					  \hline	\hline & & & \multicolumn{2}{c}{Examples} \\
					  	\rule[-2ex]{0pt}{5.5ex} Name & Colloid & Medium & Biological & Technical \\ 
					  	\hline\hline \rule[-2ex]{0pt}{5.5ex} Colloidal sol & Solid & Liquid & muddy water & Paint, Ink, Sol-Gel \\ 
					  	\hline \rule[-2ex]{0pt}{5.5ex} Emulsion & Liquid & Liquid & \begin{tabular}{c}
					  		biological membranes \\
					  		fat digestion\\(milk,butter,...)
					  	\end{tabular} & \begin{tabular}{c}
					  	Drug delivery, \\ emulsion \\ polymerization
					  \end{tabular} \\ 
					  	\hline \rule[-2ex]{0pt}{5.5ex} Foam & Gas  & Liquid & Vacuoles & Soap foam \\ 
					  	\hline \rule[-2ex]{0pt}{5.5ex} Aerosol & Solid & Gas & Smoke, Pollen & porous plastic \\ 
					  	\hline \rule[-2ex]{0pt}{5.5ex} Aerosol & Liquid  & Gas & Clouds & Hair spray, Smog \\ 
					  	\hline \rule[-2ex]{0pt}{5.5ex} Solid suspension & Solid  & Solid & Wood,Bone & Composites  \\ 
					  	\hline \rule[-2ex]{0pt}{5.5ex} Porous material & Liquid & Solid & Oil reservoir rock, Pearl & High impact plastics \\ 
					  	\hline \rule[-2ex]{0pt}{5.5ex} Solid foam & Gas & Solid & Pumice, Loofah & Styrofoam \\ 
					  	\hline 	\hline
					  	 \end{tabular} 
					  \caption{Examples for Colloids in different states of matter \cite{colloidaldomain}}
					  \label{tab:excol}
					  \end{centering}
					  \end{table}
					
\begin{figure}
		\centering
	\includegraphics[width=\linewidth]{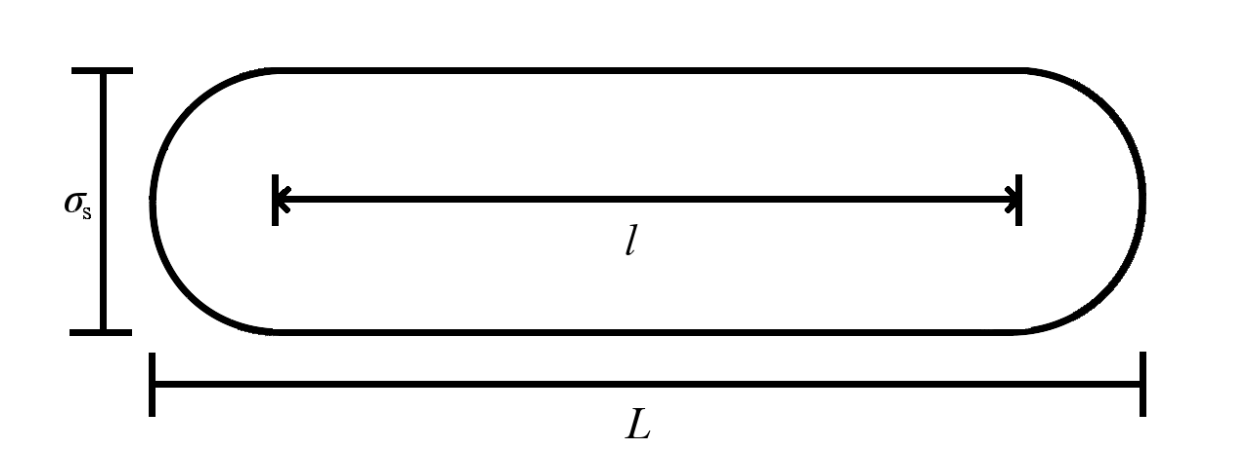}
	\caption{2-dimensional schematic depiction of a spherocylinder with total length $L$, radius $\sigma_S$ and line segment $l$. The 3-dimensional spherocylinder can be constructed by rotation the above picture around its line segment. \cite{Antonpaper}}
	\label{fig:spherschem}
	\end{figure}

					  The colloids discussed in the rest of this work will be solid spheres and spherocylinders dispersed in highly viscous fluids. The spheres are characterized by their radius, while the spherocylinders, which are cylinders with half-spheres of same the radius attached to both flat ends, are characterized by their radius and the height of the cylinder, also called the line segment. Figure \ref{fig:spherschem} shows a depiction of such a spherocylinder of total length $L$, radius $\sigma_S$ and line segment $l$.  These sizes are related by $l=L-\sigma_S$. The ratio $p=L/\sigma_S$ is called the aspect ratio. \\
					  The motion of these spheres and spherocylinders is governed mainly by Brownian motion. A phenomenon, which will be further explained in the following section.

					 \subsection{Brownian Motion\label{sec:BM}}
					 The study of Colloidal particles is in many ways rooted in the study of \textbf{Brownian Motion}.
					 This phenomenon of the irregular, seemingly inexhaustible motion of small particles suspended in fluids inherited its name from the botanist Robert Brown (1773-1858), who nowadays is often falsely said to be the first person to discover these ubiquitous fluctuations in particle trajectories. \cite{HistofBM}\\
					 In fact it appears that the Brownian motion of small grains and pollens was observed very soon after the invention of the microscope.\cite{gray} Browns main contribution was to discover that the same motions were also performed by inorganic materials, ruling out the widespread idea that living organism were causing the observed trajectories. When Brownian motion was subsequently even observed in gas bubbles in fluid reservoirs enclosed by crystals it became very clear that an explanation of this effect had to be found within physics rather than within biology.\\
					  It took several years and experiments to rule out external effects, such as vibrations or temperature gradients between the fluids and their surroundings and many others as a source of the particle motions and recognize Brownian motion as a phenomenon completely inherent to the fluid itself. Yet it was not until 1905, 78 years after Browns first experiments, that Albert Einstein(1879-1955) published his papers on Brownian motion linking the effect to the known phenomenon of diffusion. Diffusion describes the distribution of fluids within other fluids over time.\cite{HistofBM} \\
					 The effect was first examined by Thomas Graham in 1833 for the distribution of mixtures of gases\cite{Graham}. A few years later, in  1855, Adolf Fick was able to derive the following diffusion equation governing the time evolution of the concentration of a fluid:
					 \begin{align}
					 	\frac{\partial c}{\partial t}= D\Delta c \label{eq:diffusion},
					 \end{align}      
						 where $D$ is the diffusion constant and $\Delta$ the Laplace-operator in the $d$ dimensions, in which the fluid expands.\cite{Fick}\\ Einstein was able to show, that this equation does not only determine the evolution of a concentration, but also the statistic of a single particle over time.\cite{Einstein1}\cite{Einstein2} To make this connection one simply interprets the concentration $c(\vec{x},t)$ of all particle as the probability of a particle to be found at the place $\vec{x}$ at time $t$. From this it is easy to derive that the mean square displacement of a single particle is given by $\langle \vec{x} \rangle=2dDt$. \footnote{ To see this one simply integrates $\frac{d}{dt}\langle x^2 \rangle = \int_{\mathbb{R}^d} \vec{x}^2 \frac{d}{dt}c dx^d \underset{\eqref{eq:diffusion}}{=}\int_{\mathbb{R}^d} \vec{x}^2 D\Delta c dx^d \underset{2 x P.I.}{=}2dD \int_{\mathbb{R}^d}c dx^d=2dD$.} \\
						 Marian Smulochowski(1872-1917), who had been working on the topic since 1900, but not published anything till 1906, came up with the same result for the mean square displacement. He assumed that, while all particles were moving with a velocity of constant absolute value, the direction of each particle's velocity would change constantly due to collisions with the fluid molecules. These collisions happened so fast, he suggested, that they were not visible for the human eye, and all that could be observed was the seemingly completely random Brownian motion. \cite{smulo} \\
						 Since their results were in accordance with experimental findings Brownian motion could now be described as the statistic interaction of the suspended particles with the molecules of the surrounding fluid.\\
						 In 1908 Paul Langevin published an equation of motion for a Brownian particle.\cite{Langevin} Since this equation is the basis for all our simulations, we will have a closer look at its derivation and some solutions in the next section.

					 \subsubsection{Langevin-equation}
					 The following derivation is loosely based on \cite{Zwanzig}. \\
					 To derive Langevin's equation of motion for a Brownian particle, we assume that the motion of its center of mass is governed by Newtons equation of motion:
					 \begin{align}
					 	m\frac{d \vec{v}}{dt}=\vec{F}(t),\label{eq:newton}
					 \end{align}
					 where $\vec{F}(t)$ is the total force acting on the suspended particle, i.e. all forces due to external sources as well as all forces caused by the fluid itself. \\
					 For the time being we assume that there are no external forces acting on the particle, so that $\vec{F}$ corresponds solely to the forces due to the interaction between the particle and the fluid.\footnote{External forces can be reintroduced by replacing $\vec{F}$ by $\vec{F}+\vec{F}_{ext}$ in the end.}\\
					 From experimental data we know that particles in a fluid with an initial mean velocity $\vec{v_0}$ come close to rest(,up to fluctuation,) after a certain amount of time due to friction. For fluids with a small Reynolds  number ($Re\ll 1$) we further know, that this friction can be modeled by a friction force \begin{align} 	
					 \vec{F}_{sto}=-\zeta \vec{v} \label{eq:Stokesforce} \end{align} linearly depending on the negative velocity.  
					 $\zeta$ is called the friction coefficient and depends on the geometry and the orientation of the particle as well as on the viscosity $\eta$ of the surrounding fluid. Stokes' law states that for a sphere of radius $a$ this coefficient is given by $\zeta= 6\pi a\eta$. However in a more general setting $\zeta$ can be a symmetric 2-tensor \cite{tensorsym}. In accordance with standard notation we assume from here on that $\zeta$ is a scalar. \footnote{If $\zeta$ was a 2-tensor, most of the calculations below are exactly the same, if the following replacements are made:\\ $\frac{1}{A} \rightarrow A^{-1}$ for any 2-tensor $A$ and $e^{(\cdot)}\rightarrow exp_M(\cdot)$, where $exp_M$ is the matrix exponential.}\\
					 If we assume $\vec{F}_{Sto}$ was the only contribution to $\vec{F}$, the solution of equation \eqref{eq:newton} would be \begin{align}
					 	\vec{v}(t)=e^{-\frac{\zeta}{m}t}v_0
					 \end{align} and thus $\vec{v}$ as well as $\langle v^2 \rangle$ would both converge towards zero as $t$ goes to infinity. The latter is a contradiction to the equipartition theorem according to which $\langle v^2 \rangle =d\frac{k_BT}{m}$ should be true as the system goes to equilibrium.\\
					 To fix this deficit we add an additional random Force $\delta\vec{F}(t)$ to our force and obtain the Langevin-equation \begin{align}
					 	m\frac{d\vec{v}}{dt}=-\zeta \vec{v}+\delta \vec{F}(t). \label{eq:langevin}
					 \end{align}
					 Like in Smulochowki's approach this random force can be interpreted as the change in momentum due to collisions with the fluctuating fluid molecules.\\
					Experimental data suggests that on average the force is indeed given by equation \eqref{eq:Stokesforce}. To implement this in our model we demand $\langle \delta \vec{F}(t)\rangle =0$. This can be done without loss of generality, due to the fact that every non zero mean force could be interpreted as external force. We further demand that $\delta \vec{F}(t)$ is not autocorrelated for times greater than the mean time between to collisions of two fluid molecules with the particle $\tau_0$, i.e. $\langle \delta\vec{F}_i(t_1)\delta \vec{F}_j(t_2)\rangle=0$ for all times $t_1,t_2$with$|t_1-t_2|<\tau_0$.\\
					Since the relevant time scale of the system $\tau_r=\frac{m}{\zeta}$ usually exceed $\tau_0$ by multitudes ($\tau_r \gg \tau_0$), we actually assume that $\tau_0 \rightarrow 0$ and thus model $\delta \vec{F}(t)$ as general white noise with \begin{align}
					\label{eq:whitenoiseauto}	
					 \langle\delta\vec{F}_i(t_1)\delta\vec{F}_j(t_2)\rangle= 2B\delta(t_1-t_2)\delta_{ij}.
					\end{align} 
					The last simplification we introduce to $\delta\vec{F}(t)$ is that all higher momenta be zero, i.e. we assume that $\delta\vec{F}(t)$ is Gaussian white noise. Since $\delta\vec{F}(t)$ is the result of a very large amount of independent collisions of the particle with fluid molecules, whose mean square momentum is bounded by the equipartition theorem, this very strong assumption can be interpreted as a consequence of the central limit theorem and is therefor thoroughly justified.  \\
					As of now all that remains to be determined is the constant $B$ in equation \eqref{eq:whitenoiseauto}, which will be achieved by exploitation of the equipartition theorem. The general solution to equation \eqref{eq:langevin} can be obtained by variation of constants and is given by \begin{align}
						\vec{v}(t)=e^{-\frac{\zeta}{m}t}\vec{v}_0+\int_{0}^{t}e^{-\frac{\zeta}{m}(t-t')}\frac{\delta\vec{F}(t')}{m}dt'.
					\end{align}
					Now let $v_p(t)$ be the projection of $\vec{v}(t)$ onto one direction.\footnote{If $\zeta$ is a tensor, we would choose the principal axis of $\zeta$ here. As a consequence $\zeta$ hast to be replaced by its corresponding eigenvalues in the equations \eqref{eq:determB} and \eqref{eq:B} and $B$ depends on the chosen axis.} 
					If $\zeta$ does not depend on the time (or at times scales short enough to assume exactly that), inserting $v_p(t)$ into the equipartition theorem results in \begin{align}
						\label{eq:determB} \frac{k_BT}{m}=\langle \vec{v_p}^2\rangle = e^{-\frac{2\zeta}{m}t}v_p(0)^2 + \frac{B}{\zeta m}\left(1-e^{-\frac{2\zeta}{m}t}\right). 
					\end{align}
					As times goes to infinity the first and the last term in this equation go to zero. Taking this limit and solving for $B$ gives us \begin{align}
						B=\zeta k_BT \label{eq:B}.
					\end{align}
					Equation \eqref{eq:B} is often referred to as fluctuation-dissipation-theorem as it describes the equilibrium between dissipative friction and the thermal noise due to particle fluctuation.
					 \\ As stated at the beginning of this section we can reintroduce external forces simply by replacing $\vec{F}$ by $\vec{F}+\vec{F}_{ext}$. The result is the Langevin equation with external forces:\begin{align}
					 	\label{eq:langevin+ext}
					 		m\frac{d\vec{v}}{dt}=-\zeta \vec{v}+\delta \vec{F}(t)+\vec{F}_{ext}.
					 \end{align}
					 $\vec{F}_{ext}$ can depend on time, external fields and other Brownian particles.\\
					 In cases where $\zeta v \gg m\frac{dv}{dt}$ or $t\gg \frac{m}{\zeta}$ we can simplify this equation further by neglecting the inertia terms. The equation we obtain is the overdamped Langevin equation \begin{align}
					 	\vec{v}(t)= \frac{1}{\zeta}\left(\delta\vec{F}(t)+\vec{F}_{ext}\right). \label{eq:langevinod}
					 \end{align}
					 This overdamped Langevin equation is the foundation of the Brownian Dynamics algorithm used in all our simulations and further described in section \ref{sec:BD}. One can interpret this equation by assuming that due to the high friction the equilibrium velocity is reached within an infinitely small amount of time and hence there is no acceleration. Interestingly the force couples to the velocity here, unlike in standard Newtonian dynamics, where the force applied to a particle couples to the particle's acceleration.\\
					 If a particle has rotation degrees of freedom additional to the movement of his center of mass, these rotation degrees of freedom will also perform Brownian motion. In such a case the collisions with fluid molecules also cause a stochastic changes in angular momentum and thus a torque $\delta \vec{M}(t)$ that, like the stochastic force $\delta \vec{F}(t)$, can be modeled as Gaussian white noise. Exactly analogous to our considerations for the Force we obtain an overdamped Langevin equation:\begin{align}
					 	\frac{d\vec{e}}{dt}=\frac{1}{\zeta^r}\left(\delta\vec{M}(t)+\vec{M}_{ext}\right)\times \vec{e}, \label{eq:langorod}
					 \end{align} 
					 where $\vec{M}_{ext}$ is an external torque, caused by external fields or other particles, $\vec{e}$ is a unit-vector associated with the orientation of the particle and $\zeta^r$ is the friction coefficient corresponding to the rotations, which like before can be tensor-valued. \cite{Zwanzig} \\
					 Since we will usually talk about diffusion constants rather than friction coefficients from here on, we will shortly introduce the connection of the two. By calculating the mean square displacement via the diffusion equation \eqref{eq:diffusion} and via the Langevin equation \eqref{eq:langevin} and comparing the results one can see the relation \begin{align}
					 	D=k_BT\zeta^{-1}.
					 \end{align}
					 Even  though both can be 2-tensors, it will be sufficient to talk about their eigenvalues along principal axes from here on. \\
					 Throughout this chapter, that the time evolution of the orientation decouples from the time evolution of the center of mass. These assumptions are not valid in general. However it was shown in \cite{winkelent}, that the assumption is valid for all particles with orthogonal planes of mirror symmetry through their center of mass. Since all particles discussed in this work fall into this category the use of these assumptions was legitimate.

					  \subsection{Storage and loss shear modulus\label{sec:mod}}
					  As the properties of many materials, in particular gels, do highly depend on their flow properties, but also on their structural properties, methods are needed to measure the first without destroying the latter.  One possible way to do so is to apply small amplitude oscillatory shear to the probe. In experiments this is usually done in a cone-plate or a Couette geometry, where the cone/plane oscillates sinusoidally $\Omega (t)=\Omega_0\cos( \omega t)			$. This causes a shear rate $\dot{\gamma}(t)\propto \cos(\omega t)$ and  a strain $\gamma(t)\propto \frac{1}{\omega}\sin(\omega t)$ that both are also sinusoidal. Consequentially the stress response of the system can also be represented by a sinus curve. \\ For a solid body this stress response is according to Hooke's law directly proportional to the strain, i.e. $\sigma(t)\propto \sin(\omega t)$. The deformation is immediate. This behavior is called ideal elasticity.\\
					  An ideal fluid on the other hand reacts with an delay. Its internal stress is proportional to the shear rate $\sigma(t)\propto \cos(\omega t)$. The proportionality constant between $\sigma$ and $\dot{\gamma}$ is the dynamic viscosity.\\  
					  In general materials are neither ideal solid bodies nor ideal fluids. In this cases the stress, while still  sinusoidal with frequency $\omega$, does not coincide with neither the strain nor the shear rate, but is given by the following superposition \begin{align}\label{eq:moduli}
					  \sigma(t)=\gamma_0\left[ G'(\omega)\sin(\omega t)+G''(\omega)\cos(\omega t)\right],
					  \end{align}
					  where $\gamma_0$ is the maximal amplitude of the oscillatory strain. The term $G'(\omega)$ is in phase with the strain and is called the storage modulus. It is proportional to the elastic energy stored during the process of shearing. The out of phase term $G''(\omega)$ is called the loss modulus and represents elastic energy dissipated through the system. Their ratio $G''/G'=\tan(\delta)$, called the loss tangent, is low ($\ll1$) for solid like materials and large ($\gg1$) for liquid-like fluids. It can further be stated, that in solid-like materials $G'$ and $G''$ are almost independent of the shear frequency, while in liquid like materials $G'$ is proportional to $\omega^2$, while $G''$ is proportional to $\omega$. \\
					  We should note, that equation \eqref{eq:moduli} of course only makes sense as long as the stress depends linearly on the amplitude $\gamma_0$. This parameter region is called the linear viscoelastic regime. In this regime $G'$ and $G''$ should be independent of the maximum strain $\gamma_0$.
					  
					 \cite{Rheologiebuch}
					  
					 \subsection{Introduction to Gels\label{sec:gel}}
					 Jellies from cooked down meat or from fruit juice cooked with sugar are solid-like dispersion and have been part of our households for centuries. Chemists knew as far back as the 19th century of several other systems capable of forming organic and inorganic jellies. Thomas Graham studied the replacement of water with other liquids in jellies of silicic acid and coined the term gel.\\
					 For technical applications gels play a particular important role in pharmaceutics, food industry and many parts of engineering. The extraordinary properties of so called aerogels also increasingly gain significance in many branches.\\
					 It is due to the reason, that it is much easier to identify a gel than to define what a gel actually is, that many different definitions of the term gel are around. As theoretical approaches cover the hole scope of what might be practically considered a gel we chose a phenomenological approach.\cite{gelpaper}\\
					 Many definitions agree that a gel consists of a coherent dispersed percolating system  or network of at least two phases, of which at least one is liquid.
					 It is further commonly established that gels, unlike e.g. Newtonian fluids, can carry their own weight long enough for humans to observe. While not completely solid they have solid-like properties.\\ Described in terms of storage and stress moduli, we expect $\tan(\delta)\ll 1$ on time scales accessible for the human eye. On that same scale $G'$ should be close to independent of $\omega$. \\
					 Summa summarum a gel in the further context of this work will be defined by the following properties: \begin{enumerate}
					 	\item formation of percolating networks of two or more types of particles, of which at least one is fluid. 
					 	\item existence of an extended plateau in $G'$ (independence of $\omega$)
					 	\item solid-like behavior $\tan(\delta)\ll 1$ on the same interval as $G'$ exhibits a plateau.\\
					 	\cite{gelpaper}\cite{Rheologiebuch}
					 	 
					 \end{enumerate}

					 \subsection{Jeffery orbits}
\begin{figure}
					  	\centering
				\input{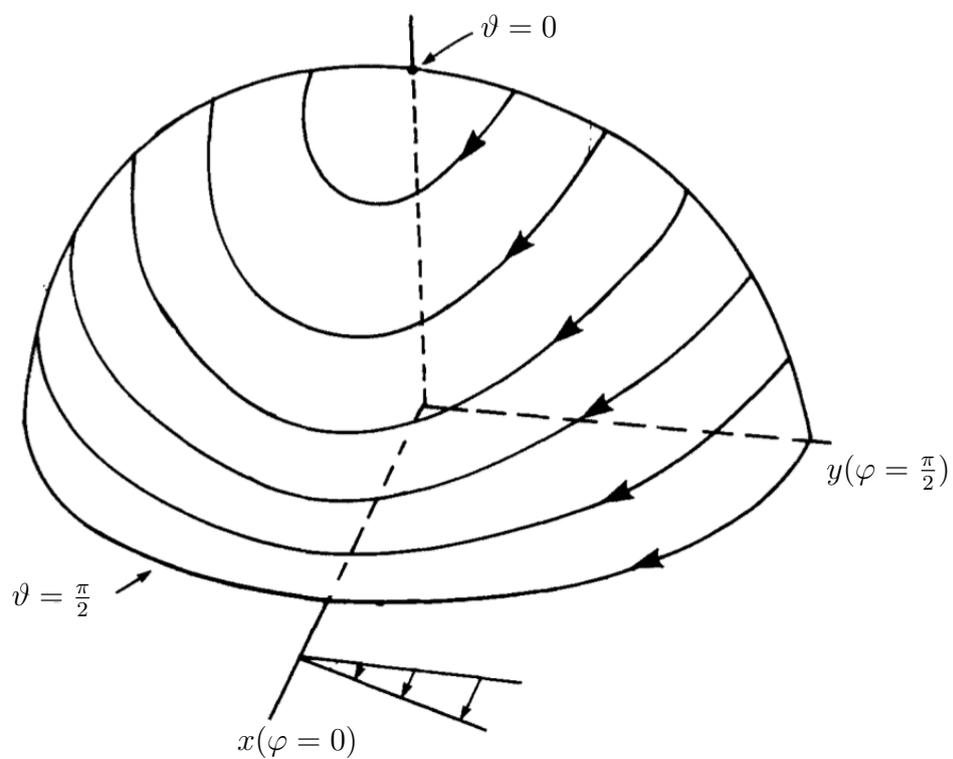}
				\caption{Depiction of Jeffery-orbits for prolate spheroids, ellipsoids, spherocylinders. The particles' center of mass is situated at the origin of the coordinate system. Its orientation is completely defined by the angles $\phi$ and $\theta$. \cite{Jeffery2}}
				\label{fig:Jeffery}
				\end{figure}

					 If a voluminous sphere is suspended in a shear flow, the flow on one side of the sphere is faster than on the other, leading to a rotation of the sphere. Since all our simulations ignore the orientation of spheres due to their symmetry, we also ignore this effect. \\
					 However lacking one of the rotational symmetries of a sphere spherocylinders actually start a precession movement due to this effect, which is why it can not be ignored so easily. This precession causes spherocylinders with fixed center of mass onto so call closed \textbf{Jeffery orbits}. These trajectories are depicted in figure \ref{fig:Jeffery}. \cite{Jeffery1}
					 Jeffery figured out formulas for these orbits for rotation ellipsoids surrounded by fluids with $Re \le 1$ in 1922. If the unique half axis of such an rotation ellipsoid has length $a$ and its orientation is described in the spherical coordinates $\varphi$ and $\vartheta$  the Jeffery orbits are completely determined by the ratio of the half axis $r_e=\frac{a}{b}$, where $b$ is the length of the two non-unique half-axis, the shear rate $\dot{\gamma}$ and the initial orientation of the ellipsoid. In formulas the orbits are given by \begin{align}
					 	\phi(t)=&\arctan\left( r_e \tan\left(\frac{t-t_0}{T}\right) \right) \nonumber\\ \label{eq:Jeff} \\
					 	\theta(t)=&\arctan\left(C(t_0)  \left(r_e^2\cos^2(\phi(t))+\sin^2(\phi(t))  \right)^{-\frac{1}{2}}\right), \nonumber	 	
					 \end{align}
					 where $C(t)=\tan(\theta(t))\sqrt{r_e^2\cos^2(\phi(t)+\sin^2(\phi(t))}$ and $T=\frac{1}{\dot{\gamma}}\left(r_e+\frac{1}{r_e}\right)$.	 \cite{Jeffery1}\cite{Jeffery2}
			
				In Jeffery's original paper another  equation describing the rotation of the ellipsoid around its symmetry axis is added to the equations \eqref{eq:Jeff}. Since our simulation does not keep track of the orientation around the symmetry axis we ignore it completely.\\
				So far we discussed Jeffery orbits only for rotation ellipsoids. To expand them also to spherocylinders we realize, that we can write $r_e=\frac{a}{b}$ in terms of the main moments of inertia $J_{max}$ and $J_{min}$ of the ellipsoid. It is then given by $r_e=\sqrt{2\frac{J_{max}}{J_{min}}-1}$. \\
				Writing $r_e$ like this one can show, that spherocylinders in a shear field also obey the equations \eqref{eq:Jeff}. This is quite plausible due the fact that spherocylinders have the same symmetries as rotations ellipsoids and the geometric differences between the two are only marginal, however it was also shown analytically by Bretherton in 1962.\cite{bretherton} \\
				In the simulations we have integrated Jeffery orbits with an Euler-integrations of the equations \eqref{eq:Jeff}. As comparison we also implemented shear flow, where the spherocylinders are treated solely as line segments, that do not perform Jeffery orbits.

					 \section{Computational Methods}
					 In this chapter we will introduce and discuss all computational and mathematical methods necessary to understand the simulations, whose results are presented over the course of this work, and the evaluation of said results. We will start off with an introduction to reduced units, right before the Brownian dynamics algorithm is introduced in section \ref{sec:BD}. After that we will give an overview of the different boundary conditions, interactions and forces and their implementation in the simulation program. Finally we will have a look some of our evaluation tools, namely the cluster-algorithm, radial and angular distribution functions and topological analysis of simplicial complexes.
					 \subsection{Reduced Units}
					 It is common practice in numerical simulations to describe systems in reduced units rather than SI- or natural units. This is usually done by picking as many independent parameters as necessary to describe a given system and normalizing the units these parameters are measured in to scales characteristic to this systems. All other units are then expressed in terms of these normalized units, usually by keeping up the relations between units known from SI-units.  The reduced unit of velocity of a system would for example be given as quotient of the reduced unit of length and the reduced unit of time.  
					  Independent in this context means that neither unit of the chosen parameters can be expressed in terms of the others. For example in a system characterized by time, length and velocity, only two of the three units are independent, because either of the three can be expressed using the other two. This means if we were to set the unit of length to be equal to the size of the system and the unit time to be the mean time it takes for a particle to move past this length, the unit of velocity would already be determined by the mean velocity of a particle in that system.\\This procedure ensures, that all physical law are invariant under the choice of units.\\
					  In this work and all our simulations we have chosen the diameter of a colloidal sphere $\sigma_k$, its friction coefficient $\zeta$ and $k_bT$ as fundamental parameters. In reduced units all three of them will be normalized to 1. All further units can be derived from them. The conversion formulas can be found in table \ref{tab:reduni}. Reduced units are furnished with an $\ast$.
					  \begin{table}[h]
					  	\begin{centering}
					  		
					  		\begin{tabular}{lp{3cm}p{5cm}}
					  		\hline\hline 	Physical variable & & Conversion rule \\ \hline
					  			\hline Diameter of a unit-sphere & & $\sigma_k^\ast=1$ \\
					  			Temperature & & $k_BT^\ast=1$ \\
					  			Friction coefficient of a unit-sphere & & $\zeta^\ast=1$ \\
					  			\hline
					  			Diffusion constant & & $D^\ast = \frac{\zeta}{k_BT}D$ \\
					  			Time & & $t^\ast = \frac{D}{\sigma_k^2}t$ \\
					  			Length & & $l^\ast= \frac{1}{\sigma_k}l$ \\ 
					  			Energy & & $E^\ast = \frac{1}{k_bT}E$ \\
					  			Force & & $F^\ast = \frac{\sigma_k}{k_BT}F$ \\
					  			Velocity & & $\vec{v}^\ast=\frac{\sigma_k}{D}\vec{v}$ \\
					  			Density & & $\rho^\ast= \sigma_k^3\rho$ \\
					\hline  		\hline 	
					  	\end{tabular} 
					  	\caption{Conversion table from non-reduced units to reduced units for the physical variables relevant to this work. Reduced units are furnished with an $\ast$.}
					  	\label{tab:reduni}
					  \end{centering}
					\end{table}
					The use of reduced units has become very common due to the fact, that a lot of different physical systems look essentially alike, when these units are used, a large system with big particles might be just a rescaled version of a smaller system. Due to this fact a single simulation could possibly describe many different systems. \\
					A further reason is of purely computational interest. Because reduced fundamental units are usually chosen to normalize characteristic scales of the system, most sizes measured in these units do not stray very far from 1. In fact most values deviate by only  2-3 orders of magnitude at most, which drastically reduces numerical errors, that occur when handling very big or very small numbers.\cite{Frenkel}\cite{Simbuchofliquids}\\
					From here on we will only use the reduced units from table \ref{tab:reduni}. For the sake of better readability  however we will drop the $\ast$.  
				
					 \subsection{Brownian Dynamics\label{sec:BD}}
					 The algorithm at the core of all simulations in this work is the so called \textbf{Brownian Dynamics} algorithm. It simulates the dynamic and time evolution of multi particle systems governed by the overdamped Langevin equation \eqref{eq:langevinod}. This is achieved by performing an Euler-Maruyama integration of this equation for every particle, which will be explained in this section. For better understanding, we will shortly and very vaguely explain how one could derive such an integrator.\footnote{Note that this has to be understood  more like a way to make the algorithm plausible, rather than a rigorous proof. The layout of argumentation is based on Anton Lüders Bachelor thesis. \cite{BAnton}}\\
					 For that numerical integration the time interval of interest is discretized and hence described by finitely many time steps $t_0,...,t_N$, where N is the number of time steps calculated. We will denote the distance between two consecutive time steps by $\Delta t$.(Theoretically $\Delta t$ can depend on the time steps we are looking at.)\\
					 We can now obtain the position of the $i-$th particle at the $j-$th time step by integrating equation \eqref{eq:langevinod}: \begin{align}
					 	\vec{r}_i(t_j)=\vec{r}_i(t_{j-1})+ \frac{1}{\zeta}\int\limits_{t_{j-1}}^{t_j} \left(\vec{F}(t')+\delta \vec{F}(t')\right)dt'.
					 \end{align}
					 The force $\vec{F}$, which in \eqref{eq:langevinod} was called $\vec{F}_{ext}$, includes all particle interactions as well as external forces. It will usually depend on the position of all other particles.\\
					 The first term under the integral is quite well-behaved and can numerically be solved with any integrator of choice. In Brownian dynamics an Euler-integrator is chosen, i.e. we approximate that term by $\int\limits_{t_{j-1}}^{t_j}\vec{F}_{ext}(t')dt'\approx \vec{F}_{ext}(t_{j-1})\Delta t$. 	The second term under the integral however is a bit more difficult due to $\delta\vec{F}$ being a random distribution. However since we know that $\delta \vec{F}(t)$ is normal distributed and uncorrelated we can  interpret this integral as an average over independent normal distributions. It is not a very hard task to show that such an average again results in a normal distribution, whose mean and variance are also given by the respective averages. Since we know the mean and variance of $\delta\vec{F}$ we can now easily figure out, that the mean of the second integral has to be zero, while its variance is given by \begin{align}
					 	\left\langle \left(  \frac{1}{\zeta ^2} \int\limits_{t_{j-1}}^{t_j}\delta\vec{F}^k(t')dt'\right)^2 \right\rangle= \frac{1}{\zeta^2}\int\limits_{t_{j-1}}^{t_j}dt'\int\limits_{t_{j-1}}^{t_j}\langle \delta {F}^k(t')\delta {F}^k(t'')\rangle dt''=2D\Delta t.
					 \end{align}
					 We used $k$ here to enumerate the components of $\delta\vec{F}$. (Note that for non isotropic particles $D$ can depend on $k$.)\\
					 Numerically this is implemented by replacing the second integral by a standard random vector $\vec{R}$ rescaled with the factor $\sqrt{2D\Delta t}$. Hence our Brownian dynamics integrator is given by
					 \begin{align}
					 \vec{r}_i(t_j)=\vec{r}_i(t_{j-1})+ \frac{D}{k_bT}\vec{F}(t_{j-1})\Delta t+\sqrt{2D\Delta t}\vec{R}. \label{eq:BDalg}.
					 \end{align}
					 We also have replaced $\frac{1}{\zeta}$ by $\frac{D}{k_BT}$ in the last step.
					 The random numbers can be generated with the Box-Muller method from equally distributed (pseudo-)random numbers.\\
					 If the simulated particles are anisotropic, like the spherocylinders used in this work, and therefore $D$ is given by a tensor, we solve equation \eqref{eq:langevinod} in the basis  given by the principal axes of $D$ and, in this basis, also obtain the integrator  apart from the fact, in every component of equation \eqref{eq:BDalg} $D$ is replaced by its corresponding eigenvalue. \\
					 For the a spherocylinder the principal axes of $D$ coincide with its rotation-axis $\vec{e}_\parallel$  and two more arbitrary vectors $\vec{e}_{\bot,1}$ and $\vec{e}_{\bot,2}$, that form an orthonormal basis with $\vec{e}_i$. We will call the two different eigenvalues of $D$  $D_\parallel$ and $D_\bot$. Then the integrator for $\vec{r}_i=(r_\parallel, r_{\bot,1},r_{\bot,2})$ can be expressed like 
					  \begin{align}
					  {r}_\parallel(t_j)&={r}_\parallel(t_{j-1})+ \frac{D_\parallel}{k_BT}{F}_\parallel(t_{j-1})\Delta t+\sqrt{2D_\parallel\Delta t}{R_\parallel} \\
					   {r}_{\bot,k}(t_j)&={r}_{\bot,k}(t_{j-1})+ \frac{D_{\bot}}{k_BT}{F}_{\bot,k}(t_{j-1})\Delta t+\sqrt{2D_\bot\Delta t}{R}_{\bot,k} & {k \in \left\{1,2\right\}}.
					  \end{align}
					  This is the standard integrator of a Brownian dynamics algorithm. \cite{Emrak}\cite{Loewen}\\
					  Starting from the overdamped Langevin equation for the orientation of an anisotropic particle \eqref{eq:langorod} we obtain a very similar integrator, which for the case of a spherocylinder then can be expressed as \begin{align}
					  	\vec{e}_\parallel(t_j)=\vec{e}_\parallel(t_{j-1})+\frac{D_r}{k_BT}\vec{M}(t_{j-1})\times \vec{e}_\parallel(t_{j-1})\Delta t+ \sqrt{2D^r\Delta t}\left(R_{\bot,1}\vec{e}_{\bot,1}+R_{\bot,2}\vec{e}_{\bot,2}  \right),
					  \end{align}
					  where $D^r$ is the rotation diffusion constant for spherocylinders, $\vec{M}$ is the torque in the spherocylinder and $R_{\bot,k}$ are standard normal distributed random numbers. \\
					  The various diffusion constants, that occurred in this section, and that are relevant in our simulations depend only on the surrounding fluid and the aspect ratio of the spherocylinders.\footnote{This is of course only true in reduced units} Approximate values for these constants can be calculated using the formulas in table \ref{tab:diffconst}.
					  
					  \begin{table}[h]
					  	\begin{centering}
					  		
					  		\begin{tabular}{lcp{6cm}}
					  			\hline\hline 	Colloid & Diffusion constant & Formula \\ \hline\hline
					  			Sphere  & D & $3\pi \eta \sigma_k$ \vspace{0.1cm}\\
					  			(Auxiliary constant)& $D_0$ & $\frac{k_BT}{\eta L}$ \\
					  			Spherocylinder(orthogonal) & $D_\bot$ & $\frac{D_0}{2\pi}\left(\ln(p)+0.839+\frac{0.185}{p}+\frac{0.233}{p^2}\right)$ \\ Spherocylinder(parallel) & $D_\parallel$ & $\frac{D_0}{2\pi}\left(\ln(p)-0.207+\frac{0.980}{p}+\frac{0.133}{p^2}\right)$ \\
					  			Spherocylinder(rotation) & $D_r$ & $\frac{3D_0}{\pi L^2}\left(\ln(p)-0.662+\frac{0.917}{p}+\frac{0.05}{p^2}\right)$\\ \hline\hline
					  		\end{tabular} 
					  		\caption{Approximate formulas for the diffusion constants for spheres and spherocylinders, as derived in \cite{Loewen}. The Diffusion constants depend on the diameter $\sigma_K$ of a sphere, the diameter  $\sigma_s$ of a rod, the aspect ratio $p=\frac{L}{\sigma_s}$ of a rod and the viscosity $\eta$ of the dispersion medium. Comparison with experimental data has shown that these formulas are valid for $p\in \left[2,30\right]$. \cite{Stabdiffusionspaper}}
					  		\label{tab:diffconst}
					  	\end{centering}
					  \end{table}
					  \subsection{Periodic Boundary Conditions}
\begin{figure}
	\centering
	\includegraphics[width=0.7\linewidth]{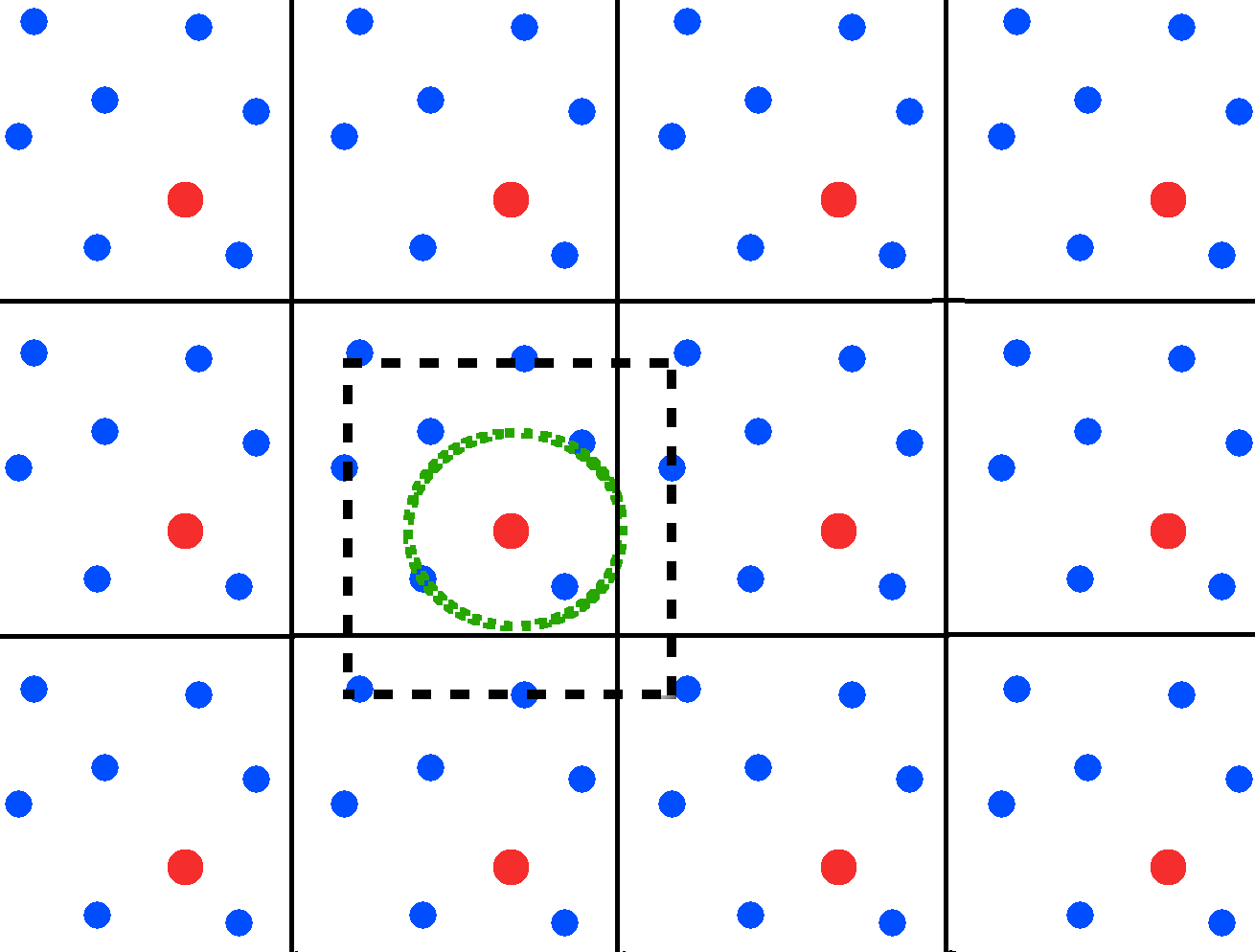}
	\caption{Illustration of periodic boundary conditions, the nearest image convention (dashed box) and a cut-off radius (green circle)}
	\label{fig:bc}
	
	\end{figure}

					  Even though the calculation force of modern computers has made impressive advances over the last few decades, their capability to simulate real word systems is still very limited, due to the shear amount of degrees of freedom in such systems. In order to still simulate system sizes comparable to the thermodynamic limit boundary conditions that mimic infinitely big systems need to be introduced. A common choice are \textbf{periodic boundary conditions}. \cite{Frenkel}\\
					  The basic idea behind periodic boundary conditions is to simulate an infinitely big space by plastering it periodically with copies of a smaller, computationally feasible simulation box, as depicted in figure \ref{fig:bc}. Thus for every particle in the simulation box there are infinitely many copies of itself in the copies of the simulation box, that cover the real space.
					  Numerically this is implemented by identifying opposite boundaries of the simulation box. A particle leaving the simulation box to the right, re-enters the box from the left at the same instant, exactly like in the real space it would leave one copy of the box to the right and enter the next copy from the left.	Periodic boundary conditions also make the adaption of the distance of particles a necessity. Most commonly in this setting the distance of two particles is taken to be the shortest real space distance between any two copies of  these particles. As a consequence no distance (in one spacial dimension ) can exceed half the box length. \\
					  With the adaption of the distance, there obviously needs to be a coherent modification in the calculation of particle interactions, which usually depend on the distance.\\
					  A naive, even though possibly complicated, approach would be to write the interaction forces as infinite series, that actually considers the interaction with all copies of the particles. However, not only is this numerically often not feasible, the series do not necessarily converge for a lot of common forces, such as for example gravity or the coulomb force. A more practical approach is the nearest image convention. If this convention is applied, the interactions are only calculated for a particle and the closest copy of every other particle. A similar idea is to introduce  of a cut-off radius $r_c$ and then simply ignore the interaction between particles, which are further apart than $r_c$. The closest image convention is a significant improvements in terms of simulations times, even more through the cut-off radius, but also come with increasing numerical errors. These errors are sufficiently small though for the short-range interactions that occur in our simulations, which is why both method were applied in this work. \cite{Frenkel}\cite{Simbuchofliquids}
					   What short-range in this context means and some examples of short-range interactions will be discussed in section \ref{sex:interac}.
					  
					   \subsubsection{Lees-Edwards boundary conditions}
\begin{figure}
					\centering
					(a)\includegraphics[height=0.4\linewidth]{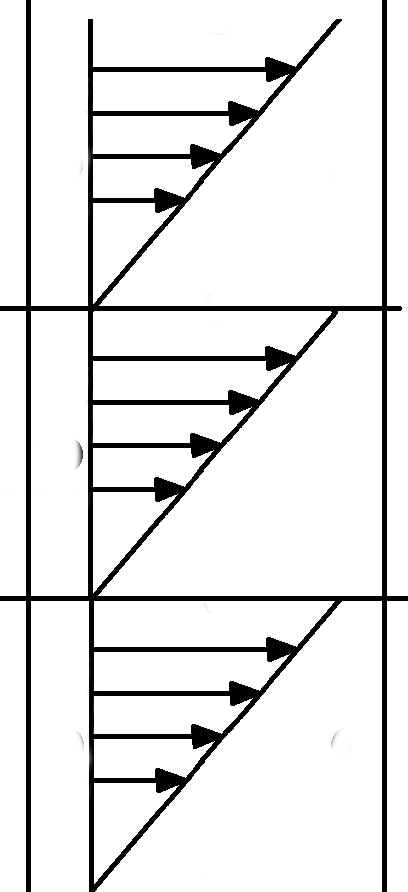}\hspace{1cm}
					(b)\includegraphics[height=0.4\linewidth]{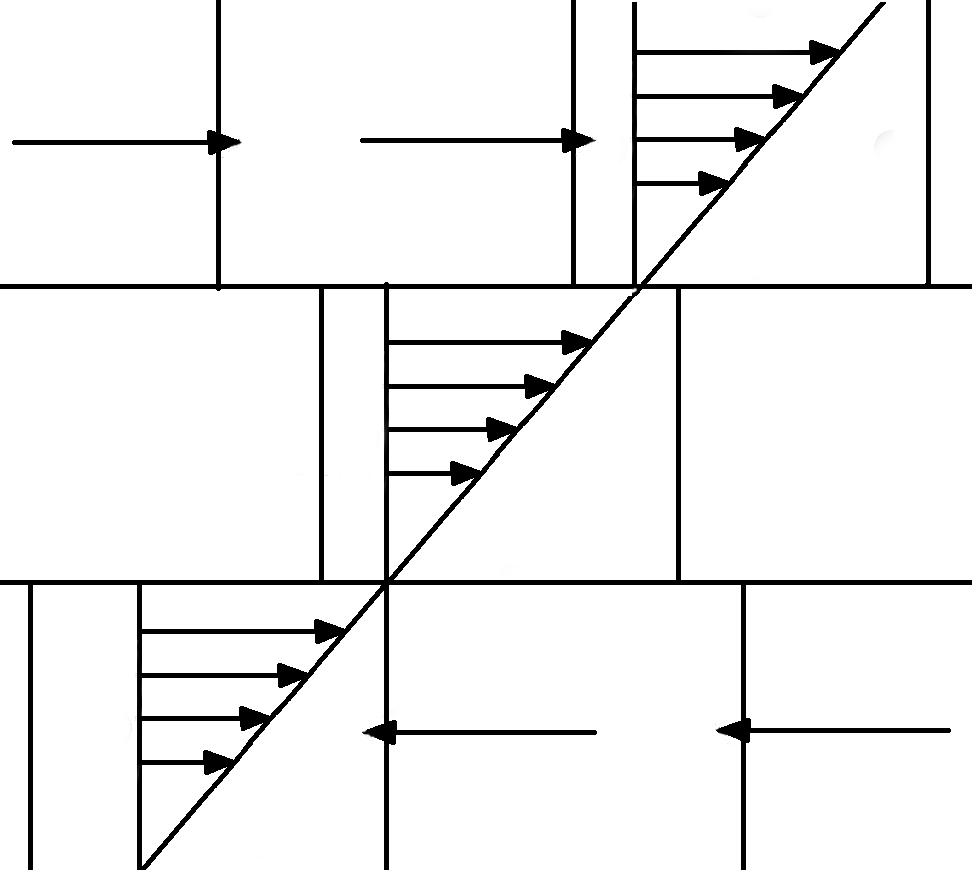}
					\caption{Illustration of Couette flow with standard periodic boundary conditions (a) and Lees-Edwards boundary conditions (b).}
					\label{fig:LE}
					\end{figure}

					   \textbf{Lees-Edwards boundary conditions} are an extension to standard periodic boundary conditions in order to account for laminar flow, as it might for example be induced, if the system is under shear stress. \cite{Lees}\cite{Simbuchofliquids}\\
					   Let us for example imagine a simulation box, exposed to a Couette flow, such that a fluid particle at the bottom of the box might remain still, while a particle at the top of the box moves with a velocity of $\vec{u}$. If we applied standard periodic boundary conditions to this system, a particle just below the top of the box would move with a velocity of almost $\vec{u}$, while a particle only slightly higher, at the bottom of the first copy of the simulation box would hardly move at all. Viewed in the real space the flow profile in this scenario looks like a sawing blade rather than the profile of a Couette flow, as is illustrated in figure \ref{fig:LE} (a). Lees-Edwards boundary conditions fix this nuisance by giving the copies of the simulation box a velocity equal to $k\vec{u}$, where $k$ is a whole number, that depends on the distance of the copy to the simulation box.\footnote{Only the distance orthogonal to planes of constant flow velocity need to be considered to determine $k$.} \cite{Simbuchofliquids} An illustration of the moving copies of the simulation box, as well as the resulting Couette-flow profile is depicted in figure \ref{fig:LE}(b). In our simulations Lees-Edwards boundary conditions were applied, whenever shear stress was applied to the system.
					
					 \subsection{Forces and Interactions \label{sex:interac}}
					 In this section we will discuss all the forces particles were exposed to, during simulations related to this work. All relevant forces in this context are either due to particle-particle interaction or due to a periodic shear flow. We will discuss both cases separately. 
					 \subsubsection{Particle-Particle Interactions} 
					 All the particle-particle interactions simulated in this work can be derived from a pair potential function $V(r)$, that only depends on $r=|r_i-r_j|$ the distance of the two particles in question.
					 In order to use periodic boundary conditions consistently, all these interactions need to be short ranged, which means $\int\limits_{\mathbb{R}^3\setminus B_r(0)} V(|r|) d\vec{r}  < \infty $ for an $r>0$. $B_r(0)$ is here an open Ball of radius $r$ with center at the origin. The fact that this integral is finite implies that the numerical error caused by the closest image convention and the introduction of a cut-off radius are negligibly small, if the box, or respectively the cut-off radius, are sufficiently big. \\
					 The basis for all pair potentials used here is the Lennard-Jones potential with a cut-off radius $r_c=2.5$. It is given by 
					 \begin{align}
					 V_{LJ}(r)=\left\{  
					 \begin{matrix} 
					 4\epsilon\left( \left(\frac{\sigma}{r}\right)^{12}-\left(\frac{\sigma}{r}\right)^6\right) & \text{for } r\le 2.5\sigma\\ 0 & \text{for } r > 2.5\sigma\end{matrix} \right. . \label{eq:LJ}
					 \end{align} 
					 This potential has an repulsive part ($r<2^{1/6}\sigma$), that prevents particle overlap, as well as an attractive part ($ 2^{1/6}\sigma <r < 2.5 \sigma$). Particles interacting through such a potential can be interpreted as attractive spheres with a radius of approximately $\sigma$. The spheres are not completely hard, yet an overlap decreasing the distance far below $\sigma$ is rater rare. The possible overlap and the power of the attraction is proportional to $\epsilon$. \\
					 From the Lennard-Jones potential we derive another potential, a hard- core Lennard-Jones potential \cite{kugelgel}, for sticky spheres by combining a hard-core potential with the Lennard-Jones potential. The hard-core potential is used for radii smaller than an inner radius $r_0$. For radii between $r_0$ and $r_0+2.5$ a standard Lennard-Jones potential is used. The resulting potential is given by:
					 \begin{align}
				 V_{HCLJ}(r)=\left\{  \label{eq:hclj}
				 \begin{matrix} \infty & \text{for } &r< {r_0}\\
				 V_{LJ}\left({r-r_0}\right) & \text{for }& {r_0}\le r\le {2.5+r_0}\\ 0 & \text{for }& r > {2.5+r_0}\end{matrix} \right., \end{align} 
				
					 and can be interpreted as spheres with a hard core, covered in a soft sticky shell. Compared to the hard core radius the region allowing overlap, as well as the attractive region of the Lennard-Jones potential have become much smaller, hence the picture of a sticky spheres becomes more accurate then that of  attractive spheres.  We interpret the radius of spheres interacting with this potential to be $r_0+1$, which is the distance where the potential is zero. This makes the unit of length in our simulations $\sigma=r_0+1$ as well.\\
					 The last potential we will need during this work is a Kihara-Lennard-Jones potential  for spherocylinders. \cite{RevModPhys.25.831} The potential itself looks exactly like the Lennard-Jones potential introduced in equation \eqref{eq:LJ}. Instead of inserting the center of mass distance $r$ of spherocylinders, as we did for spheres, we will insert the shortest distance between the surface of spherocylinders $d$: \begin{align}
					 	V_{KLJ}(d)=V_{LJ}(d). \label{eq:KLJ}
					 \end{align}
					   Since the calculation of the distance between the surface of two spherocylinders can be quite involved a short discussion of the topic can be found in the appendix in section \ref{App:distspher}.\\
					   Due to the fact that this interactions acts along the line of shortest distance and that the cut off radius is short compared to the aspect ratios of spherocylinders, we also interpret spherocylinders with this pair potential as sticky rods, rather than attractive ones.\\
					   The hard core Lennard-Jones potential could of course be used as basis for a Kihara-like potential as well.
					 \subsubsection{Small amplitude oscillatory shear \\ (Determination of storage and loss modulus) \label{sec:SAOS}}
					 In order to determine the storage and loss modulus of a system of colloids we need to apply small amplitude oscillatory shear. We do this by applying an external force in $x$-direction linearly depending on the distance in y direction to the middle of the box   :\begin{align}
					 	F_{saos}=\zeta \dot{\gamma}(t)(y-\bar{y})\vec{e}_x.
					 \end{align}
					$\bar{y}$ is here the $y$-coordinate of the middle of the simulation box and $\gamma=\gamma_0\sin(\omega t)$ the periodic strain with maximum amplitude $\gamma_0$ and frequency $\omega$. Together with Lees-Edwards boundary conditions this force leads to a strain, as it would be caused by a periodically changing Couette-flow.\cite{Heptner}\\
					To determine the storage and loss modulus, we have a look at the $xy$-components of the virial stress:
					\begin{align}
						\sigma_{xy}=\frac{-1}{2N}\sum\limits_{k,l=1}^N (x^k-x^l)\vec{F}_y^{kl},
					\end{align}
					where $x^j$ is the $x$-coordinate of the $j$-th particle and $\vec{F}_y^{ij}$ is the $y$-component of the particle-particle interaction between the $i$-th and the $j$th particle. (Since we assume to be in an overdamped system any further velocity dependent terms in the Virial tensor can be neglected.)\cite{stresstens}\\
					Since we expect a periodic response to the periodic shear the resulting stress data will be fitted using a sinusoidal function \begin{align}
						\sigma(t)=\sigma_0\sin(\omega t+ \delta). 
					\end{align}
					Comparing this function with the equations \eqref{eq:moduli} we can see quickly (trough application of the addition theorems for trigonometric functions) that the storage and loss modulus is respectively given by \begin{align}
						G'(\omega)= \frac{\sigma_0}{\gamma_0}\cos(\delta)\nonumber \\ \\
						G''(\omega)=\frac{\sigma_0}{\gamma_0}\sin(\delta). \nonumber
					\end{align}
					 \subsection{Cluster algorithm}
					 	 The percolating networks, which gels consist of, consist of connected \textbf{clusters} of particles. While, of course, not every bigger cluster is such percolating network, we can definitely rule out the existence of such a network if there are no big clusters to be found. In this way clusters of a certain size are a necessary, yet not sufficient, condition for the formation of gels. \\
					 	 As one indicator for percolating networks the size of clusters, and in particular the mean cluster size $\bar{c}$ and the maximum cluster size $c_{max}$, were central observables in this work. \\
					 	 For the course of the work we define clusters in the following way: \\\\ \textit{We call two colloids $a$ and $b$ \textbf{connected} if their distance is smaller than $1.5\sigma$, \\i.e. $d(a,b)<1.5\sigma$ or if other colloids $c_1,...,c_{n}$  $(n\in \mathbb{N}\cup \{\infty\})$ exist such, that the distance between $a$ and $c_1$, the distance between $c_i$ and $c_{i+1}$ and the distance between $c_n$ and $b$ are each smaller than $1.5\sigma$, i.e. $d(a,c_1)<1.5\sigma,d(c_i,c_i+1)<1.5\sigma$ and $d(c_n,b)<1.5\sigma$. The connected components in this definition of connectednessed are called \textbf{clusters}.} \\\\ 
					 	 The algorithm we use to determine the clusters in our systems works as follows. At first a random particle $a$ is chosen, marked as " sorted " and added to a new cluster list. Then all particles within a distance of $1.5\sigma$ or less are also marked as " sorted " and then appended to the same cluster list as $a$. We then move on to ne next particle $b$ in this cluster list and mark and  append all unmarked particles, that are closer to $b$ than $1.5\sigma$. We keep going to the next particle on the list, adding and marking as " sorted " unmarked particle within cluster distance, till the end of the list is reached. At this point this list contains all the particles of one cluster. \\
					 	 After one cluster list is finished, a new one is created, starting with an unmarked colloid, if there are any. The whole process is repeated until all particles are marked as " sorted " and thus belong to a cluster. \cite{Haun} \\
					 	 The mean cluster size $\bar{c}$ can now be calculated as the average length of the cluster lists, while the maximum cluster size $c_{max}$ corresponds to the maximum length of all cluster lists.
					 \subsection{Radial distribution and angular correlation}
					 To get further insight into the structure of the clusters in our system we calculated  the \textbf{radial distribution} function $g(r)$, which is defined as \begin{align}
					 	g(r)=\frac{1}{\rho N} \sum\limits_{i,j=1; i\neq j}^N \delta(r-|r_{ij}|)\label{eq:distdis}.
					 \end{align} 
					 In this definition $\rho=\frac{N}{V}$ is the particle density in the simulation box, and $r_{ij}$ is the distance vector between particle $i$ and particle $j$.
					 To make sense of this numerically, we counted the number of particles with distance $r+dr$ to a particle $i$ normalized the result with respect to the volume of the shell around that particle with radius $r+dr$ and then averaged over all particles. Or as formula \begin{align}
					 	g(r,dr)=\frac{1}{\rho N 2}\sum\limits_{i}^N \frac{N_k^i}{4\pi k^2dr^3} &   \text{ for }  (k-1)dr\le r <(k)dr, \label{eq:disthist}
					 \end{align}
					 where $N_k^i$ is the number of particles within a shell of inner radius $(k-1)dr$ and outer radius $kdr$ around the particle $i$. Thus the term $4\pi (k^2 dr^2)dr$ is the volume of this shell (for small $dr$). This spherical shell we are looking at here is the $k$-th shell from the center. The additional factor $2$ in the first term is needed because otherwise we would count every particle twice in equation \eqref{eq:disthist} compared to equation \eqref{eq:distdis}.  \\
					 For spherocylinders we used the distance of the centers of mass to calculate the radial distribution. Since, other than spheres, spherocylinders are anisotropic we further calculated the \textbf{angular correlation} function for those systems. It is given by\begin{align}
					 	g_a(r)=\frac{1}{N\rho g(r)}\sum\limits_{i,j=1;i\neq j}^N P_2(\cos(\vartheta_{ij}))\delta(r-|r_{ij}|) \label{eq:angdis},
					 \end{align}
					 where $\vartheta_{ij}$ is the enclosed angle between the spherocylinders $i$ and $j$ and $P_2(x)=\frac{3}{2}x-\frac{1}{2}$ is the second Legendre-polynomial. 
					 Numerically this is calculated via \begin{align}
					 	g_a(r,dr)=\frac{1}{N_k}\sum\limits_{i,j=1;i\neq j}^N {P_2(\cos(\vartheta_{ij}))}\mathbb{1}_{\left.\left[(k-1)dr,kdr\right.\right)}(r) &   \text{ for }  (k-1)dr\le r <(k)dr, \label{eq:anghist}
					 \end{align} where $N_k=\frac{1}{2}\sum\limits_{i=1}^N N^i_k$ is the total number of pairs of particles with a distance between $(k-1)dr$ and $kdr$ and $\mathbb{1}_{\left.\left[(k-1)dr,kdr\right.\right)}(r)$ is the characteristic function of $ \left.\left[(k-1)dr,kdr\right.\right) $, i.e it is 1 of $r$ is within the interval and zero otherwise. \\
					 Since $P_2(x)$ is positive for $x$ close to $\pm1$ and negative for $x\approx 0$ $g_a(r)$ is positive, if spherocylinders with center of mass distance of $r$ are on average oriented parallel. If spherocylinders of this distance are on average perpendicular on the other hand $g_a(r)$ becomes negative. Due to the fact that the integral from $-1$ to $1$ over $P_2(x)$ is zero, $g_a(r)$ is also around zero, if there is no preferred orientation between spherocylinders of distance $r$. 
					\cite{Loewen}\cite{Simbuchofliquids}

					 \subsection{Topological analysis\label{sec:topo}}
\begin{figure}
	\centering
	\rotatebox{270}{\includegraphics[width=0.5\linewidth]{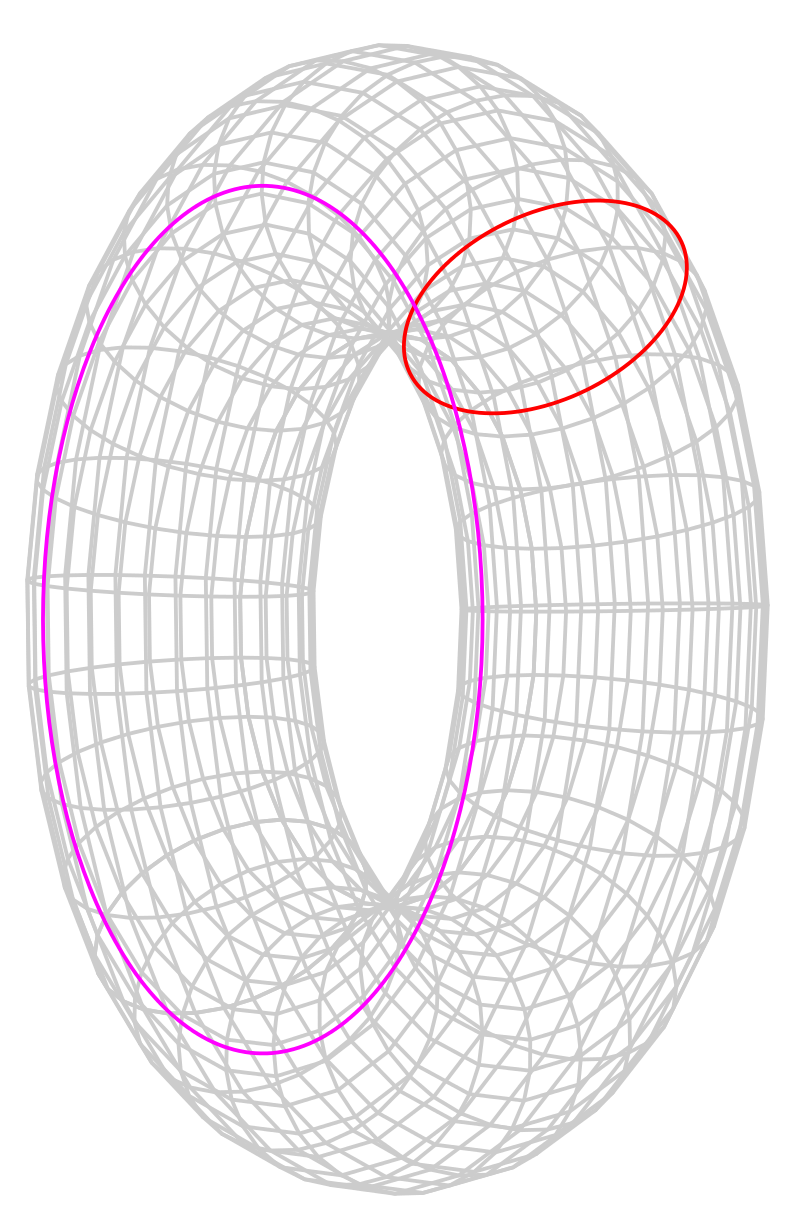}}
	\caption{Illustration on how to count 2-d holes in a torus. The two circles (red and violet) can not be continuously retracted, i.e. the torus must exhibit non trivial topology. Picture from \url{https://de.wikipedia.org/wiki/Torus\#/media/File:Torus_cycles.svg}}
	
	\label{fig:holesex}
	\end{figure}

\begin{figure}
	\centering
	\includegraphics[width=0.7\linewidth]{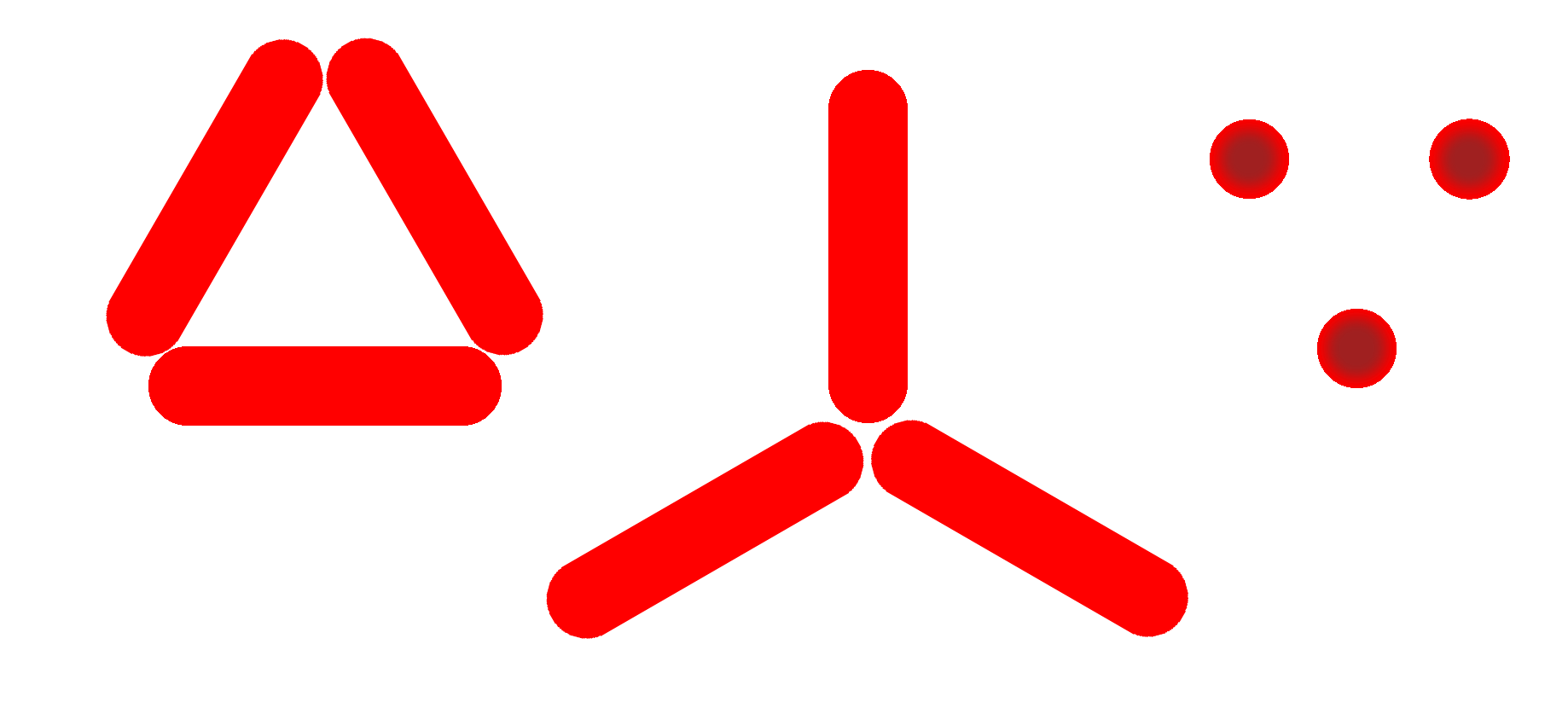}
	\caption{Examples why neither the shortest distance nor the center of mass distance between spherocylinders is suited to properly capture the topology of systems of spherocylinders. With regard to their center of mass distance, all three of the systems above are equivalent (spherocylinders on the very right are perpendicular to the paper plane.), despite having completely different topology. With regard to the shortest distance between spherocylinders the first two stay equivalent.  }
	\label{fig:exmetrsphero}
	\end{figure}

					 To determine the porosity of the percolating networks we used a method called \textbf{persistent homology} to analyze the topology of these networks. While we merely used it to count differently sized holes in our networks, the method is ,theoretically, capable of far more detailed analysis of the topology of data sets. \\
					 We will give a short description of the ideas involved. More detailed explanations can be found in \cite{TDA2}  \cite{TDA1} and \cite{Javaplex}. The latter also is the tutorial for the matlab tool "javaplex", which is an implementation of the algorithm described below. \\
					 It is a common problem in algebraic topology to determine the topological invariants of simplicial complexes. A simplicial complex is a set of, possibly different dimensional, simplices. A n-dimensional simplex is a polytope with n+1 corners, i.e. a 0-dimensional simplex is a dot, a 1-dimensional simplex a line segment, a 2-dimensional simplex a triangle  and a 3-dimensional simplex is a tetrahedron and so on. The topological invariants of interest are $n$-dimensional holes. For our purposes it is enough to picture 0-dimensional holes as connected components, 1-dimensional holes as literal holes and 2-dimensional holes as enclosed volumina. As a example we can look at a 2-dimensional sphere and a 2-dimensional torus. Both have one connected component, i.e. one 0-dimensional hole. The sphere has no holes in its surface, all circles we put in the surface can be retracted to a dot without breaking them. The torus in the other side has two 1-dimensional holes, since there are two classes of circles that can neither be smoothly transformed into each other nor into a dot, as is depicted in figure \ref{fig:holesex}. Both the sphere and the torus have one enclosed volume.\\
				Counting the holes of a simplicial complex is done using so called homology groups \footnote{The dimension of the n-th homology group can roughly be thought of as the number of n-cycles modulo the n-boundary of a given structure} and reduces to determining the dimension of the kernel of a matrix. This procedure is readily implemented in javaplex and will not be discussed in further detail. \\
				The question for us is now how to turn our data into a simplicial complex, without losing to much of the structural information. \\
				For the sphere systems this was quite simple. We simple used the centers of the spheres as nodes for the simplicial complex and added edges between two nodes, if their euclidean distance was smaller then a certain length $d$. To count the number of holes of an diameter of around $2$, $d=2$ was chosen, the complex constructed and its holes counted with javaplex. For spheres this method is quite accurate. \\
				For spherocylinders neither the shortest nor the center of mass distance are very well suited for this kind of procedure, because in both cases we would count the wrong number of holes as is visualized in figure \ref{fig:exmetrsphero}. \\ 
				 Instead we placed equally spaced spheres along every spherocylinder and then applied the same algorithm as for spheres. Since the computing time of the algorithm highly depends on the number of dots and the distance $d$ up to which edges are added, we had to reduce the number of spheres along a spherocylinder, when calculating the number of holes for bigger $d$. To count the number of holes, with a diameter around half the aspect ratio of the rods for example, we could only approximate a rod with a chain of three spheres. For $d=1.5$ we used a sphere every unit-length, i.e. the number of spheres equals the aspect ratio, and for $d=10$ we placed a sphere every 10 unit length, i.e. the number of spheres is $p/10+1$.\\
				 Obviously some information is lost, when the rods are replaced by spheres this way, which is why the number of holes calculated for the spherocylinder systems has to be taken with a grain of salt. We expect them to be in the right order of magnitude, but not exact results. \\
				
				   	 	\cleardoublepage
						\section{Simulations}
						To enable attempts to reproduce our studies we will use this chapter to discuss all simulated systems with regard to all necessary parameters.\\
						All simulations were carried out with the simulation program initially written by Ullrich Siems. \cite{Ullidiss} The program is written in C++ and is object oriented. The program is still part of current work and while by far the greatest contribution still is Ullrich Siems', many small additions and minor corrections have been implemented by Anton Lüders, Jacob Holder and myself, during the time I have been working with it.\\
						
						The program starts by initializing a simulations box, which can be chosen to be $2-$ or $3-$dimensional and periodic or non-periodic in any direction. (For this work all boxes were $3-$dimensional and periodic in all directions.) In the next step a fixed number of particle positions (spheres or spherocylinders) are randomly distributed in this box or read from a file to create an initial state. In this procedure only initial states with negligible overlap of particles are allowed. Starting from this initial state the Brownian Dynamics algorithm runs for a set number of time steps with a fixed time step width $\Delta t$.\\
						
						As already mentioned we have done simulations with spheres and spherocylinders.  In the first ones a hard-core-Lennard-Jones  \eqref{eq:hclj} potential was used to describe the interactions between spheres. We performed simulations with varying volume fraction $\Phi$ and varying hard-core radius $r_0$. The size of the simulations box was kept at $20\times 20 \times 20$ for all parameters. A complete list of all non-constant parameters can be found in table \ref{tab:spar} .\\
						For the spherocylinder systems a Kihara-like Lennard-Jones \eqref{eq:KLJ} potential was used to calculate the particle-particle interaction. We varied the volume fraction $\Phi$ and the aspect ratio $p$ of the spherocylinders. The simulations boxes had a size of $100\times 100 \times 100$ for all aspect ratios  lower or equal to $20$ and $250\times 250 \times 100$ for all greater aspect ratios.A complete list of all non-constant parameters can be found in table \ref{tab:scpar}.\\
						After an appropriate relaxation time the equilibrated systems all were subjected to small amplitude oscillatory shear with several maximum strains $\gamma_0$ and a frequencies $\omega$. In these last simulations Lees-Edwards boundary conditions were used rather than periodic boundary conditions.\\
						In all simulations the interactions strength was set to $\epsilon=5$ and the cut-off radius remained at $r_c=2.5$. To calculate the volume of a particle, we used its effective diameter following the Weeks-Chandler-Anderson theory: \begin{align}
							\nonumber \sigma_{eff}=\int\limits_0^\infty \left[1- e^{-V_{rep}(r)/k_bT}\right]dr,
						\end{align}
						where $V_{rep}$ is the repulsive part of the pair potential, with its minimum shifted to coincide with zero.\\
						\begin{table}[h]
							\begin{centering}	  		
								\begin{tabular}{l|c}
									\hline	\hline 
									\rule[-2ex]{0pt}{5.5ex} Colloids & Spheres  \\ 
												\hline 
												$\Phi\left[\%\right]$ &  $6.5$ \hspace{3cm} $13$ \\
												\hline
												$r_0$ & $0$\hspace{2.3cm}\hspace{2.3cm} $9$   \\
												$\Delta t \left[\frac{\sigma^2}{D}\right] $ & $7.5e-5$\hspace{1cm}  \hspace{1cm} $1.0e-6$ \\
												\hline
											    $\gamma_0\left[\% \right]$ & \hspace{0.5cm} $0.1 $ \hspace{0.6cm}$ 0.21 $ \hspace{0.6cm}$0.46$ \hspace{0.6cm}$ 1$ \hspace{0.6cm}$ 2.15$ \hspace{0.6cm}$ 4.62$ \hspace{0.6cm}$ 10$ \hspace{0.6cm}$ 21.5$ \hspace{0.6cm}$ 46.2$\\
											    
											    \hline	
											     $\omega \left[\frac{D}{\sigma^2}\right]$ & \hspace{0.5cm} $1.0e3$\hspace{0.3cm}$ 2.15e3 $\hspace{0.3cm}$4.62e3$\hspace{0.35cm}$ 1.0e4$\hspace{0.35cm}$ 2.15e4$\hspace{0.35cm}$ 4.62e4$\hspace{0.35cm}$ 1.0e5$\hspace{0.35cm}$ 2.15e5$\hspace{0.35cm}$ 4.62e5$ \\
								\hline 	\hline
							\end{tabular} 
							\caption{All values for non constant parameters used in simulations related to this work. In rows not separated by a line only the parameter pairings actually shown in the corresponding column were used. Between rows separated by a line all possible combinations of parameters were simulated.}
							\label{tab:spar}
						\end{centering}
					\end{table}
						\begin{table}[h]
							\begin{centering}	  		
								\begin{tabular}{l|c}
									\hline	\hline 
									\rule[-2ex]{0pt}{5.5ex} Colloids & Spherocylinders  \\ 
									\hline 
									$\Phi\left[\%\right]$ &\hspace{0.3cm} $0.26$\hspace{0.3cm} $0.31$ \hspace{0.3cm}$0.41$\hspace{0.3cm} $0.52$\hspace{0.3cm} $0.63$\hspace{0.3cm} $0.73$\hspace{0.3cm} $0.84$\hspace{0.3cm} $0.94$\hspace{0.3cm} $1.05$\hspace{0.3cm} $1.57$\hspace{0.3cm} $2.09$ \\
									\hline
									$p$ & $10$\hspace{2.1cm}$20$\hspace{2.1cm} $30$ \hspace{2.1cm}$40$\hspace{2.1cm} $50$   \\ \hline
									$\Delta t \left[\frac{\sigma^2}{D}\right] $ & $1.0e-5$ \\
									\hline
									$\gamma_0\left[\% \right]$ & \hspace{0.5cm} $0.1 $ \hspace{0.6cm}$ 0.21 $ \hspace{0.6cm}$0.46$ \hspace{0.6cm}$ 1$ \hspace{0.6cm}$ 2.15$ \hspace{0.6cm}$ 4.62$ \hspace{0.6cm}$ 10$ \hspace{0.6cm}$ 21.5$ \hspace{0.6cm}$ 46.2$\\
									
									\hline	
									$\omega \left[\frac{D}{\sigma^2}\right]$ & \hspace{0.5cm} $1.0e3$\hspace{0.3cm}$ 2.15e3 $\hspace{0.3cm}$4.62e3$\hspace{0.35cm}$ 1.0e4$\hspace{0.35cm}$ 2.15e4$\hspace{0.35cm}$ 4.62e4$\hspace{0.35cm}$ 1.0e5$\hspace{0.35cm}$ 2.15e5$\hspace{0.35cm}$ 4.62e5$ \\
									\hline 	\hline
								\end{tabular} 
								\caption{All values for non constant parameters used in simulations related to this work. All possible combinations of parameters were simulated.}
								\label{tab:scpar}
							\end{centering}
						\end{table}

							\cleardoublepage
					
						\section{Results\label{sec:results}}
						In this section we will present and discuss all our simulations and the extracted data. We will start off with the simulations concerning spherical particles interacting via a Lennard-Jones potential or a hard core Lennard-Jones potential respectively. The chapter will follow closely the paper by Santos, Campanella and Carignano \cite{kugelgel}. We reproduced their main results in order to proof the functionality of our program and methods.\\
						After that the systems of spherocylinders are presented and evaluated in similar fashion.					
						\subsection{Evaluation of the simulations of sticky spheres \label{sec:kugelintro}}
\begin{figure}[H]
		\centering
		$\Phi=6.5\%$\hspace{4,5cm} $\Phi=13\%$\\
		\rotatebox{90}{\hspace{2.7cm}$r_0=0$}\includegraphics[width=0.4\linewidth]{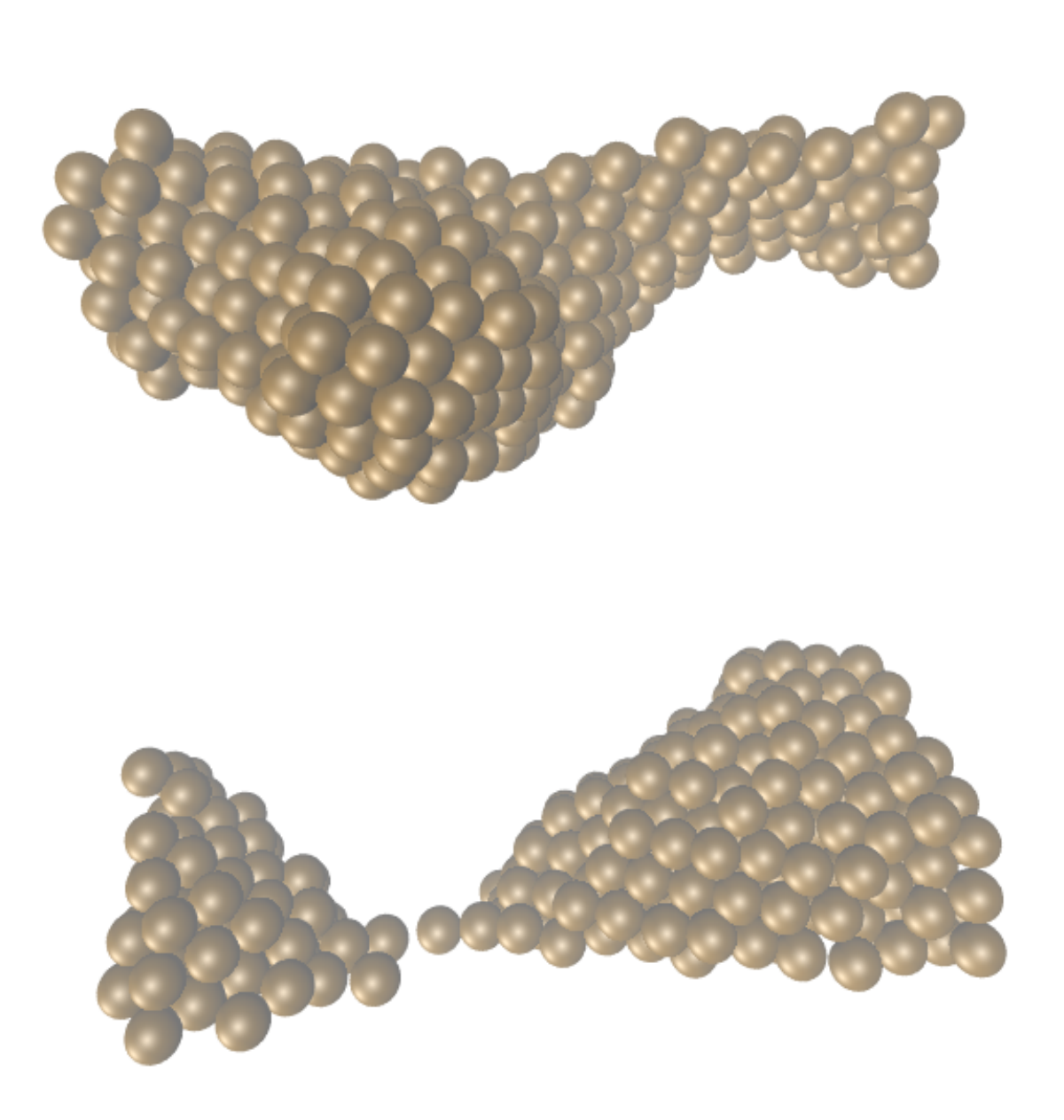}
		\rotatebox{90}{\hspace{-0.5cm}\rule{7.5cm}{0.05cm}}
		\includegraphics[width=0.4\linewidth]{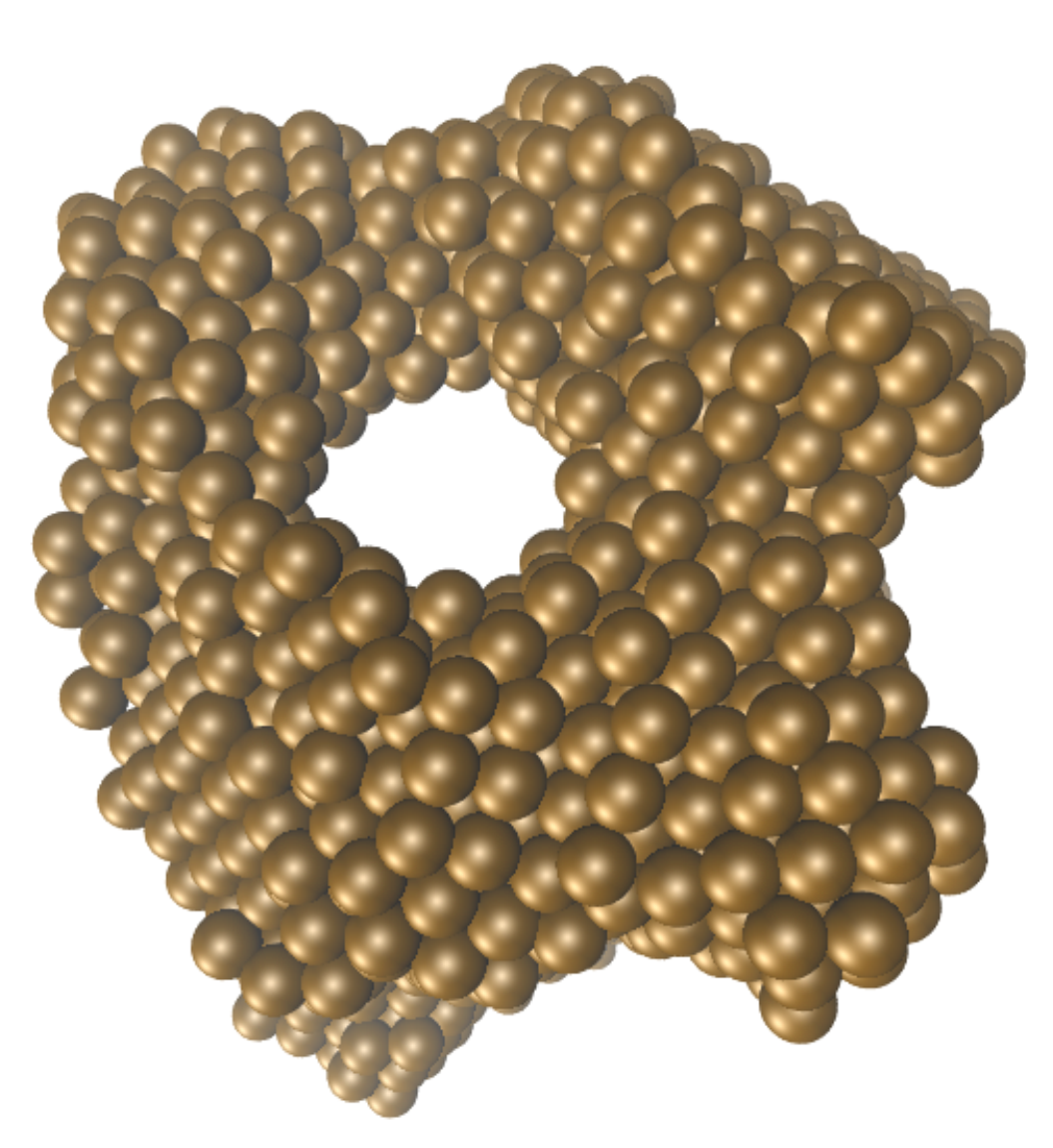}
		\hspace{1cm}\rule{13cm}{0.05cm}
		\rotatebox{90}{\hspace{3.5cm}$r_0=9$}\includegraphics[width=0.4\linewidth]{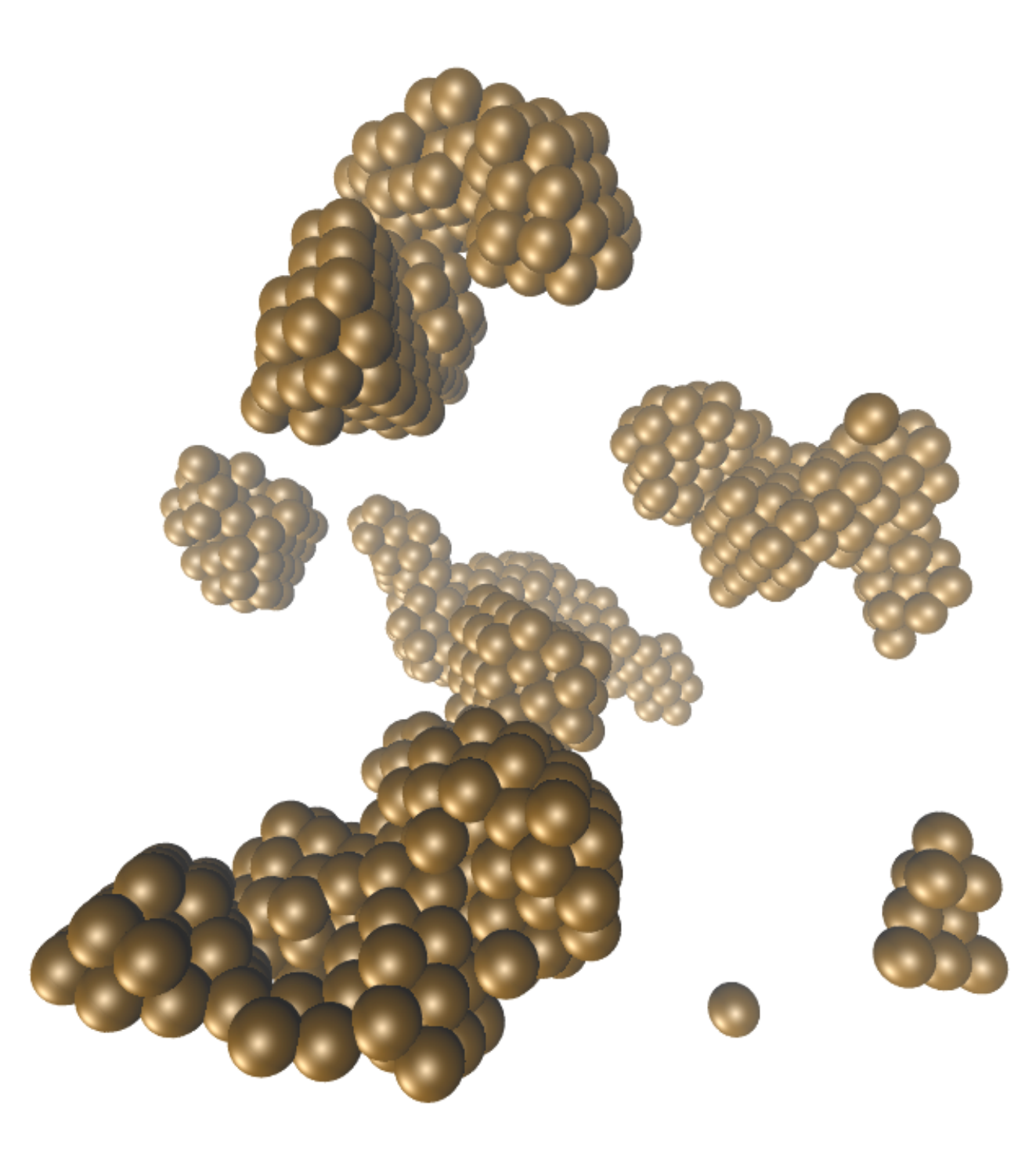}
		\rotatebox{90}{\rule{8cm}{0.05cm}\hspace{-0.5cm}}
		\includegraphics[width=0.4\linewidth]{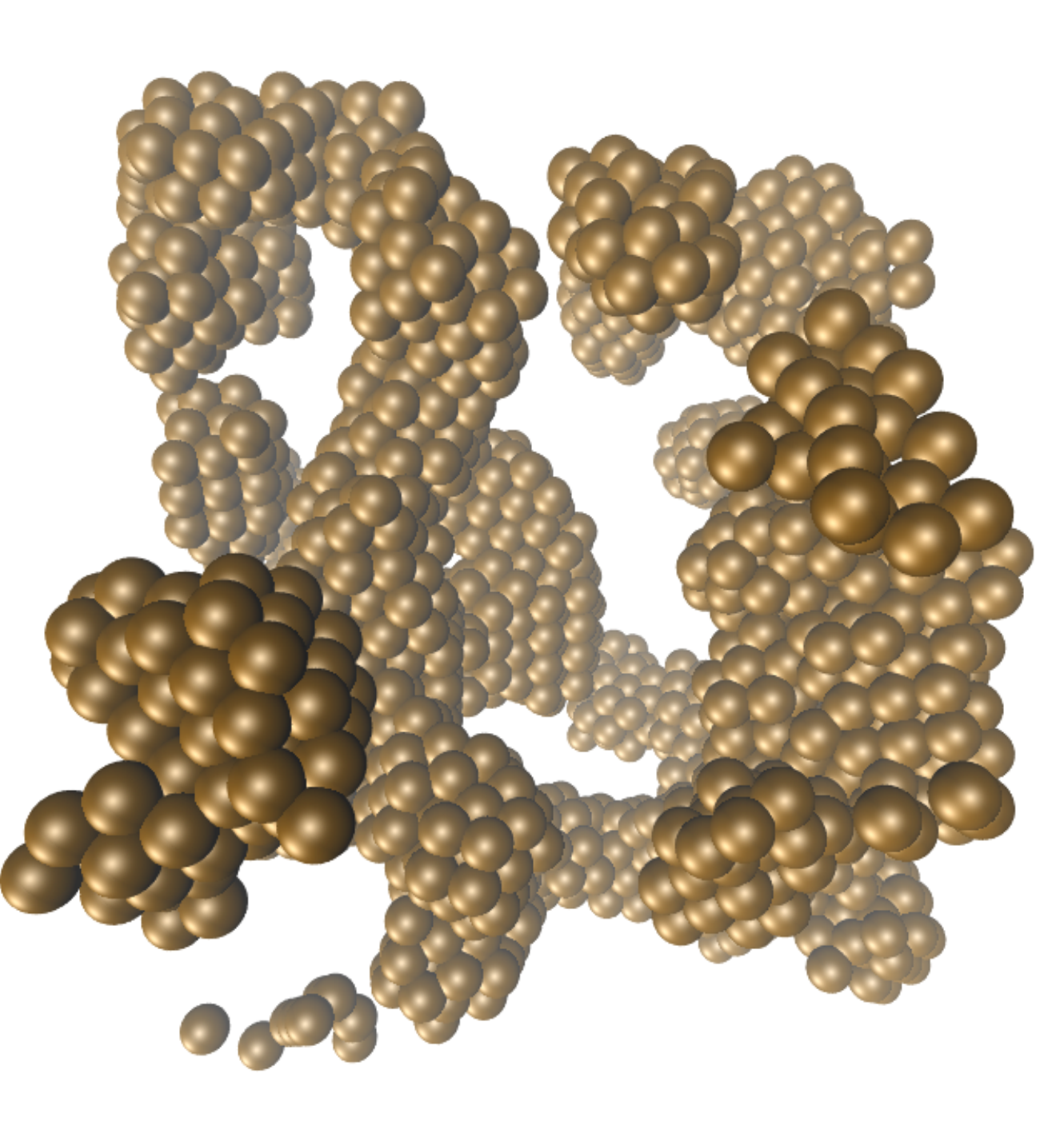}
		\caption{Systems of spherical particles (1000 particles on the left and 2000 on the right) with different interaction potentials (Lennard-Jones potential, $r_0=0$ top hard-core-Lennard-Jones  with $r_0=9$ bottom) after a simulation time of $t=300$. All  remaining simulation parameters are equal for each picture and can be found in table \ref{tab:spar}.}
		\label{fig:kugelbilder}
		\end{figure}

						As just mentioned this section will discuss the results of our simulations concerning spherical colloids. After a short qualitative description of the final systems we will dive deeper and have a look at the evolution of the potential energy of the system, the mean and max cluster sizes, the radii of gyration and the topological properties of the systems. In the last subsection we have a look at the storage and loss moduli of one of the systems.\\
						In total we consider four systems of spherical particles, as was already described in table \ref{tab:spar}. They differ in their volume fraction, as for two systems the spheres cover $13\%$ of the volume while in the other two only $6.5\%$ are occupied, and in their hard core radius, i.e. two systems interact with a standard Lennard-Jones potential while the other two interact via a hard core Lennard-Jones potential with $r_0=9$. Of course the Lennard-Jones potential can be interpreted as having a hard core radius of $r_0=0$. The interaction strength was kept at $\varepsilon=5$ in all four systems.\\ All simulations ran for a time of at least $t=300\frac{\sigma^2}{D_0}$, which also is the time at which we applied the small amplitude oscillatory shear and at which all further analysis of final states was conducted.\\
						A picture of these final states can be seen in figure \ref{fig:kugelbilder}. 
					    In these pictures we can also already spot quite easily the differences between the four systems. In both systems with a volume fraction of only $\Phi=6.5\%$ we can see that the spheres do not percolate throughout the system. At least not in all spacial directions. In fact the system with $r_0=0$ is still connected in one of three dimensions. The clusters in the system with lower volume fraction and $r_0=9$ on the other hand are all well separated.\\
					    The systems with $\Phi=13\%$ show higher connectivity. In fact the system with $r_0=9$  percolates in every spacial direction, while the $r_0=0$ system still percolates in two directions. Comparing these pictures with the results in \cite{kugelgel} we would expect the latter system to break its bonds in the remaining two dimensions and assemble in a more a less sphere-like geometry. Due to the periodic boundary conditions the state depicted in figure \ref{fig:kugelbilder} is rather stable and we did not see these percolation breaks on time scales as large as $t=1800\frac{\sigma^2}{D_0}$. Compared to the system with $r_0=9$ and $\Phi=13\%$  we can definitely see a tendency of the $r_0=0$-spheres to assemble in as little space as possible, while the $r_0=9$-spheres concentrate rather locally, thus making up a porous, sponge-like network spanning throughout the whole system. \\ 
						Considering its percolation properties and overall appearance the system with $\Phi=13\%$ and $r_0=9$ appears to be our most promising candidate to be considered a gel. A look at its storage and loss modulus in section \ref{sec:rheokugel} will confirm this first impression.

						\subsubsection{Evolution of potential Energy}
\begin{figure}[H]
							\centering
						\scalebox{0.8}{	 \input{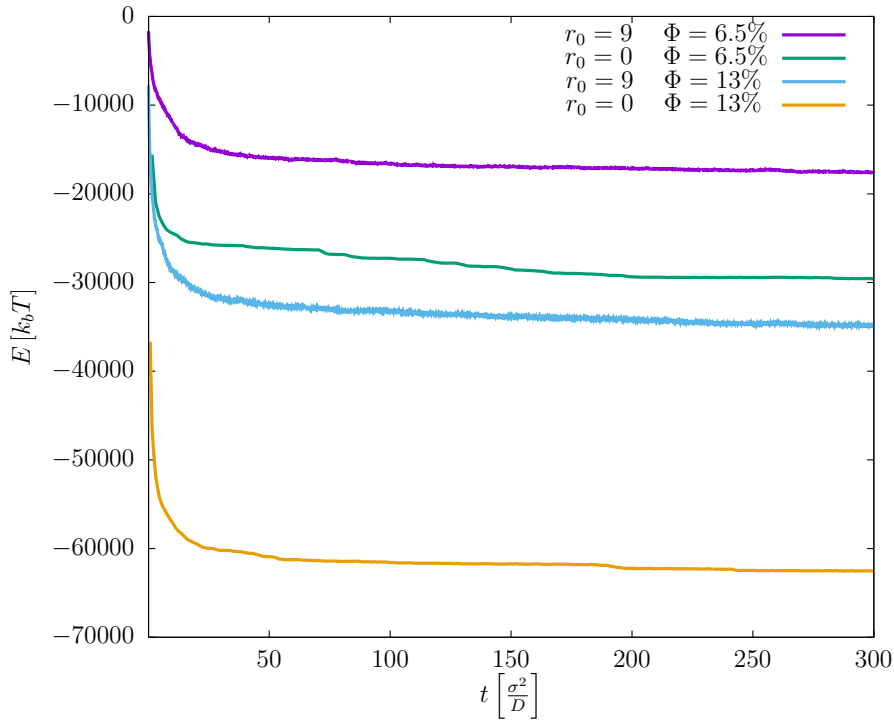}}
						
							\caption{Potential Energy of spheres interacting with a Lennard-Jones ($r_0=0$) or a hard-core-Lennard-Jones potential depicted over time. For both potentials systems of different volume fraction $\Phi$ were simulated. The standard Lennard-Jones particles were simulated for much longer than depicted. The sparsity in their data points arises from bigger time steps used in these simulations  }	\label{fig:kugelenergie}
						\end{figure}

						In this chapter we have a closer look at the potential energy of the four systems over the course of a time period of $t=300\frac{\sigma^2}{D_0}$. The actual time evolution is depicted in figure \ref{fig:kugelenergie}. \\
						We can see more or less an exponential decay in the energy for all systems. While this means, that the energy of the systems is not completely constant yet and might never be, we believe the fact that all changes in energy for the later times depicted to be insignificant enough to consider the systems at equilibrium, i.e. all analysis conducted on the final state is not significantly influenced by the particles relaxing to an equilibrium state. We are aware, that if a system would stop to percolate in any direction, consequently there would be a sudden drop in the overall energy not fitting to this exponential decay. Since the simulations of the systems, that are still percolating in some direction were running even longer then the time depicted in figure \ref{fig:kugelenergie} without this happening, we confidently consider this event rare enough to still consider the systems sufficiently equilibrated.\\
						The differences in the energy of the four systems can quite easily be explained: There are twice as many particles in the systems with $\Phi=13\%$ then there are in the systems where the volume fraction is $\Phi=6.5\%$. Since the energy of the system is the sum over all pair-potentials, we expect the energy to be approximately proportional to $n^2$, hence the factor between the low density systems and the high density ones should be around 4. We actually see a factor between 3 and four. It seems plausible, that in the higher density cases more particles do not interact at all, because there is only limited space in direct proximity to every particle, compared to the low density cases, which could be the reason why our factor is slightly lower than expected.\\
						The differences in energy of the systems with different hard core radius $r_0$ are likely due to the fact that the systems with $r_0=9$ tend not to concentrate in one place as much as do the other systems, hence less of the particles are actually interacting.

						\subsubsection{Evolution of maximum and mean cluster size}
\begin{figure}[H]
								\centering
								\scalebox{0.7}{\input{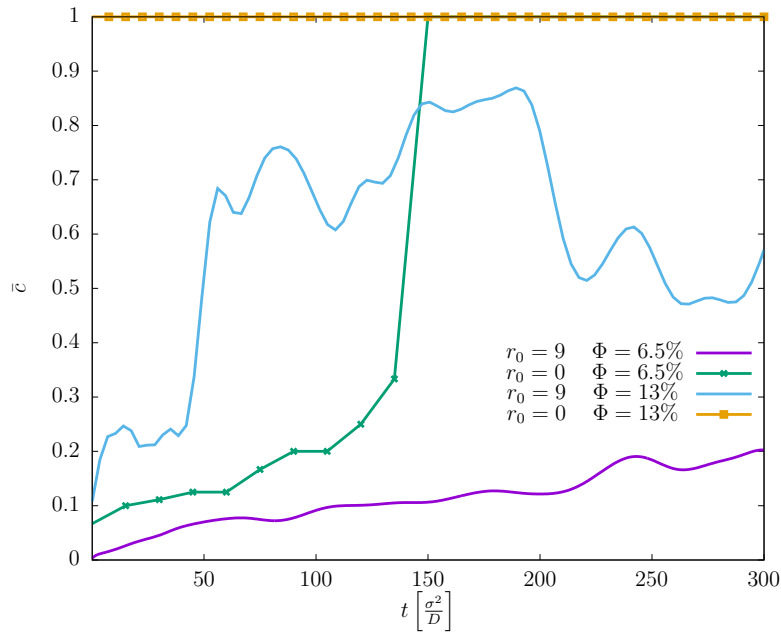}}
							
								\caption{Mean Cluster of spheres interacting with a Lennard-Jones ($r_0=0$) or a hard-core-Lennard-Jones potential ($r_0=9$) depicted over time. For both potentials systems of different volume fraction $\Phi$ were simulated. The standard Lennard-Jones particles were simulated for much longer than depicted. The sparsity in their data points arises from bigger time steps used in these simulations. The blue and violet curve were smoothed using Gnuplot's "smooth bezier" function. }	\label{fig:kugelclustermean}
							\end{figure}

\begin{figure}[H]
								\centering
								\scalebox{0.7}{\input{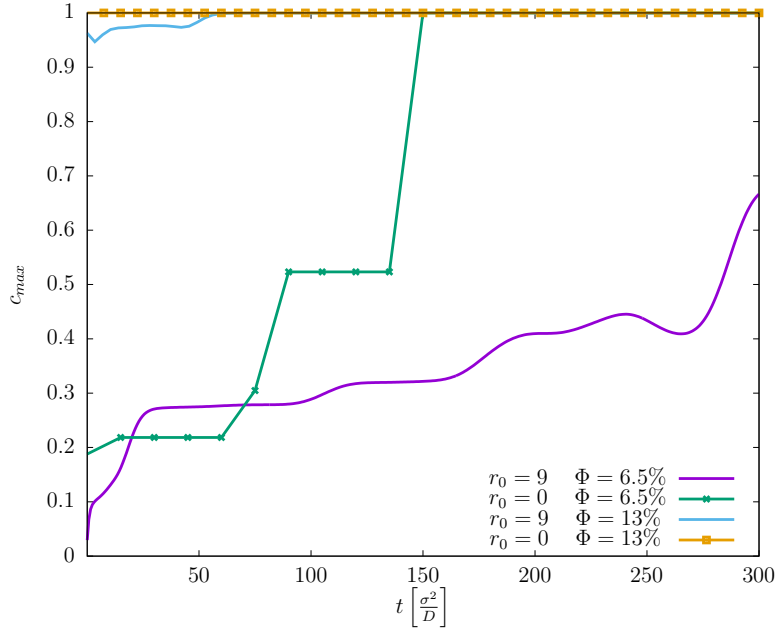}}
							
								\caption{Max Cluster $c_{max}$ of  spheres interacting with a Lennard-Jones ($r_0=0$) or a hard-core-Lennard-Jones potential ($r_0=9$) depicted over time. For both potentials systems of different volume fraction $\Phi$ were simulated. The standard Lennard-Jones particles were simulated for much longer than depicted. The sparsity in their data points arises from bigger time steps used in these simulations. The blue and violet curve were smoothed using Gnuplot's 'smooth bezier' function.  }
									\label{fig:kugelclustermax}
							\end{figure}

					    Since gels usually consist of percolating networks we expect the particles of the gel to be mostly in one, or not more than a few clusters. To figure out which of the sphere system match this description we had a closer look at the time evolution of the normalized mean cluster size $\bar{c}$ and the normalized maximum cluster size. We normalized with respect to the particle number, i.e. the product $\bar{c}N$ is the actual mean cluster size and analogous for $c_{max}$.\\
					    The results for the mean cluster size are depicted in figure \ref{fig:kugelclustermean}, while the time evolution of $c_{max}$ can be seen in figure \ref{fig:kugelclustermax}.\\
					    We realize in both plots, that the system with $r_0=0$ and $\Phi=13\%$ reaches $\bar{c}=c_{max}=1$ after so little simulation steps that our data output intervals were to large to even capture a state were these values are smaller. \\
					    Comparing in particular the curves for the system with $r_0=9$ and $\Phi=13\%$ (the blue curve in both figures) we get a feeling for the fact that the mean cluster size is a lot more susceptible for fluctuation than the maximum cluster size. This appears quite plausible if we imagine all particle to be in one cluster, i.e. both $\bar{c}$ and $c_{max}$ would be 1. If now a single particle were to leave the cluster, $c_{max}$ would still be around 1, if there are sufficiently many particles in the system, while $\bar{c}$ would drop by half.\\
					    This is exactly what happens in the system described by the blue curve. After a relatively short time almost all particles are joined in one cluster, making $c_{max}\approx 1$. Over time single particles however leave and rejoin this main cluster making $\bar{c}$ oscillate between $1$ and $0.5$, which gives the resulting curve between these two values, after smoothing over a few time steps.\\
					    Looking at the green curves we can see that the system with $r_0=0$ and $\Phi=6.5\%$ takes a quite long time to assemble all particles in clusters. What in fact happens is, that smaller clusters are formed quite quickly. Then these cluster move very slowly on their own until they randomly collide and merge with each other.\\
					    The last system did not proceed further than the first described step within the simulated time, hence it is still separated into several clusters.\\
					    We are aware that cluster size itself is not a sufficient indicator for percolation. However the other way around systems of insufficient cluster sizes cannot percolate.

	%
						\subsubsection{Topological analysis of the final state}
						As gels are often described as spongy or porous networks we conducted a topological analysis on the final states as described in section \ref{sec:topo}. We counted the number of holes $n_{1.5}^h$ in our networks that have an approximate radius of $1.5\sigma$			and the number of holes $n_3^h$ with an approximate radius of $3\sigma$. The results are depicted in table \ref{tab:kugeltop}. These results show the trend, that the system with $\Phi=13\%$ and $r_0=9$ tends to be more porous as the one with the same volume fraction, but $r_0=0$. 
						
						\begin{table}[h]
							\begin{centering}	  		
								\begin{tabular}{l c | c c}
									$r_0$ & $\Phi$ & $n^h_{1.5}$ & $n^h_{3}$ \\
									\hline \hline $0$ & $6\%$  & $2$& $0$ \\
									$9$ & $6\%$ & $2$& $0$\\
								    $0$ &  $13\%$ & $1$&$1$\\
									$9$ & $13\%$& $10$& $2$
								
								\end{tabular} 
								\caption{Number of holes of radius $1.5\sigma$ and $3\sigma$ in the final state of the simulations of  particles with different density and different $r_0$.}
								\label{tab:kugeltop}
							\end{centering}
							
							We are well aware, that these results are only of minor interests, if only so few systems of such small box size are viewed, due to lack of statistics. This analysis will be a little more interesting in the spherocylinder systems, since those are significantly larger, which is why we wanted to shortly discuss these results here anyway. 
						\end{table}
						\subsubsection{Small amplitude oscillatory shear\label{sec:rheokugel}}
\begin{figure}[H]
								\centering
								\scalebox{0.7}{\input{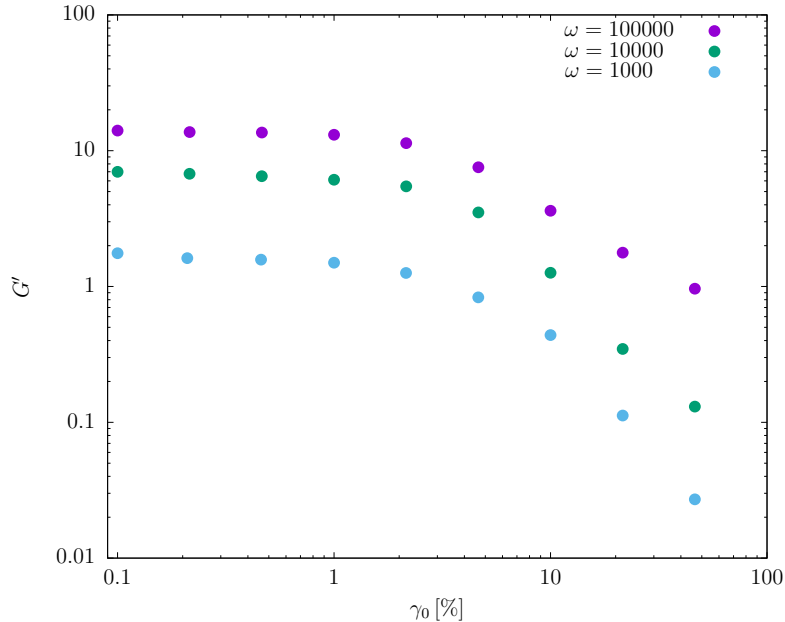}}
							
								\caption{Storage modulus $G'$ dependency on maximum strain $\gamma_0$ for the system with $r_0=9$ and $\Phi=13\%$ for frequencies $\omega=100000$ (violet), $\omega=1000$(green) and $\omega=1000$(blue). The simulations from which $G'$ was derived were performed at $t=300$ . }	\label{fig:kugelgammamod}
							\end{figure}

\begin{figure}[H]
								\centering
								\scalebox{0.7}{\input{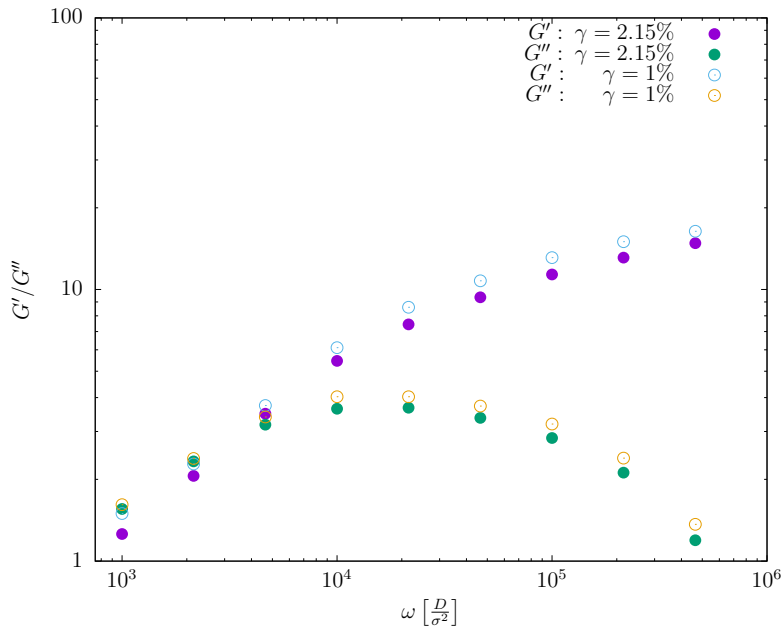}}
							
								\caption{  Storage modulus $G'$ and loss modulus $G''$ dependency on the frequency $\omega$ for the system with $r_0=9$ and $\Phi=13\%$ for  $\gamma_0=1\%$ (empty points) and $\gamma_0=2.15\%$(filled points) . The simulations from which $G'$ was derived were performed at $t=300$ . }	\label{fig:kugelomegamod}
							\end{figure}

						In the last view sections we figured and confirmed, that out of our simulated systems of spheres only the system with a volume fraction $\Phi=13\%$ and a hard-core radius of $r_0=9$ actually is a porous percolating network, and hence a potential candidate for a gel. To determine whether this system is actually gelated we have a look at its storage and loss modulus.\\
						The storage and loss modulus were calculated by applying small amplitude oscillatory shear and determining the phase difference between the strain and the stress as explained in section \ref{sec:SAOS}
						To properly compare $G'$ and $G''$ we first need to find the linear viscoelastic region, i.e. the regime of $\gamma_0$, in which the moduli depend linearly on the strain.\\
						To do so we plotted the dependency of $G'$ on the maximum strain $\gamma_0$ for various frequencies in figure \ref{fig:kugelgammamod}. We can see there that for small $\gamma_0$ up to $1\%$ or $2.5\%$ $G'$ starts out constant and for greater values decays, hinting towards the fact, that for $\gamma_0\gtrsim2.5\%$ the strain starts to destroy the structure of the system. To determine its rheological properties before that happens we have a look at the frequency dependence of $G'$ and $G''$ for strains of $1\%$ and $2.15\%$ in the plot depicted in figure \ref{fig:kugelomegamod}. \\
					
\begin{figure}[H]
								\centering
								\scalebox{0.7}{\input{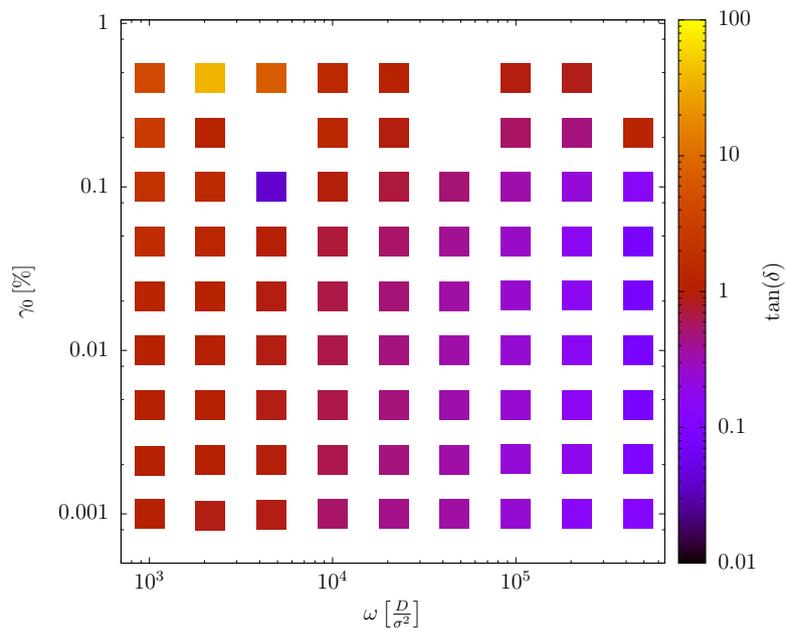}}
							
								\caption{Loss tangent $\tan(\delta)$ dependency on frequency $\omega$ and maximum strain $\gamma_0$for the system with $r_0=9$ and $\Phi=13\%$.The simulations from which $G'$ was derived were performed at $t=300$.}	\label{fig:kugeltiles}
							\end{figure}

								We can see there that for small frequencies and both depicted strains the systems start out with $G'(\omega)<G''(\omega)$, suggesting that on very large timescales we would expect rather liquid-like than solid-like behavior.Even though due to the fact that $\tan(\delta)=G''/G'$) is still very close to $1$ in this regime we can assume that liquid-like in this case would still be far from an ideal fluid. Somewhere between $\omega=2.15\cdot10^3$ and $\omega=4.62\cdot10^3$ there exists a gel-point $\omega_0$ where the inequality between $G'$ and $G''$ flips, i.e. $G'(\omega)>G''(\omega)$ for all $\omega>\omega_0$. From that point on the distance between $G'$ and $G''$ steadily increases for higher frequencies, giving us an actual time scale, where our systems behaves predominantly solid-like.
								\\
								These results match well with measurements of weak gels in experiments and also the results shown in \cite{kugelgel}, where more detailed analysis of similar systems was conducted. \\
								We take this as evidence that our implementation of small amplitude oscillatory shear, as well as the Lees-Edwards boundary conditions work well and the general method can be applied to systems of spherocylinders as well.  \\
								For the final part of the discussion of spherical particles we depicted the loss tangent of the system in dependency of the maximum strain $\gamma_0$ and the frequency $\omega$ in figure \ref{fig:kugeltiles}. Keeping in mind that the linear viscoelastic regime ranged up to $\approx 2.5\%$ we can see similar behavior to the previous plot for all strains within that regime, i.e. $\tan(\delta)\lessapprox 1$ for frequencies around $4\cdot10^3$ and solid like behavior on all smaller time scales.
								
							\newpage
						\subsection{Evaluation of the simulations of sticky rods\label{sec:evalstab}}
						After having tested our simulation program, in particular the periodic shear forces and the implementation of Lees-Edwards boundary conditions added by myself to Ullrich Siems program, by reproducing the results for spheres in \cite{kugelgel} as discussed in the previous sections, we now went on to a vaster study of systems of spherocylinders. \\
						The contents of this chapter will be ordered in the same way they were in the discussion of the results of our simulations with spheres, i.e. we will start out in this section by qualitatively discussing some of the final states of the simulations, then move on to have a closer look at the time evolution of the potential energy and the maximum and mean cluster sizes of the systems. After that we proceed to investigate whether there is any order in the orientation of the spherocylinders in the final state.
						Finally we have a look at the rheological properties of the systems to see, whether any of them have gel character at all.\\ 
						All in all we started simulations with spherocylinders for 55 combinations of aspect ratio $p$ and volume fraction $\Phi$. Most of these simulations ran for a time of $t=5000\frac{\sigma^2}{D_0}$. All of them interacted via a Kihara-like Lennard-Jones potential with interaction constant $\epsilon=5$. The aspect ratios ranged between $p=10$ and $p=50$ and were incremented in steps of $10$. The volume fractions were picked to be between $\Phi=0.26\%$ and $\Phi=2.09\%$. The exact values can be found in table \ref{tab:scpar}. The spherocylinders with $p\le 20$ were simulated in a periodic box of size $100\times 100\times 100$, while all systems with $p\ge 30$ were simulated in a larger periodic box of size $250\times 250\times 100$.\\
						As stated in \cite{Loewen} and \ref{sec:BM} the diffusion constants we used for spherocylinders only match experimental results for $p\le 30$. Unfortunately this makes some of our system rather a toy model and less comparable to experimental data.

						\begin{figure}[H]
							\centering
							$p=10$\hspace{4,5cm} $p=30$\\
							\rotatebox{90}{\hspace{1.5cm}$\Phi=0.26\%$}\includegraphics[width=0.3\linewidth]{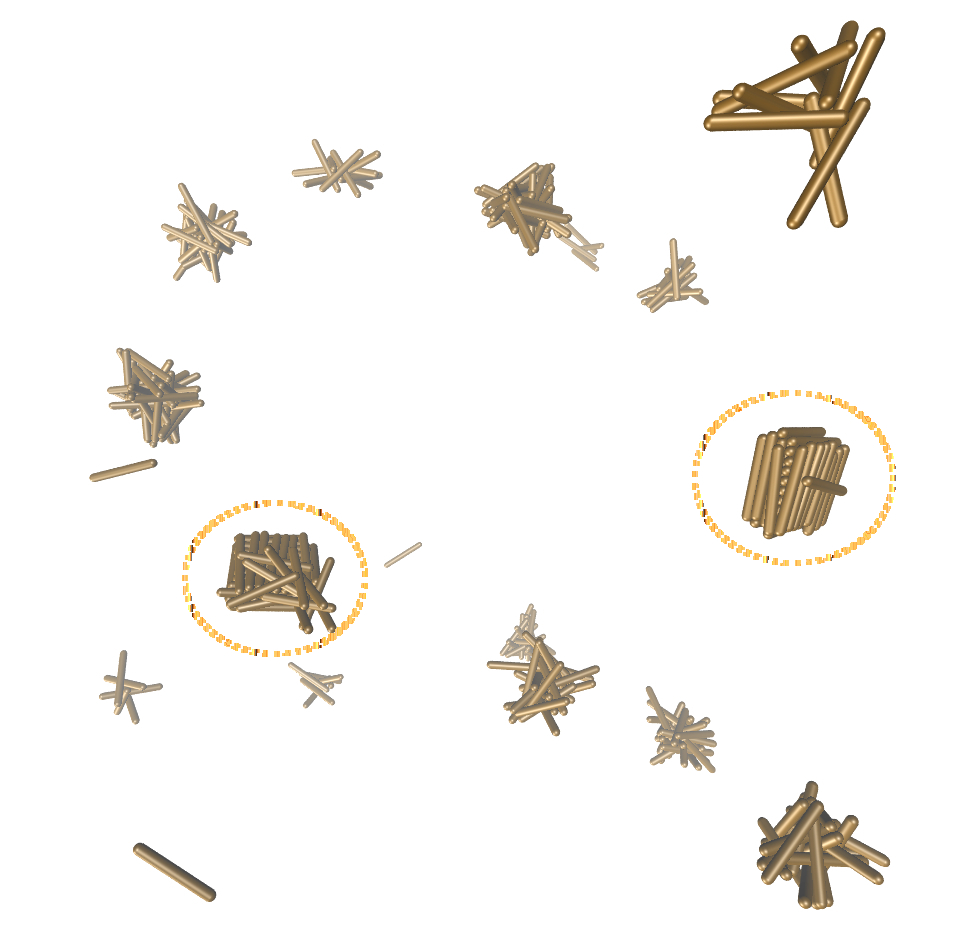}
							\rotatebox{90}{\hspace{-0.5cm}\rule{5.5cm}{0.05cm}}
							\includegraphics[width=0.35\linewidth]{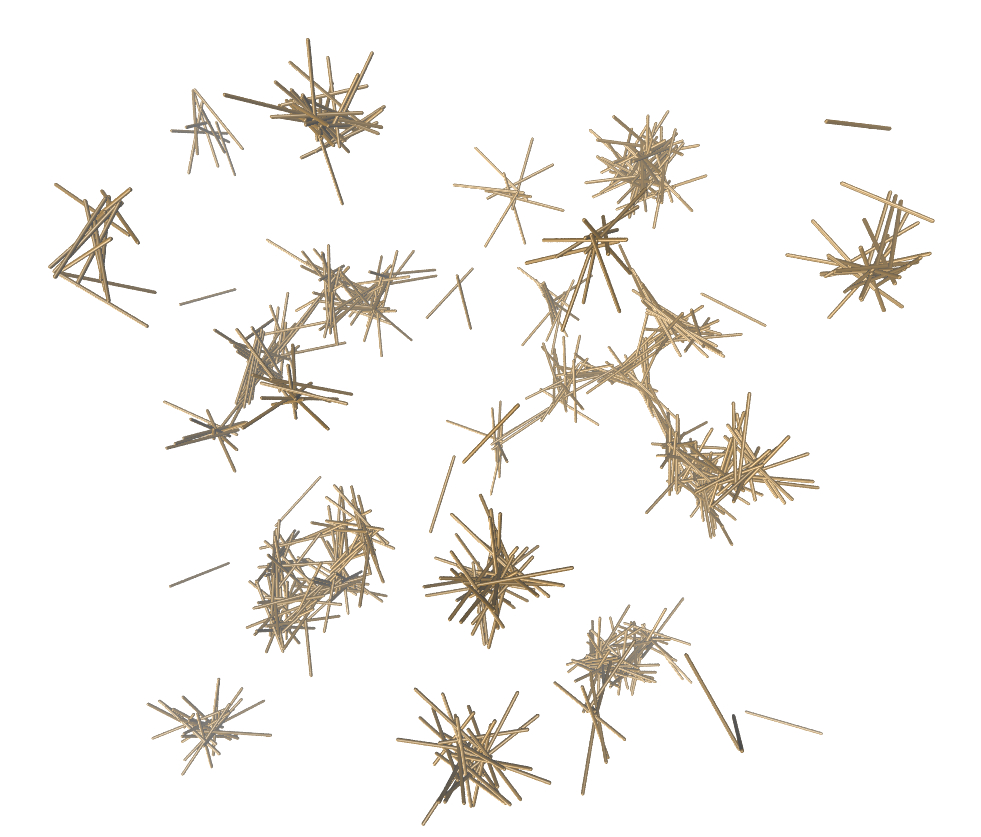}
							\hspace{1cm}\rule{13cm}{0.05cm}
							\rotatebox{90}{\hspace{1.5cm}$\Phi=0.52\%$}\includegraphics[width=0.3\linewidth]{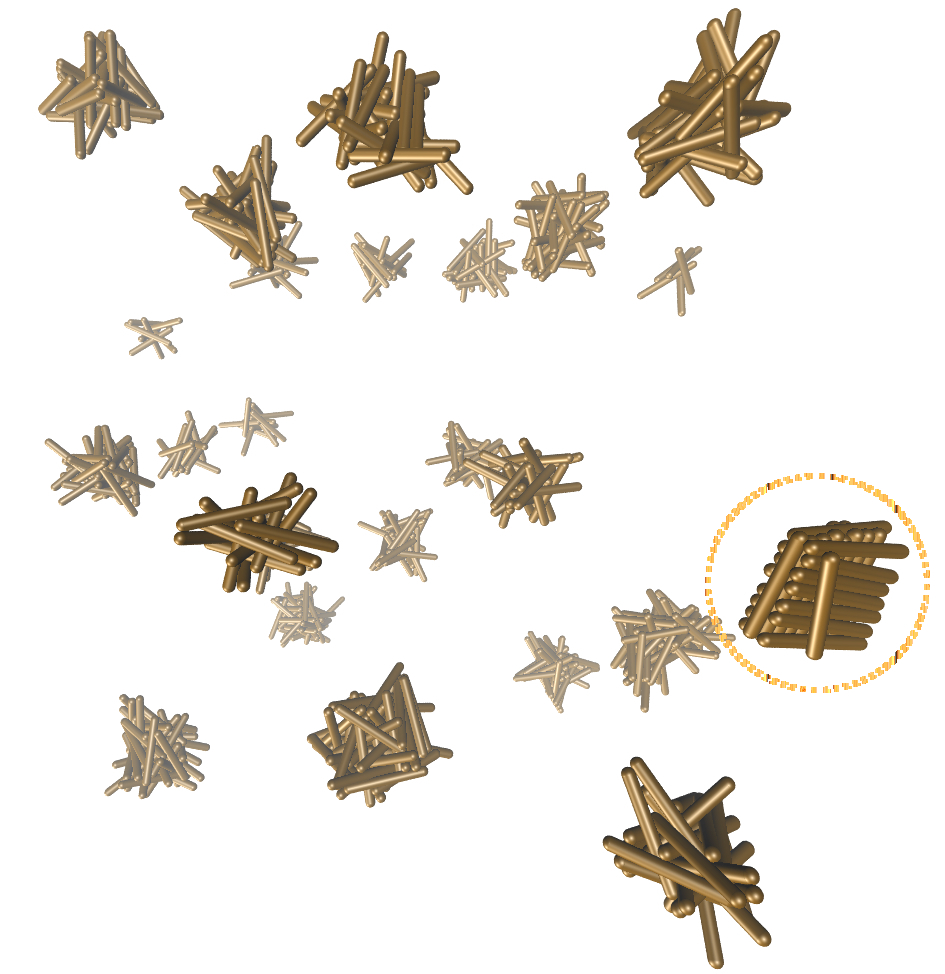}
							\rotatebox{90}{\rule{6cm}{0.05cm}\hspace{-0.5cm}}
							\includegraphics[width=0.35\linewidth]{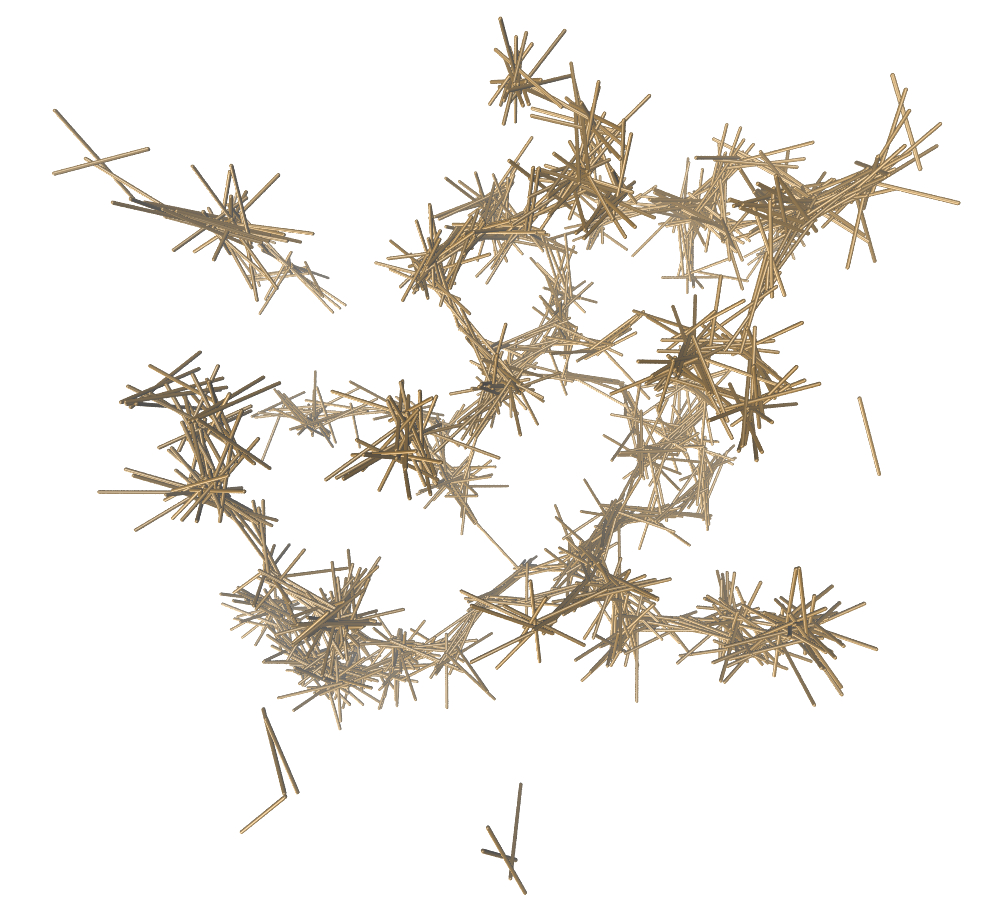}\vspace{-0.5cm}
								\hspace{1cm}\rule{13cm}{0.05cm}
								\rotatebox{90}{\hspace{1.5cm}$\Phi=1.05\%$}\includegraphics[width=0.3\linewidth]{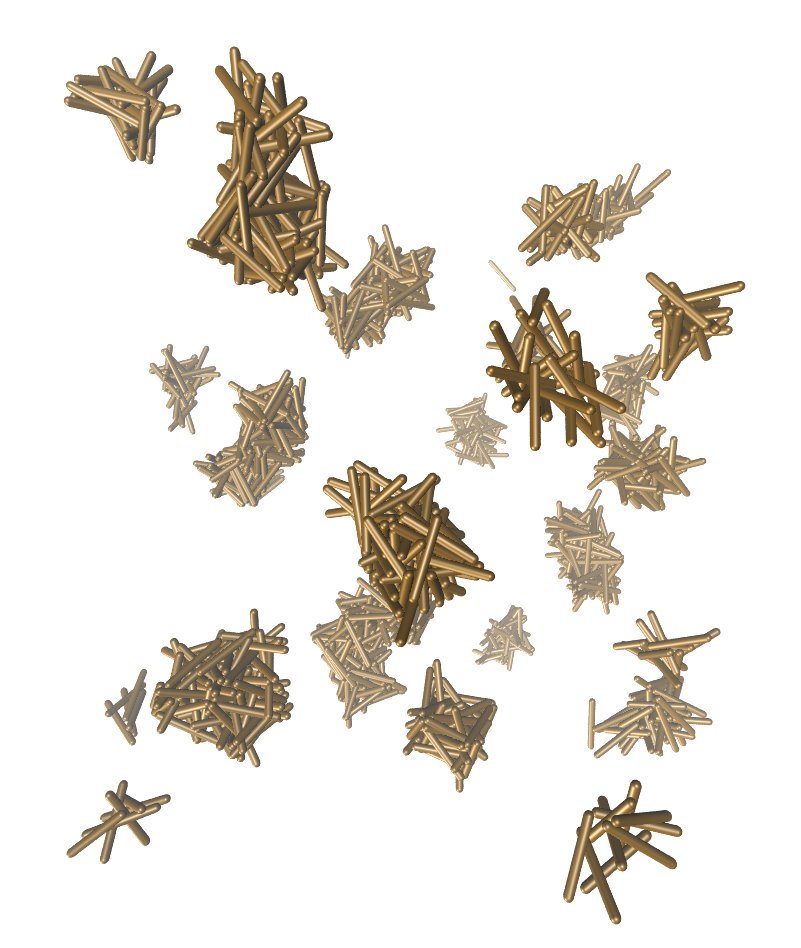}
								\rotatebox{90}{\rule{6cm}{0.05cm}\hspace{-0.5cm}}
								\includegraphics[width=0.35\linewidth]{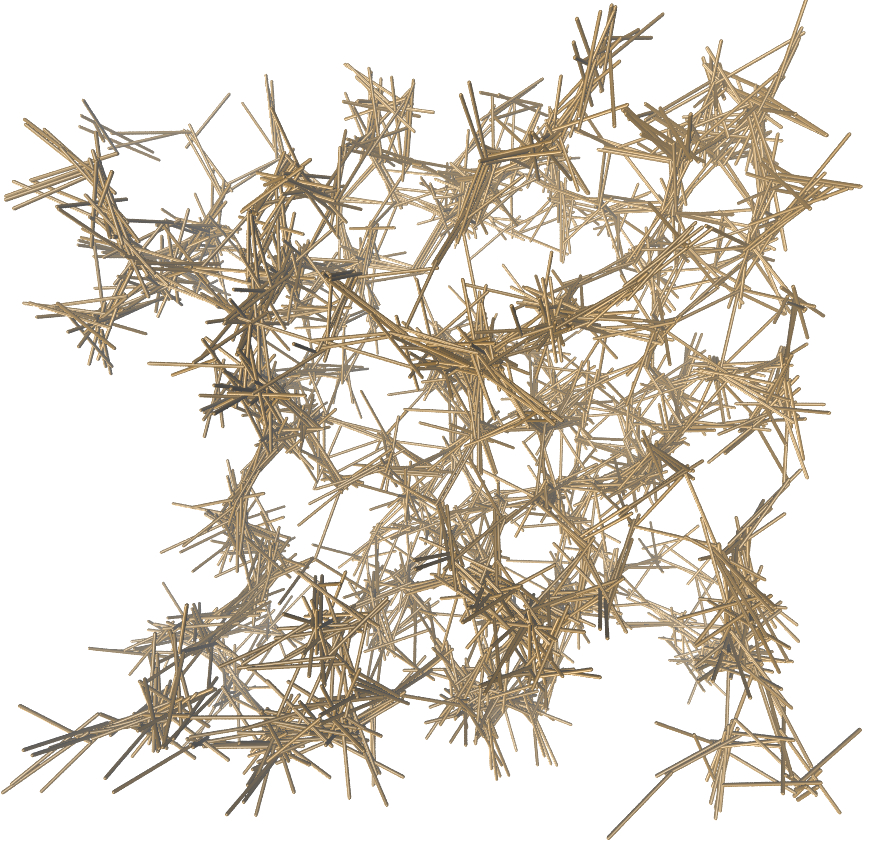}
							\caption{Systems of spherocylinders of different aspect ratio ($p=10$ on the left and $p=30$ on the right) and volume fraction ($\Phi=0.26$ at the top,$\Phi=0.52$ in the center and $\Phi=1.05$ at the bottom) interacting  via a Kihara-like Lennard-Jones potential after a simulation time of $t=5000$. Further similar simulations with the parameters shown in table  \ref{tab:scpar} were conducted as well.}
							\label{fig:rad1}
						\end{figure}
\begin{figure}
	\centering
	\includegraphics[width=0.4\linewidth]{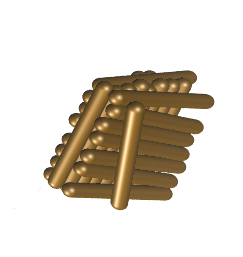}
	\caption{Cuboid-like stack of spherocylinders as sometimes witnessed in systems with $p\le20$.}
	\label{fig:cuboid}
	\end{figure}

						In figure \ref{fig:rad1} we depicted a few samples of pictures of the final states of these simulations. In the left column pictures of spherocylinders with an aspect ratio of $p=10$ can be seen. From top to bottom the volume fractions are $0.26\%$, $0.52\%$ and $1.05\%$. In the right column we see picture of spherocylinders with an aspect ratio of $p=30$. Note that even though these rods appear to be of similar length than the ones in the left column, they would be 3 times longer if the simulation boxes were to scale. The volume fractions in the right column are the same as in the left one. \\
						While these six pictures are just a small sample of all the simulations we ran, they are nicely suitable to notice and discuss a few trends in the formation of networks in these systems, that will be quantified later on. \\
						We can see for example quite easily how the size of the connectedness components in the systems increase with increasing volume fraction. The same effect can be seen, when going from smaller to larger aspect ratio. The cluster sizes in the systems with $p=30$ are way bigger than in the ones with $p=10$. \\
						The extend of this effects goes so far that the spherocylinders with $p=30$ start from a percolating network, with only few particles not being a part of it, already at densities as low as $\Phi=0.52\%$. For the volume fraction of $\Phi=1.05\% $ this networks is spans throughout the whole simulation box and includes every particle. For spherocylinders with $p=10$ we only witnessed almost all particles as part of one cluster for the greatest volume fraction $\Phi=2.06\%$, as can be seen later. This gives us already a strong hint towards the fact, that also in our systems gelation is more likely to occur with increasing aspect ratio.\\
						 Before moving on to the quantitative discussion we want to guide attention to the cuboid-like structures, consisting of a few layers of parallel spherocylinders, that can be seen in the two lower density pictures with $p=10$ encircled by an orange line. A more detailed picture of one of these can be found in figure \ref{fig:cuboid}. We found formations like this not in great number, but quite regularly in systems with an aspect ratio of $p=10$ and a few times in systems with aspect ratio of $p=20$. If this trend were to be confirmed in further studies, further simulations could be run on how to separate these cuboids from the bulk of other clusters and later how such cuboids behave as colloids themselves.

						\subsubsection{Evolution of potential Energy}
\begin{figure}[H]
							\centering
							\scalebox{0.7}{\input{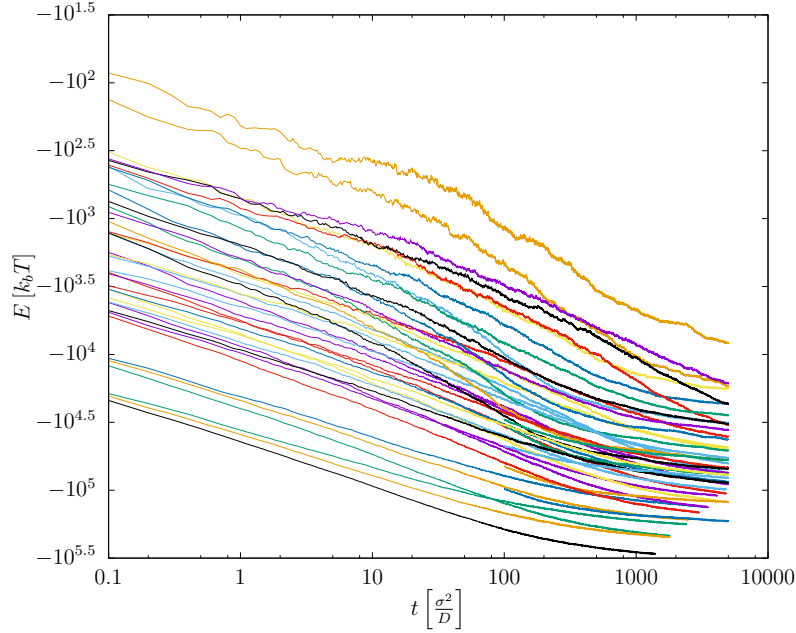}}
							
							\caption{Time evolution of the potential energy of several systems of spherocylinders with aspects ratios $p$ between $10$ and $50$ and volume fraction $\Phi$ between $0.26\%$ and $2.06\%$ over a time of $t=5000\frac{\sigma^2}{D_0}$}
							\label{fig:stabenergie}
						\end{figure}

\begin{figure}[H]
							\centering
							\scalebox{0.7}{\input{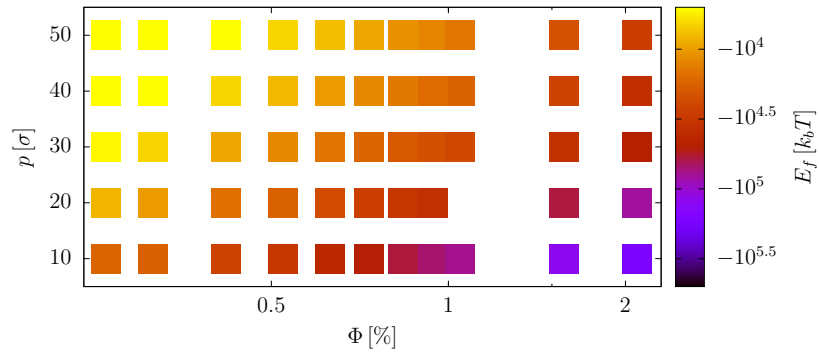}}
													\caption{Final potential energy $E_f$ of spherocylinder systems for some $\Phi\in \left[ 0.26\%, 2.06\% \right]$ and $p \in \left[10,50\right]$ after a simulated time of $t=5000\frac{\sigma^2}{D_0}$.}							 
								 \label{fig:Energietiles}
						\end{figure}

						Like for spheres we will now have a look at the potential energy of the spherocylinder systems over the course of the simulation time, which for most systems was $t=5000\frac{\sigma^2}{D_0}$. The actual time evolution is depicted in figure \ref{fig:stabenergie}. While some of the depicted energy evolution seem already  to be close to constant a few unfortunately still change quite rapidly, so that we can hardly confidently state, that all these system are at equilibrium.\\ There are two main reasons why we nevertheless proceeded our further analysis with the final states of the simulation runs belonging to these partly not completely equilibrated systems. The first of these reason is the fact that visual inspection of these states did not exhibit major structural changes over most of the simulation time, so that we feel sufficiently safe to say, that the analysis  of cluster size, topology, and angular ordering will not be significantly altered by further time evolution. We are aware, that this is not the optimal scenario, but the second reason might make clear, why we went with it anyway.
						This second reason is the excessive simulation time, needed to simulate many of these systems. While the simulated time $t=5000$ appears to be only a little more than $15$ times the time simulated for spheres, we actually had to calculate around $200$ times more simulation steps for spherocylinders than for spheres, due to the fact that the latter require around $10$ times smaller time steps to be properly simulated. Apart from this some of the spherocylinder systems also contained more than twice as many particles as the largest systems with spheres. All put together this lead to simulation times between four to ten weeks.
						As this high run times were a hindrance for us not only in obtaining well equilibrated systems, but also in some other steps of this work, the end of this chapter will contain some discussion about attempts and ideas to solve this problem.  Most unfortunately we were not able to properly detect a pattern, which systems would not sufficiently equilibrate. \\
						To find out how the final potential energy $E_f$ of a system is related to its density and aspect ratios we actually plotted this final energy as a function of these two parameters in figure \ref{fig:Energietiles}. We can quite easily make out the pattern, that the final energy rapidly decreases as the density increases and the aspect ratio decreases. Both trends are hardly surprising since higher volume fraction, as well as smaller aspect ratio with constant volume fraction both directly translate into an increase in overall particles, hence the potential energy, that overall has negative sign decreases correspondingly. 
						

						\subsubsection{Evolution of maximum and mean cluster size}
\begin{figure}[H]
							\centering
							\scalebox{0.7}{\input{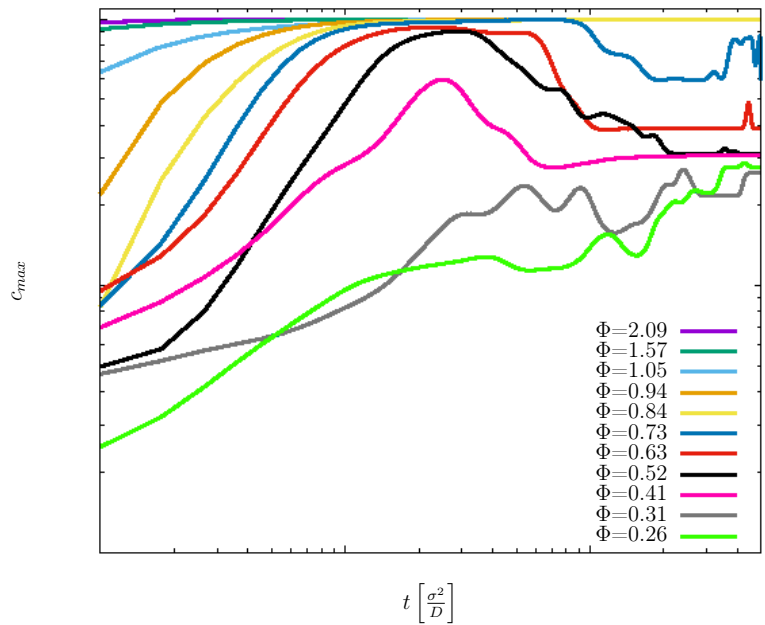}}
	
							\caption{   Evolution of the normalized maximum cluster size $c_{max}$ between $t=0$ and $t=05000\frac{\sigma^2}{D_0}$  . The aspect ratio is constant at $p=20$, while the volume fraction ranges between $\Phi=0.26\%$ and $\Phi=2.09\%$. The curves are smoothed to suppress short term fluctuations. }						\label{fig:maxclust20}
						\end{figure}

\begin{figure}[H]
						\centering
						\scalebox{0.7}{\input{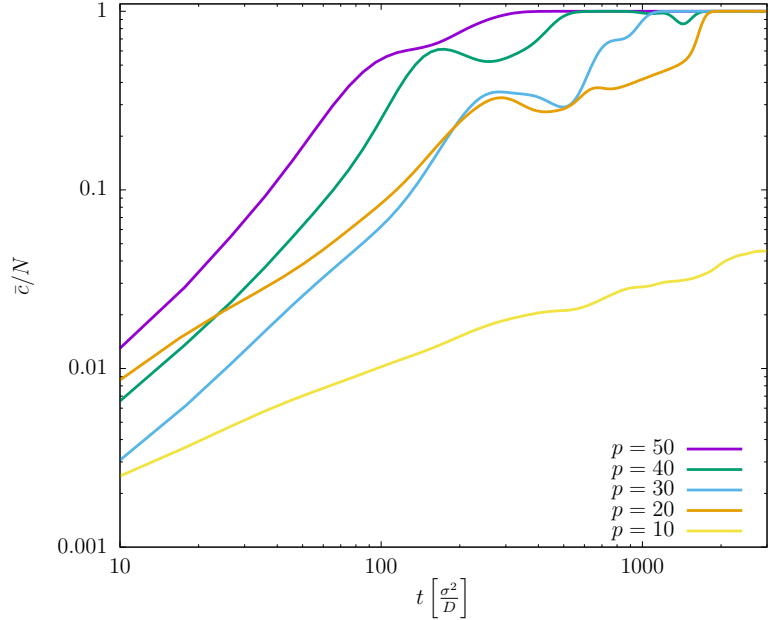}}

							\caption{   Evolution of the normalized mean cluster size $\bar{c}$ between $t=0$ and $t=5000\frac{\sigma^2}{D_0}$. The volume fraction is constant at $\Phi=1.05\%$, while the aspect ratio  ranges between $p=10$ and $p=50$. The curves are smoothed to suppress short term fluctuations. }	\label{fig:meanclust20}
					\end{figure}

						To find possible candidates for gels among our simulations we first of all have a look at the cluster sizes within the systems. The qualitative discussion of a few sample pictures at the beginning of this chapter suggested, that there is a dependency linking greater cluster size to a greater aspect ratio and a greater volume fraction. \\
						This first impression turns out to be right, when we have a closer look at the data we collected. To give a first impression of this phenomenon we depicted a plot of the time evolution of the normalized maximum cluster size  of the simulations of spherocylinders with aspect ratio $p=20$ in figure \ref{fig:maxclust20}. We can see there how all curves with a volume fraction greater than $\Phi=0.73\%$ quickly and monotonously go towards $1$, indicating that almost all particles of the systems are part of the greatest cluster. All curves below this density also rise, however do so much slower and come hardly as close to one, which means in these systems there is still a significant number of particles, that do not belong to the greatest cluster.\\
						All in all it seems reasonable to say, the initial slope of this evolution as well as the final value are both grow with increasing volume fraction.\\
						Similar observations can be made, when looking at figure \ref{fig:meanclust20}, where the evolution of normalized mean cluster size over the same time interval is depicted. In this figure the volume fraction is $\Phi=1.05\%$ for all the curves, while the aspect ratio ranges between $p=10$ and $p=50$. It is again quite nicely to see how the growth rate and the final value depend on the aspect ratio, as for example the system with $p=10$ never reaches $\bar{c}=1$, all the other curves reach that value in the same order, that corresponds to the inverse order of their aspect ratio, i.e. $p=50$ first, then $p=40$,etc..
						
\begin{figure}[H]
							\centering
							\scalebox{0.7}{\input{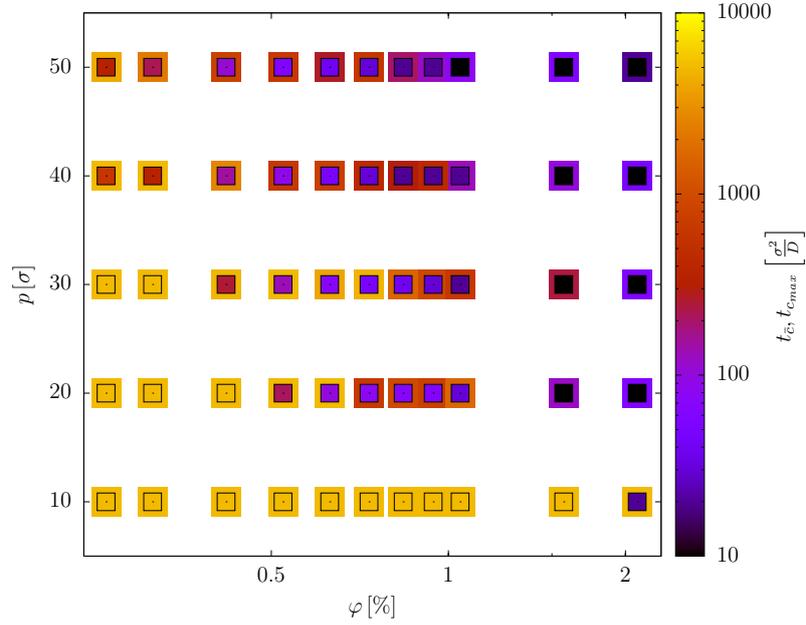}}
						
							\caption{$t_{c_{max}}$ (first time $c_{max}$ exceeds $0.95$) in dependency of the aspect ratio $p$ and the volume fraction $\Phi$ depicted in the inner square and $t_{\bar{c}}$ (first time $\bar{c}$ exceeds $0.95$) in dependency of the aspect ratio $p$ and the volume fraction $\Phi$ depicted in the outer square. in systems, where either time equals $5000$ the respective cluster size was never reached.}	\label{fig:Sclustertiles}
						\end{figure}

					As last proof of these tendencies we determined the times $t_{\bar{c}}$ and $t_{c_{max}}$ for all simulated systems. The points in time $t_{\bar{c}}$ and $t_{c_{max}}$ are respectively the first time $\bar{c}$ and $c_{max}$ exceed the value $0.95$ for the first time. Since very few systems fluctuate around this value for a bit, it is not an optimal parameter to describe the point, where all particles are in one cluster, yet it is a strong indicator. \\ The results for both times are depicted in figure \ref{fig:Sclustertiles}. The previously suggested trends can again be seen quite easily, as both cluster times decrease with growing aspect ratio and growing density. We should say, that of course neither of these relations is particularly surprising. Since a higher volume fraction implies that there are more particle in the same volume, there are obviously also more particles in close vicinity to each other. Even though greater aspect ratio on the other hand means less particles in the system, if keeping a constant volume fraction, their effective volume (volume they can cover via translation and rotation) grows with $p^3$. Apart from that one can of course interpret a spherocylinder with aspect ratio $20$ as two connected rods with $p=10$, which means more particles are already part of a cluster by default.\\
					This overall trend is well-known for rods \cite{Stabpercol} though to our knowledge was never shown for the example of rods interacting via a Kihara-Lennard-Jones potential.\\

						While in general a sufficient cluster size and percolation do not necessarily imply each other after visual inspection of the final states of our systems
						we can confidently say, that every system where mean or max cluster size  one was reached within the simulated times are actually wide-spread enough throughout the system to also be talking of percolation in these cases.

	%
	%
	%
	%
	%
						\subsubsection{Topological analysis of the final state}  
\begin{figure}[H]
							\centering
							\scalebox{0.7}{\input{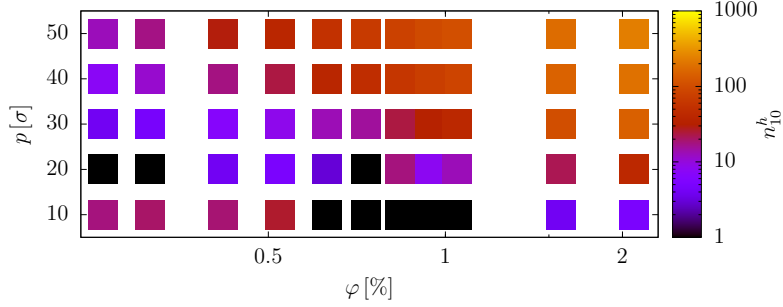}}
						
							\caption{Number of holes of radius $\approx 10$  $n^h_{10}$ in the final states in dependency of the aspect ratio $p$ and the volume fraction $\Phi$. }	\label{fig:sctopo}
						\end{figure}

Like earlier for spheres we calculated the number of holes the clusters in our final sates exhibit to have a quantitative measure how porous these networks actually are. We viewed holes of a radius of approximately $10$. The results are depicted in figure \ref{fig:sctopo}.
						The results for the systems with $p=10$ have to be handled with caution, due to the fact that the algorithm here cannot differentiate between a single spherocylinder and the endpoints of two spherocylinders. This leads to scenarios, where the algorithm could connect totally disconnected clusters and count holes in them. For $p\ge 20$ this risk decreases fast, partly because the algorithm approximates longer rods with more spheres and partly because less disconnected clusters exist in the first place.\\
						For all other systems we see a general increase in the number of holes $n^h_{10}$ as the density and the aspect ratios rise. This increase in porosity is a phenomenon we already witnessed, when qualitatively discussing the sample pictures in the introduction of this chapter, hence the overall trend is little surprising. Before moving on we would like to point the attention towards the fact, that the number of holes  in a cluster of spherocylinders can also be seen as an indicator for how far throughout the system this cluster is spread, since tightly packed spherocylinders do not allow any holes in their structure.\\
						Combined with the knowledge of cluster sizes derived from figure \ref{fig:Sclustertiles} the distribution of the number of holes in figure \ref{fig:sctopo} makes the statement, that systems with only few cluster actually percolate, at least plausible and confirms this exact claim we made after visually inspecting these systems. 
						
						\subsubsection{Angular correlation histograms}
					
\begin{figure}[H]
								\centering
								\scalebox{0.7}{\input{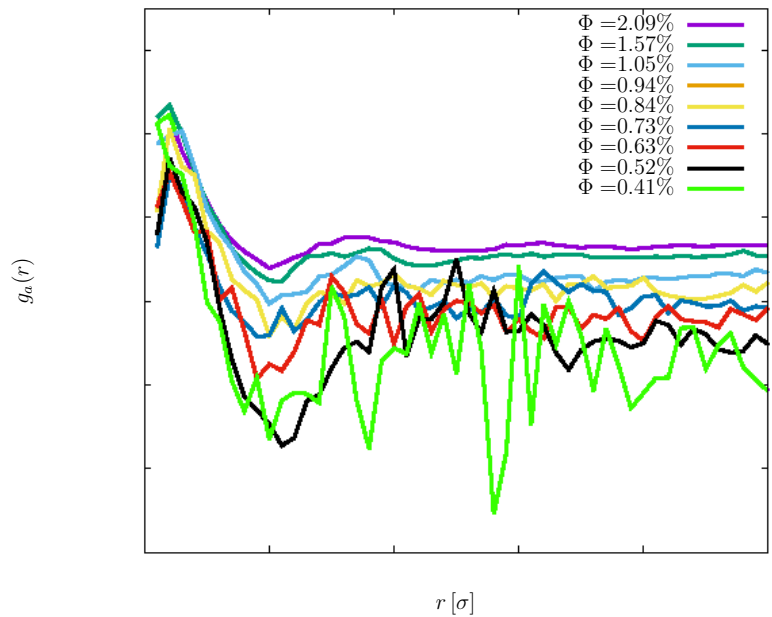}}
								
								\caption{Angular correlation function $g_(r)$ depending on the center of mass distance $r$ of spherocylinders with aspect ratio $p=20$ for volume fraction ranging between $\Phi=0.26\%$ and $\Phi=2.09\%$. }\label{fig:Winkel20}
							\end{figure}

\begin{figure}[H]
								\centering
								\scalebox{0.7}{\input{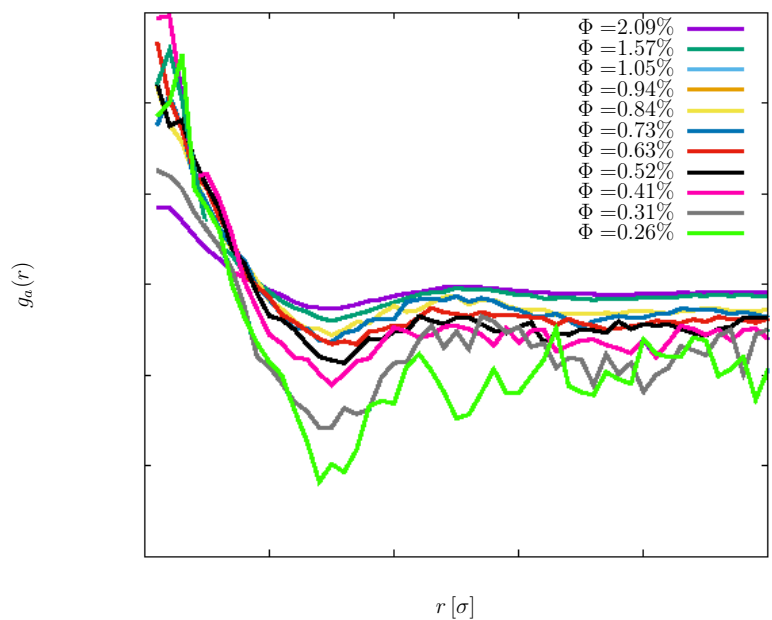}}
							
								\caption{Angular correlation function $g_(r)$ depending on the center of mass distance $r$ of spherocylinders with aspect ratio $p=30$ for volume fraction ranging between $\Phi=0.26\%$ and $\Phi=2.09\%$ }
								\label{fig:Winkel30}
							\end{figure}

	%
							Now, after having a rough overview of which system form porous percolating networks, we want to examine, whether any of these networks exhibit any orientational ordering or whether the spherocylinders are just randomly distributed. To do so we had a look at the angular distribution of the orientation of spherocylinders in dependence of the distance of the center of mass in our systems. \\
							Since we only have one sample for each pair of parameters we can, once again, only describe some tendencies in the data, due to lack of statistics. Two exemplary plots for the systems with $p=20$ and $p=30$ can be seen in figure \ref{fig:Winkel20} and figure \ref{fig:Winkel30} respectively.\\.

							 
							 The first thing that strikes the eye, when looking at these plots are the peak at $r=1$ and the low point at $r=p/2$, which every curve seems to have to some extend. These imply, that particles at a center of mass distance of $r=1$ tend to be more parallel then spherocylinders with greater distance and that particles with a distance of $p/2$ are more likely to be orthogonal to each other. Both make perfect sense. The first phenomenon because for spherocylinders with a center of mass distance of $1$ one of two degrees of freedom of the relative orientation are frozen, hence parallel orientation becomes more likely compared to pairs of particles, whose relative orientation can still change in two dimensions. The second one, because at a distance of $p/2$ parallel orientation would either cause overlap of rods or no contact between them at all, whereas orthogonal orientation is still likely to give contact between the rods without overlapping, hence making it energetically favorable. \\
							 Apart from these two extrema the angular distribution seems pretty independent of the distance. Except for some fluctuations all curves seem to settle in on a value around or slightly below zero, as the distance increases. The mentioned fluctuation are larger for low density systems, suggesting that they are rooted in lack of statistics rather than any ordering phenomenon.\\
							 In general it appears that systems of lower density tend towards a values slightly below zero as the center of mass distance grows, while the systems with greater volume fraction actually appear completely unordered at larger distance, hence go to zero. This could be explained as being an en tropically more favorable state, since the effective volume between particles is reduced this way, but since the values are already quite small, lack of proper statics in these system might as well be a just as suitable answer, till further simulations were conducted. \\
							 All in all it feels safe to say that, particularly in the percolating systems, there is no extraordinary ordering apart from effects dictated by the geometry of the spherocylinders.
	%
	%
						\subsubsection{Small amplitude oscillatory shear with and without Jeffery orbits}
\begin{figure}
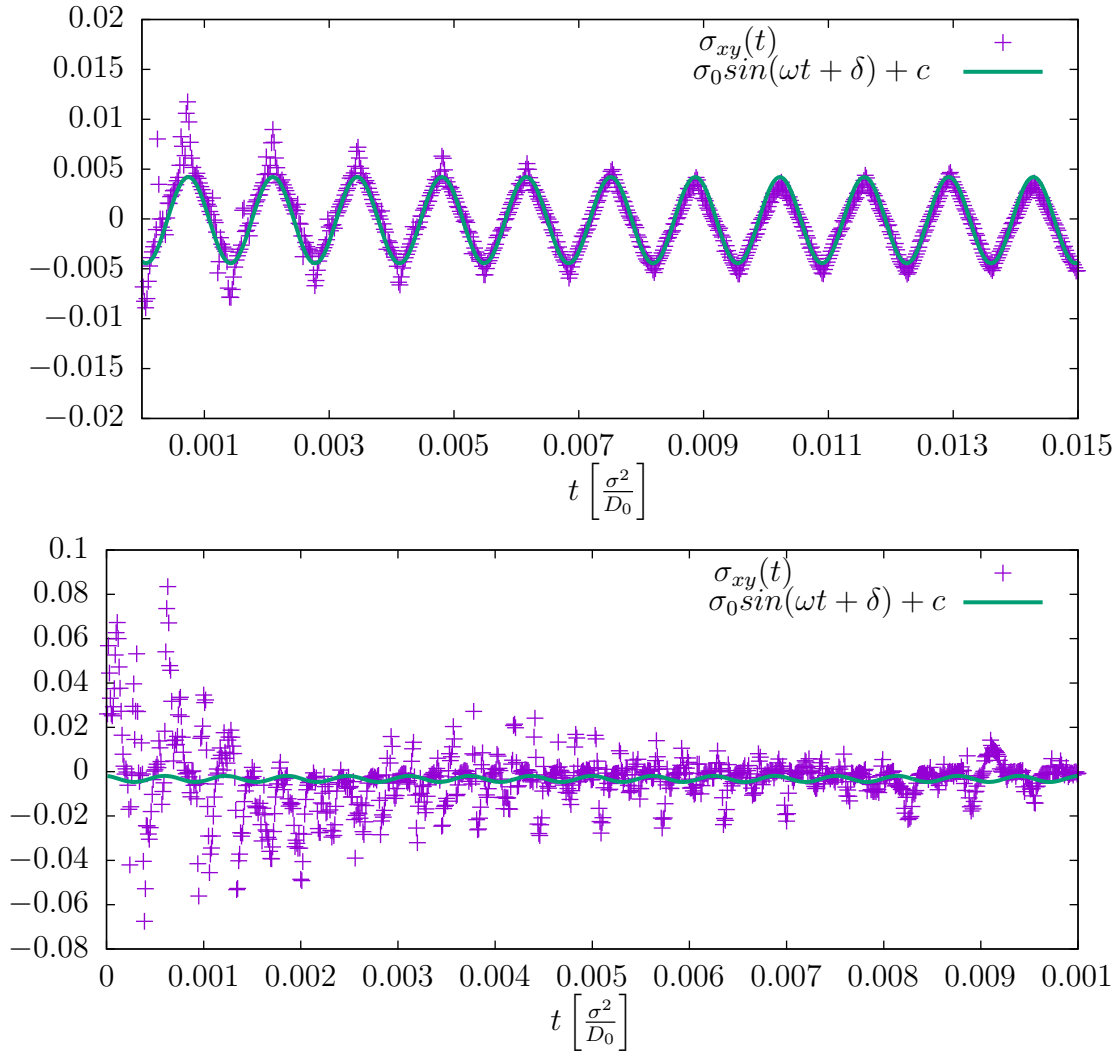

							\centering
							\input{gutfit.tex}\\
						\input{schlechtfit.tex}
							\caption{Example of a good and a bad fit for the determination of $G'$ $G''$ and $\tan(\delta)$}
							\label{fig:fitexp}
						\end{figure}

						In the last few sections of this chapter we suggested and confirmed the impression that porous percolating networks consisting of spherocylinders are favored by increasing volume fraction and aspect ratio. (At least in the low volume fraction regime we are looking at. It is quite clear that there are densities in which there is no more space for pores and every cluster necessarily percolates.)\\
						Now we want to see, whether any of these system exhibit the rheological properties of a gel, i.e. whether there exist timescales, on which the storage modulus is significantly higher than the loss modulus within the linear viscoelastic regime. \\
						 To do so we applied small amplitude oscillatory shear with several frequencies $\omega$ and maximum strains $\gamma_0$ to all systems. The exact values of $\gamma_0$ and $\omega$ use can be found in table \ref{tab:scpar}. We ran simulations, where the spherocylinders reacted as voluminous object, i.e. would perform Jeffery-orbits in a linear shear, and also simulations, in which the spherocylinders reacted like mere line segments to the shear force. We will refer to these two types of simulations as "with Jeffery-orbits" and "without Jeffery-orbits" from here on.\\
						 To find possible candidates for gel-like networks we first calculated the loss tangent $\tan(\delta)$ for all sets of parameters. The results of this calculations are plotted in figure \ref{fig:platzhalterjeff0} with the actual figures being spread throughout the pages \pageref{fig:sinuskurve} to \pageref{fig:sinuskurveend} for the simulations without Jeffery-orbits and through the pages  \pageref{fig:cosinuskurve} to \pageref{fig:nsinuskurveend} for the simulations with Jeffery-orbits. \\
						 It turns out that the results with Jeffery-orbits are  not significantly different from the results without Jeffery-orbits, which is why for the sake of not having eight consecutive pages of just pictures only plots for the latter are included in this chapter, while the other ones can be found at the very and of this work. The minor differences between simulations with and without Jeffery-orbits are hardly astonishing since the shear force did not exceed small amplitudes and maintained direction only for small amounts of times, hence the changes in orientation due to this force are also not expected to build up a lot. Due to the differences between the two kinds of simulations we will focus our attention on simulations without Jeffery-orbits from here on, even though every argument stays true if applied to the simulations with Jeffery-orbits, which can be easily seen by comparing the respective plots in figure \ref{fig:platzhalterjeff0} .\\
						 Probably the very first thing that strikes the eye, when looking at the plots belonging to figure \ref{fig:platzhalterjeff0} is the vast lack of data for aspect ratios of $p=40$ and $p=50$. Since in these plots all points, where the error of the fit required to calculate the loss tangent exceeded $20\%$, were left blank, this means that sinusoidal fit to the virial stress did not work very well in these systems. A closer look into the data exhibits indeed, that the Virial stress for these system is mostly dominated by random noise, rather than the periodic response to the shear stress. An exemplary plot for one of these cases compared to a case where the fit worked properly can be seen in figure \ref{fig:fitexp}. 
						 
						\begin{figure}[H]
							\centering
							(a)\boxed{\text{ Actual figures to be found on pages \pageref{fig:sinuskurve} to \pageref{fig:sinuskurveend}.} }\\ \vspace{1cm}
							(b)\boxed{\text{ Actual figures to be found on pages \pageref{fig:cosinuskurve} to \pageref{fig:nsinuskurveend}.} }
							\caption{The figures on the respective pages show the loss tangent's $\tan(\delta)$ dependency on frequency $\omega$ and maximum strain $\gamma_0$ for systems with aspect ratios ranging from $p=10$ to $p=50$ and volume fractions ranging from $\Phi=0.26\%$ to $\Phi=2.09\%$. Strain was applied without implementation of Jeffery-orbits in (a) and with such an implementation of Jeffery orbits in (b). \\ Every sub-plot shows the full range of frequency and strain for one set of aspect ratio and volume fraction. White tiles represent parameter sets, for which the error in the fit necessary to determine $\tan(\delta)$, was higher than $20\%.$}
							\label{fig:platzhalterjeff0}
						\end{figure}
						
					{	\begin{figure}[H]  
							\centering
							$\Phi=0.26\%$\hspace{2.7cm}$\Phi=0.31\%$\hspace{2.5cm}$\Phi=0.41\%$\\
							\rotatebox[]{90}{\hspace{5cm}$p=10$}	\input{Jeff0/links5000stickystick510.tex}\hspace{-1.25cm}
							\input{Jeff0/6000stickystick510.tex}\hspace{-1.25cm}
							\input{Jeff0/8000stickystick510.tex}\vspace{-3.8cm}\\
							\rotatebox[]{90}{\hspace{5cm}$p=20$}    	\input{Jeff0/links5000stickystick520.tex}\hspace{-1.25cm}
							\input{Jeff0/6000stickystick520.tex}\hspace{-1.25cm}
							\input{Jeff0/8000stickystick520.tex}\vspace{-3.8cm}\\
							\rotatebox[]{90}{\hspace{5cm}$p=30$}	\input{Jeff0/links5000stickystick530.tex}\hspace{-1.25cm}
							\input{Jeff0/6000stickystick530.tex}\hspace{-1.25cm}
							\input{Jeff0/8000stickystick530.tex}\vspace{-3.8cm}\\
							\rotatebox[]{90}{\hspace{5cm}$p=40$}
							\input{Jeff0/links5000stickystick540.tex}\hspace{-1.25cm}
							\input{Jeff0/6000stickystick540.tex}\hspace{-1.25cm}
							\input{Jeff0/8000stickystick540.tex}\vspace{-3.8cm}\\
							\rotatebox[]{90}{\hspace{5cm}$p=50$} \input{Jeff0/untenlinks5000stickystick550.tex}\hspace{-1.25cm}
							\input{Jeff0/unt6000stickystick550.tex}\hspace{-1.25cm}
							\input{Jeff0/unt8000stickystick550.tex}\vspace{-3.8cm}\\
							\input{cb.tex}
							\label{fig:sinuskurve}
						\end{figure}
							\begin{figure}[H]
								\centering
								$\Phi=0.52\%$\hspace{2.cm}$\Phi=0.63\%$\hspace{2.5cm}$\Phi=0.73\%$\\
								\rotatebox[]{90}{\hspace{5cm}$p=10$}	\input{Jeff0/links10000stickystick510.tex}\hspace{-1.25cm}
								\input{Jeff0/12000stickystick510.tex}\hspace{-1.25cm}
								\input{Jeff0/14000stickystick510.tex}\vspace{-3.8cm}\\
								\rotatebox[]{90}{\hspace{5cm}$p=20$}    	\input{Jeff0/links10000stickystick520.tex}\hspace{-1.25cm}
								\input{Jeff0/12000stickystick520.tex}\hspace{-1.25cm}
								\input{Jeff0/14000stickystick520.tex}\vspace{-3.8cm}\\
								\rotatebox[]{90}{\hspace{5cm}$p=30$}	\input{Jeff0/links10000stickystick530.tex}\hspace{-1.25cm}
								\input{Jeff0/12000stickystick530.tex}\hspace{-1.25cm}
								\input{Jeff0/14000stickystick530.tex}\vspace{-3.8cm}\\
								\rotatebox[]{90}{\hspace{5cm}$p=40$}
								\input{Jeff0/links10000stickystick540.tex}\hspace{-1.25cm}
								\input{Jeff0/12000stickystick540.tex}\hspace{-1.25cm}
								\input{Jeff0/14000stickystick540.tex}\vspace{-3.8cm}\\
								\rotatebox[]{90}{\hspace{5cm}$p=50$} \input{Jeff0/untenlinks10000stickystick550.tex}\hspace{-1.25cm}
								\input{Jeff0/unt12000stickystick550.tex}\hspace{-1.25cm}
								\input{Jeff0/unt14000stickystick550.tex}\vspace{-3.8cm}\\
								\input{cb.tex}
							\end{figure}
							\begin{figure}[H]
								\centering
								$\Phi=0.73\%$\hspace{2.7cm}$\Phi=0.74\%$\hspace{2.5cm}$\Phi=0.94\%$\\
								\rotatebox[]{90}{\hspace{5cm}$p=10$}	\input{Jeff0/links14000stickystick510.tex}\hspace{-1.25cm}
								\input{Jeff0/16000stickystick510.tex}\hspace{-1.25cm}
								\input{Jeff0/18000stickystick510.tex}\vspace{-3.8cm}\\
								\rotatebox[]{90}{\hspace{5cm}$p=20$}    	\input{Jeff0/links14000stickystick520.tex}\hspace{-1.25cm}
								\input{Jeff0/16000stickystick520.tex}\hspace{-1.25cm}
								\input{Jeff0/18000stickystick520.tex}\vspace{-3.8cm}\\
								\rotatebox[]{90}{\hspace{5cm}$p=30$}	\input{Jeff0/links14000stickystick530.tex}\hspace{-1.25cm}
								\input{Jeff0/16000stickystick530.tex}\hspace{-1.25cm}
								\input{Jeff0/18000stickystick530.tex}\vspace{-3.8cm}\\
								\rotatebox[]{90}{\hspace{5cm}$p=40$}
								\input{Jeff0/links14000stickystick540.tex}\hspace{-1.25cm}
								\input{Jeff0/16000stickystick540.tex}\hspace{-1.25cm}
								\input{Jeff0/18000stickystick540.tex}\vspace{-3.8cm}\\
								\rotatebox[]{90}{\hspace{5cm}$p=50$} \input{Jeff0/untenlinks14000stickystick550.tex}\hspace{-1.25cm}
								\input{Jeff0/unt16000stickystick550.tex}\hspace{-1.25cm}
								\input{Jeff0/unt18000stickystick550.tex}\vspace{-3.8cm}\\
								\input{cb.tex}
							\end{figure}
							\begin{figure}[H]
								\centering
								$\Phi=1.05\%$\hspace{2.7cm}$\Phi=1.57\%$\hspace{2.5cm}$\Phi=2.09\%$\\
								\rotatebox[]{90}{\hspace{5cm}$p=10$}	\input{Jeff0/links20000stickystick510.tex}\hspace{-1.25cm}
								\input{Jeff0/30000stickystick510.tex}\hspace{-1.25cm}
								\input{Jeff0/40000stickystick510.tex}\vspace{-3.8cm}\\
								\rotatebox[]{90}{\hspace{5cm}$p=20$}    	\input{Jeff0/links20000stickystick520.tex}\hspace{-1.25cm}
								\input{Jeff0/30000stickystick520.tex}\hspace{-1.25cm}
								\input{Jeff0/40000stickystick520.tex}\vspace{-3.8cm}\\
								\rotatebox[]{90}{\hspace{5cm}$p=30$}	\input{Jeff0/links20000stickystick530.tex}\hspace{-1.25cm}
								\input{Jeff0/30000stickystick530.tex}\hspace{-1.25cm}
								\input{Jeff0/40000stickystick530.tex}\vspace{-3.8cm}\\
								\rotatebox[]{90}{\hspace{5cm}$p=40$}
								\input{Jeff0/links20000stickystick540.tex}\hspace{-1.25cm}
								\input{Jeff0/30000stickystick540.tex}\hspace{-1.25cm}
								\input{Jeff0/40000stickystick540.tex}\vspace{-3.8cm}\\
								\rotatebox[]{90}{\hspace{5cm}$p=50$} \input{Jeff0/untenlinks20000stickystick550.tex}\hspace{-1.25cm}
								\input{Jeff0/unt30000stickystick550.tex}\hspace{-1.25cm}
								\input{Jeff0/unt40000stickystick550.tex}\vspace{-3.8cm}\\
								\input{cb.tex}
								\label{fig:sinuskurveend}
							\end{figure}
						}
						
\begin{figure}[H]
	\centering
	\includegraphics[width=0.45\linewidth]{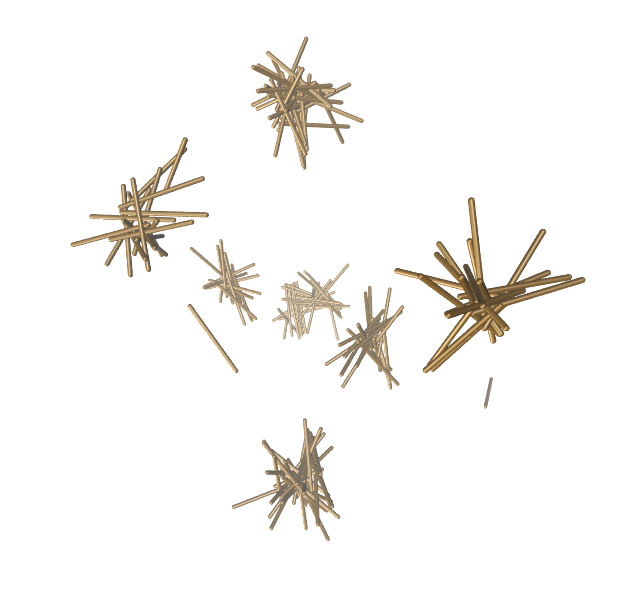}\rotatebox{90}{\hspace{-0.5cm}\rule{8cm}{0.05cm}}
	\includegraphics[width=0.45\linewidth]{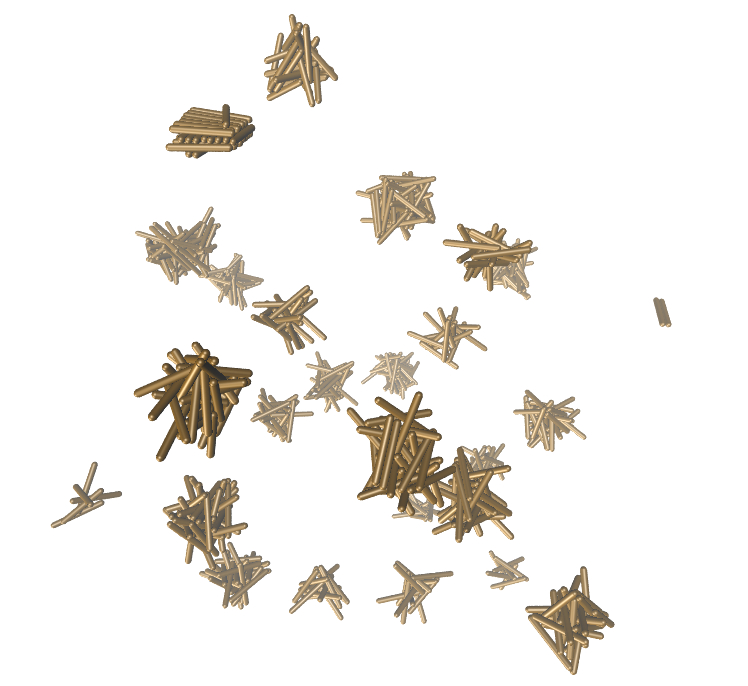}
	\rule{16cm}{0.05cm}
	\includegraphics[width=0.8\linewidth]{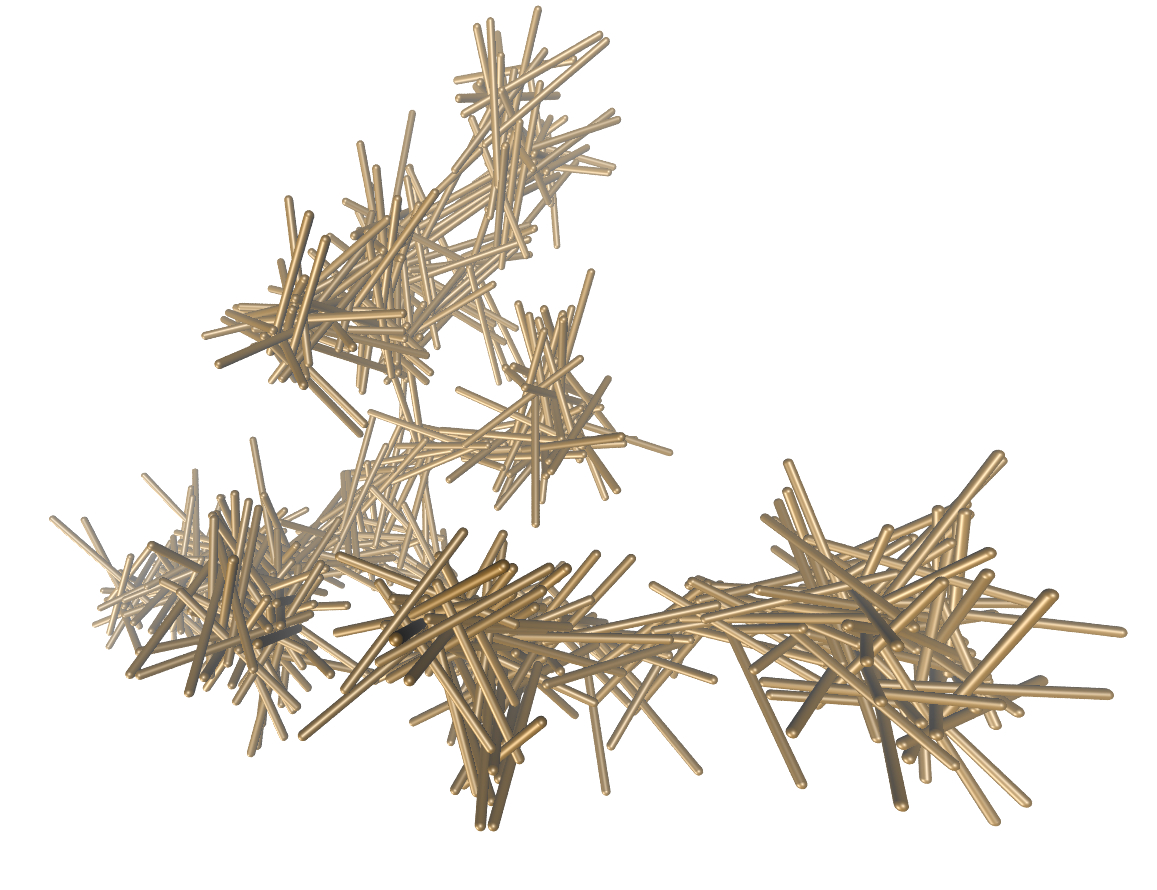}
	\caption{Three systems with the most pronounced solid-like behavior Depicted are the system with $\Phi=0.26\%$ and $p=20$ (top left), the system with $\Phi=0.52\%$ and $p=10$ (top right) and the system with $\Phi=0.74\%$ and $p=20$ (bottom).  }
	\label{fig:candgel}
	\end{figure}

						One might suggest that the reason for this noise is, because these systems are not completely equilibrated yet. This argument does not apply to all systems with $p\ge 40$ though and at the same time the method worked fine for systems with $p\le 30$, which were not at equilibrium either. (Note that it was stated earlier, that there was no clear pattern, which systems equilibrated sufficiently.)\\
						It might be possible that the angular momentum caused by the shear force, which becomes greater with bigger aspect ratios, added up so much, that greater rearrangements in these systems occurred, leading to a greater amount of noise. It might also play a role that our diffusion constants for spherocylinder are only correct for aspects ratios smaller than or equal to $p=30$, hence making all systems with $p=40$ and $p=50$ a mere toy model anyway.	 Being aware of this flaw in our data we move on to discuss the remaining cases.\\
						The remaining data follows more or less one pattern: the loss tangent is almost constant $\lessapprox 1$ for $\gamma_0 \gtrsim 1\%$ and frequencies $\lessapprox 10^4$. This implies that, in these parameter ranges the systems are all more or less behaving like a viscoelastic fluid slightly more solid-like than fluid-like on average. For $\gamma_0\lessapprox 1\%$ and $\omega\gtrsim 10^4$ most systems are likely to have a loss tangent of less than $0.01$, hence clearly exhibit solid-like character.\\
						It is very pleasant to see, that this is, up to fluctuations, the exact same picture we witnessed earlier for spheres in figure \ref{fig:kugeltiles}. If anything the solid-like regions are more distinct for spherocylinders then they were earlier for spheres.\\
						In the context of this work it is kind of baffling, that neither form nor size of these two parameter regions seem to depend on the aspect ratio or the volume fraction in any obvious way. In fact the systems where the solid-like parameter scope is most pronounced are the system with $\Phi=0.26\%$ and $p=20$, the one with $\Phi=0.52$ and $p=10$ and finally the one with $\Phi=0.74\%$ and $p=20$. \\
						Since these systems are particularly distinct in their solid-like parameter region we are going to have a closer look at them here.
						The respective final state of these systems are depicted in figure \ref{fig:candgel}.
	
\begin{figure}
	 	\centering
	 	\input{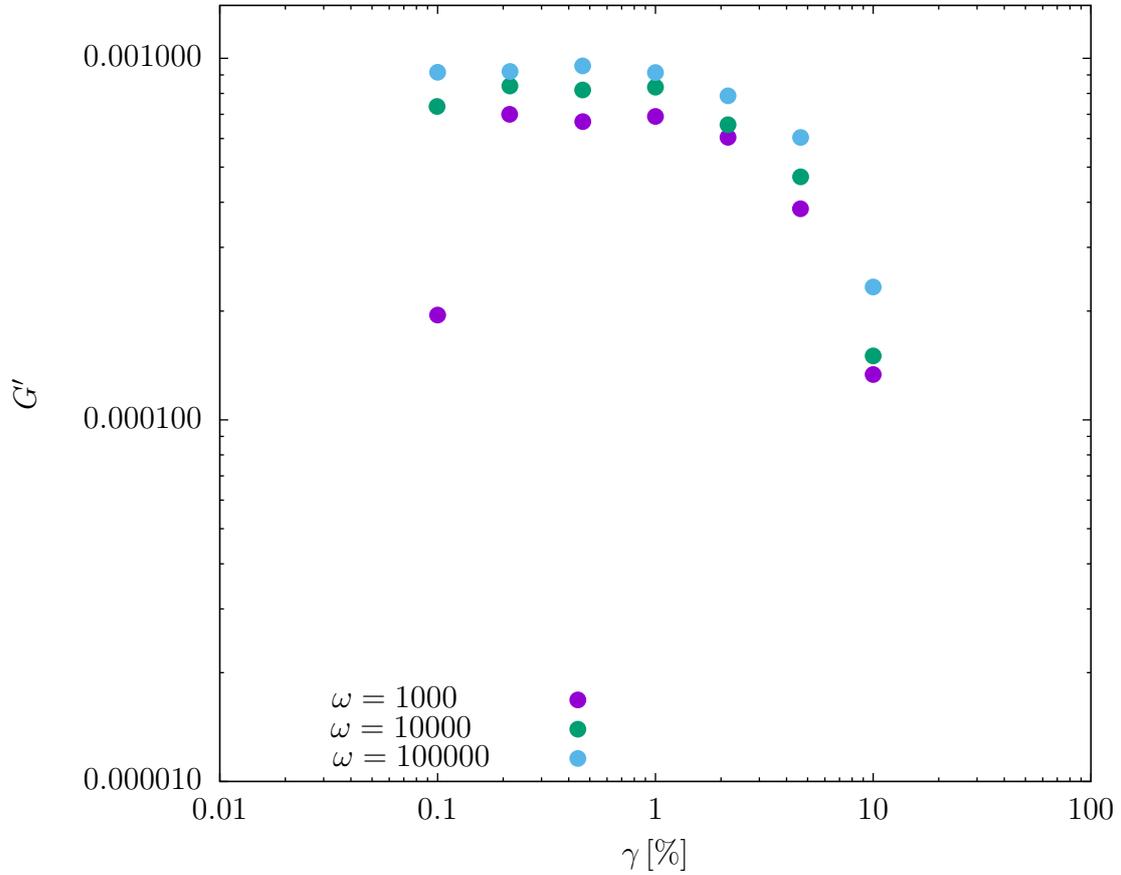}
	 	\caption{Storage modulus $G'$ dependency on maximum strain $\gamma_0$ for the system with $\Phi=0.74\%$ and $p=20$ for the frequencies $\omega=1000$(violet),$\omega=10000$(green)and $\omega=100000$(blue). The simulations from which G' was derived were performed at $t=5000$ and without Jeffery orbits.}
	 	\label{fig:candgamma}
	 \end{figure}

	While all of their loss tangents exhibit the same characteristics as our system of gelated spheres only 
	the last one ($\Phi=0.74\%$, $p=20$) is a percolating network. Hence of the three it appears to be the only and overall the most promising candidate for a gelated system. We therefore will have a closer look at its storage and loss moduli in order to determine, whether the linear viscoelastic regime of this system coincides with the parameter range of solid like behavior. \\
	The storage modulus $G'$ for this system in dependency of $\gamma_0$ is depicted in figure \ref{fig:candgamma} for a few samples of $\omega$ between $1000$ and $100000$. Apart from one runaway point for $\omega=100$ (violet) at $\gamma_0=0.1\%$ $G'$ appears to be more or less constant up to values of $\gamma_0$ around $2\%$ or $3\%$, suggesting that for amplitudes below these values we can assume to be in the linear viscoelastic regime. We should also note that $G'$ is about two magnitudes smaller than, when we calculated it earlier for the sphere system. This can be explained by the density of this system, that is about 20 times lower then earlier in the sphere system.
	
\begin{figure}
		
		\hspace{-1cm}\input{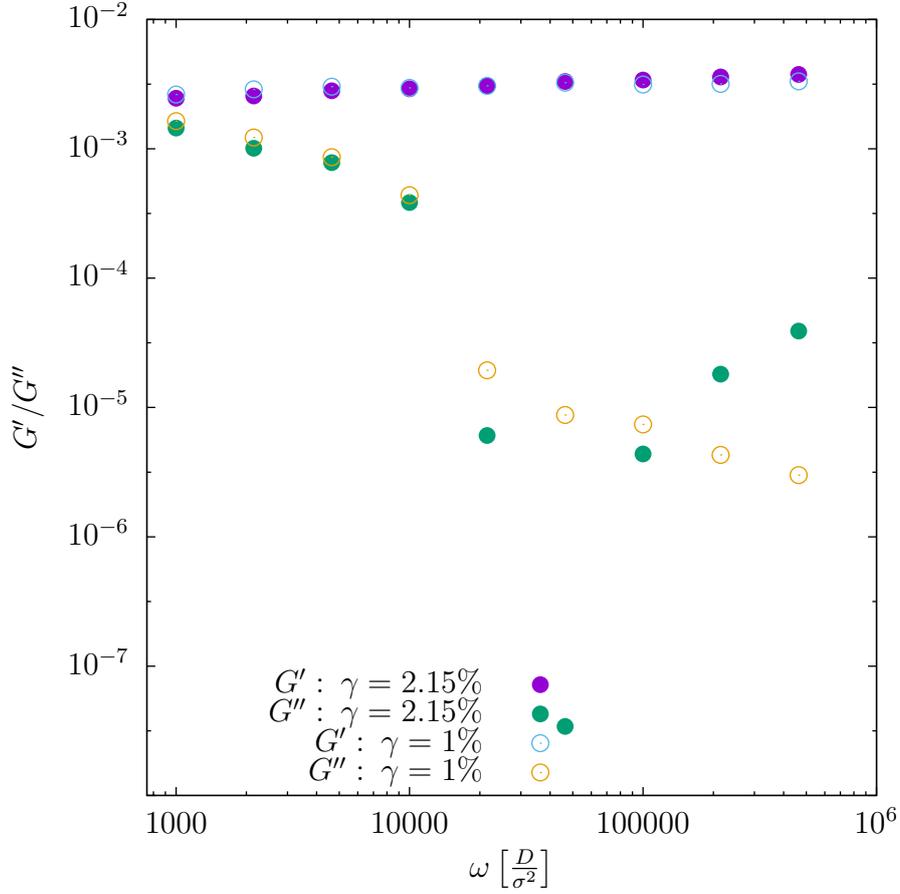}
		\caption{Storage modulus $G'$ and loss modulus $G''$ dependency on the frequency $\omega$ for the system with $\Phi=0.74\%$ and $p=20$ for the maximum strains $\gamma_0=1\%$ and $\gamma_0=2.15\%$ and without Jeffery orbits}
		\label{fig:candomega}
	\end{figure}

	 The frequency dependency of $G'$ and $G''$ finally is depicted in figure \ref{fig:candomega}. We depicted it for $\gamma_0=1\%$ and $\gamma_0=2.15\%$. We cannot see any gel point in this figure, i.e. the loss modulus is for all frequencies we could get noise free data smaller than the storage modulus. Starting from frequencies around $10^4$ and higher the loss modulus starts to drop rapidly, so that we can definitely speak of solid like behavior on these time scales. \\
	 While this in theory would speak for a gelated system, we should nevertheless point out, that the behavior of $G''$ in this sample appears to be rather odd compared to the spherical system, and does jump a lot over many orders of magnitude. 

						\subsection{Discussion}
						Throughout this last chapter we could confirm that the existence of percolating porous networks in systems of spherocylinders correlate with the density of the particles and their aspect ratio. Within these networks we could not find any particular order of orientation. \\
						After the first few sections one would assume that the occurrence of systems with the rheological properties of a gel would also increase with rising volume fraction and aspect ratio. Unfortunately our method appeared to produce mainly noise for $p\ge 40$. Due to the diffusion constant only being realistic for $p\le30$ it would have been  better to ignore systems with $p>30$ all together and sweep the range of $p=2$ to $p=30$ more thoroughly. \\
						Nonetheless we managed to actually find several systems that fit the criteria of a gel, thus proofing, that our model  can generate gel-like structures based on percolating networks of spherocylinders. 
						This is a successful extension of the results of \cite{kugelgel} from spheres to the more complicated systems of sticky rods.\\ It would have been nice to actually figure out a pattern according to which the rheological properties of our systems depend on the volume fraction and the aspect ratio. Due to the rather high fluctuations in the data depicted in figure \ref{fig:platzhalterjeff0} more statistics for our results is needed, and hence more simulations should be run in the future.\\ 
						In general a little more statistics would have been desirable for all results shown above. Since we deem the results very promising so far we suggest that the corresponding simulations will be conducted in the near future. Since some of the systems included in this work require simulation times of two months and more we propose that further attempts to decrease these run times are undertaken beforehand.

					 As conclusion of this chapter we will discuss a few attempts in that direction , that were already made or that we suggest to be tested in the future. \\
						\subsubsection{Performance enhancement for spherocylinder simulations}
						Before I started using the  simulation program written by Ullrich Siems, already used linked-cells and Verlet-lists in order to speed up its performance. (The latter needed slight modifications to be used in the context of spherocylinders.) The two methods combined lead to the simulation times described a few sentences before. Without them these times would be even higher.  \\
						Linked-cells can be used, when a cut-off radius for the interaction of particles is used. In that case the simulation box can be divided into smaller boxes, with edges at least as long as the cut-off radius, and interactions need only to be calculated for particles in the same box or neighboring boxes. If the cut-off radius is sufficiently smaller than the simulation box, this gives a significant speed up for the calculation of interaction. $\mathbb{O}(n^2)\rightarrow \mathbb{O}(n)$.\\
						When using Verlet-lists one keeps a list for every particle in the system, on which all particles within a distance of $r_{cut}+\delta$ are kept track of.($\delta$ is the Verlet constant and can be chosen at will.) Only after a particle has moved a distance greater than $\delta$ these lists need to be updated in order to calculate the correct interactions. This procedure also reduces the number of particle-particle interactions that actually have to be calculated on average drastically.\\
						While both methods are very effective the more noteworthy contribution usually comes from the linked cells. However for them to work effectively each cell together with its direct neighbors needs to  be smaller than the whole simulation  box, i.e. in ever direction of the box there need to be more than three linked cell. Unfortunately this very fact is was makes them less useful, when applied for spherocylinders. The reason being that in these cases the length of linked cells needs to be at least as long as $p+r{_{cut}}$. So for linked-cells to be an essential speed up when simulating long spherocylinders the simulation  box needs to be sufficiently big, which assuming a constant volume fraction of particles requires an increased number of spherocylinders. For long spherocylinders this increase in number can eat up the initial increase in speed quite fast. \\
						Obviously one standard way to increase performance would be an attempt to parallelize the program and let multiple processors do the work, that currently one does. One standard way to do this, which also is already implemented in the program, are, once again, linked cells. Here each processor would calculate a system consisting of the particle in a block of several linked cells. The processor only need to share the information about particles in the boundary of this block. However if there are too little linked cells, every linked cell becomes a boundary cell of its respective block. In this case the communication between processors can need so much time compared to the actual calculations that one ends up without a net gain in performance. This again is a problem for spherocylinders, since their increase in numbers, which is needed to have sufficiently many linked cells for this kind of parallelization to work, can eat up the performance gain very quickly again. (It might be possible to get around this problem, when GPUs rather than CPUs are used to deal with the vast amount of extra particles.)\\
						In another attempt to parallelization we tried to let each Verlet-list be handled by a different processor using OpenMP. However also in this attempt inter-processor communication ate up the gain in performance.\\
						 We believe the reason for the failure in the last attempt to be mainly rooted in the simulation program itself. While it is a very well-written program capable of many different tasks, off all things this versatility might be its doom in this case. Grown over the years, through work of a few different people the program grew more and more complex. Checking out optimization reports from its compilation reveals, that there is little to none automated vectorization or parallelization, due to the fact that the optimizer has difficulties to identify, whether variables depend on each other, which makes very clear why so much interprocessor communication was needed in previous parallelization attempts.   \\
						 Further this explains to great extend how
						  Anton Lüders and Jacob Holder could realize some significant speed ups by writing new programs, based on the same original program, but specialized for only one of its task. With this in mind we would suggest the same to be done for spherocylinder simulations as well and starting from there to try again its parallelization through CPU or GPU, before many $3d$-simulation of many spherocylinders are run with the original program again. 
						Obviously several other optimization methods could also be tested. An attempt, that I personally deem promising, but never tested would be a shift from the Brownian dynamics algorithm to the very similar Smart Monte Carlo algorithm.\cite{SmartMC} This kind of algorithm is particularly designed for systems that need a long time to reach equilibrium and above all could be trivially parallelized.

						\cleardoublepage

						\section{Comparison to experiments\label{sec:exp}}
\begin{figure}[H]
	\centering
	\includegraphics[width=0.7\linewidth]{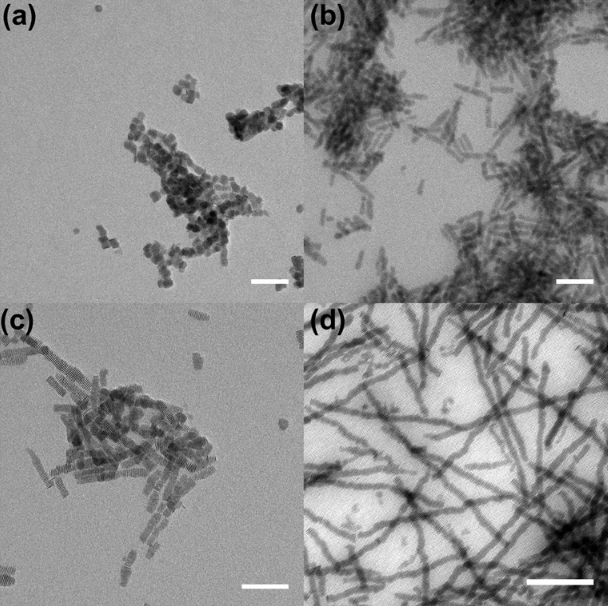}
	\caption{Transmission electron microscopy micrographs of $Eu_2O_3$-benzoate particles (a; scalebar 50nm b; scalebar 50nm c; scalebar nm d; scalebar
	100nm)\cite{Euro}. }
	\label{fig:eurogel}
	\end{figure}

\begin{figure}[H]
	\centering
 \scalebox{0.8}{   \input{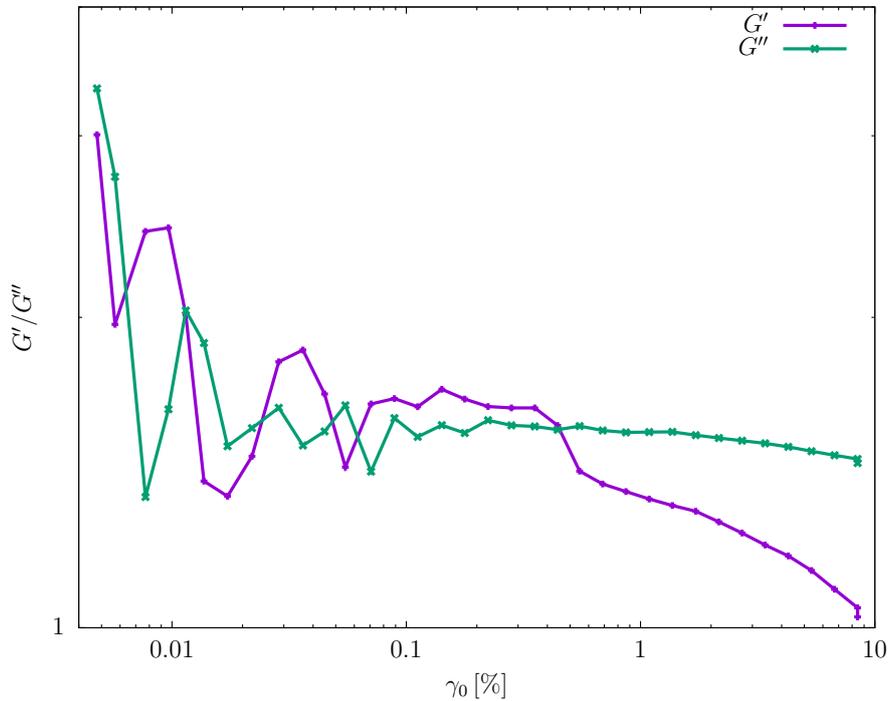}}
	\caption{Experimental results of the storage modulus (violet) and stress modulus (green) of a gelated solution of EuO-based nanrods in dependency of the maximum shear amplitude $\gamma_0$ for a  constant frequency of $\omega=10\frac{Rad}{s}$.}
	\label{fig:gammaexp}
	\end{figure}

\begin{figure}[H]
		\centering
\scalebox{0.8}{	\input{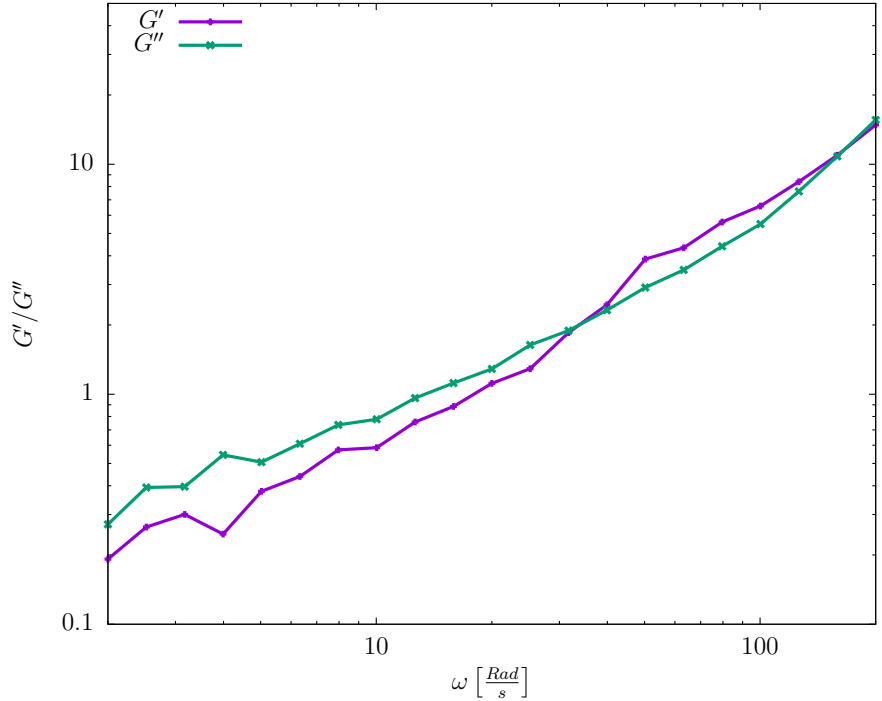} }
		\caption{Experimental results of the storage modulus (violet) and stress modulus (green) of a gelated solution of EuO-based nanrods in dependency of the shear frequency $\omega$ for constant maximum shear amplitude of $\gamma_0=2\%$}
		\label{fig:omegaexp}
	\end{figure}

						As stated earlier in the introduction Bastian Trepka and the group of Professor Polarz were able to grow EuO-based nanorods in a solution of benzyl alcohol, while controlling the aspect ratio of these rods. \cite{Euro}. It was further realized that these solutions of EuO-rods gelate as soon as the aspect ratio increases beyond $p=20$.  A picture of their structure can be found in figure \ref{fig:eurogel}. 
	As stated before we were able to observe comparable network formation in dependence of the aspect ratio. Even though this behavior is well established \cite{Stabpercol} these results make us optimistic that our model could be appropriately to describe their experimental systems. \\
	Thanks to Jacob Steindl  and the group of Clemens Bechinger it was also possible to conduct a few rheological measurements on the gels grown by Bastian Trepka. The results can be seen in the figures \ref{fig:gammaexp} and \ref{fig:omegaexp}. In figure \ref{fig:gammaexp} we see  the storage and stress modulus of the gel during an maximum amplitude sweep. We see quite well, that we are in the linear viscoelastic regime, as both are quite constant with respect to $\gamma_0$(,up to some noise for small amplitudes). We further see, that apparently both moduli are very close to each other, i.e. we are dealing with a very weak gel, if the term is even applicable here.  
	Figure \ref{fig:omegaexp} shows us the frequency-sweep for the two moduli. We do see an approximately linear growth of both moduli with frequency. There are two points were the moduli curves intersect. The loss tangent does not deviate far from 1. While we simulated a few systems that have a loss tangent of approximately 1 over a large spectrum of parameters, we have to say that we did not observe this kind of linear behavior of the moduli in any of our systems. \\
	I appears the model with the parameters we tried out is unfit to describe this particular systems' rheological properties. Retrospectively this result does not come unexpected. Knowing close to nothing about the experimental system except the aspect ratio of the rods it would have been very unlikely to get all of the important parameters right even if the model were suitable to describe the experiment. Since we did not vary the interaction strength of the rods in any of our simulations, there are still plenty of possibilities for the model to work in this particular case.

						\cleardoublepage
						\section{Outlook}
						At the end of chapter \ref{sec:results} we already explained how long simulation times limited our capability to produce a lot of statistics for our model. We also made a few suggestions how the performance of the program might be. \\
						The task for anyone diving deeper into this problem would be to implement these, and possibly more,  improvements to a code and after that recreate the simulations discussed in this work in greater number to obtain sufficient statistics. The tools to analyze and compare these results are ready at hand and free to use. Since we showed that gelated rod networks can be obtained our method, we are sure that this would be a worthwhile path to go down. As stated in chapter \ref{sec:exp} one could extend these parameter studies  to different interaction strengths and forces. We would even further suggest to also increase the parameter density for the parameters we already examined.\\
						During the course of this work we did stumble across many ideas one could look at in the future, which were not carried out so far. We would like to mention just of few of them here as further suggestions for future research.\\
						We already mentioned the cuboid-like systems of spherocylinders depicted in figure \ref{fig:cuboid} and discussed a little in section \ref{sec:evalstab}. These interesting formations  are formed in comparably low numbers in the lower density systems with aspect ratio $p=10$ and rarely in systems with $p=20$. We could imaging that more simulations are made to see how often these cuboids actually appear and how exactly this is related to the aspect ratio and the volume fraction. Maybe some external forces or different potential could be implemented to favor the growth of cuboids. \\
						If it would proof difficult or impossible to fabricate these cuboids in higher densities compared to more randomly oriented clusters, one could test whether they could be separated from the other clusters. As there effective volume per spherocylinders included in in there formation is likely to  be smaller than in other random clusters, it could for example be possible to separate them from the other particles by forcing them through some kind of mesh-geometry with a driving force.\\
						As there existence in general is already shown by our simulations, one could also move straight on to simulating these formations in greater number and analyze their properties as colloids (mean square displacement,distribution functions, ordering phenomena,...).
						\\ A further idea that occurred during the course of this work was to use the resulting networks of spherocylinders as membranes and analyze the properties of spheres moving through these networks. \\
						 Finally it would of course be very tempting to move on from Brownian dynamics to more realistic models, including hydrodynamic effects. This could be obtained by conducting molecular dynamics simulating of spherocylinders surrounded by much smaller spherical particles, i.e. actually simulating the surrounding fluid rather than approximating it via the Langevin-equation. While this of course blows up the number of particles that need to be simulated a system like this would be a little easier to parallelize compared to the pure spherocylinder systems. The reason for this is that the lion's share of computation time would be needed to calculate the movement of the fluid particles. Since these can be assume to be spherical, one could use the well known and very effective standard methods in order to speed up most of the calculations.	\\
						 All in all there are many ideas to be followed, tools and a program, that with small adaptations,  can be used to explore any of them.  Based on our results we are generally very optimistic that there are plenty of interesting results and phenomena to be found in future research related to this work.

		\cleardoublepage
		
		\section[Kurze Zusammenfassung auf Deutsch (Short summary in German)]{\color{white}}
					  \section*{\vspace{-3.7cm}\hspace{2.5cm}\Large Kurze Zusammenfassung auf Deutsch \\\vspace{-0cm}\hspace{2.5cm}(Short summary in German)}
					  
Die hier vorliegende Arbeit befasste sich mit Computersimulationen von Kolloidsystemen. Genauer wurden Systeme bestehend aus klebenden Kugeln bzw. klebenden Stäbchen mit einem Brownsche Dynamik Algorithmus simuliert und dahingehend untersucht, ob sich in ihnen poröse, perkolierende Netzwerke mit den rehologischen Eigenschaften von Gelen bilden. Ursprüngliches Ziel der Arbeit war mit Hilfe eines Programms, das in großen Teilen von Ullrich Siems \cite{Ullidiss} geschrieben wurde, die experimentellen Ergebnisse von Bastian Trepka zu reproduzieren. Bastian Trepka züchtet EuO-basierte Nanostäbchen, die, sobald ihr Aspekteverhältnis einen Wert von etwa $20$ überschreitet, ein Gel mit ihrem Lösungsmittel bilden.\cite{Euro}\\
Diese Arbeit enthält zunächst eine kurze Heranführung an die physikalischen und numerischen Grundlagen, die nötig sind um die benutzten Simulationsmethoden sowie die Interpretation der Resultate zu verstehen. Es folgt ein Überblick über alle  Parameter, mit denen wir Simulationen durchgeführt haben, um größtmögliche Reproduzierbarkeit zu gewährleisten. Im Anschluss daran werden unsere Ergebnisse detailliert vorgestellt. Erst werden die Simulationen zu klebenden Kugeln evaluiert. Mit diesen versuchten wir, die Ergebnisse von Santos, Campanella und  Carignano \cite{kugelgel} zu reproduzieren, um nachzuweisen, dass unsere Software und unsere Methoden einwandfrei funktionieren. Es folgen die Ergebnisse der Simulationen von klebenden Stäbchen inklusive der Analyse der Rheologie und Topologie der entstandenen Systeme. In Kapitel 6 werden die letzteren Ergebnisse mit den experimentellen Daten verglichen, die von Jabcob Steindl zu Bastian Trepkas Gelen gesammelt wurden. Abschließend wird ein Überblick über Vorschläge und Ideen gegeben, wie ausgehend von den Ergebnissen dieser Arbeit in Zukunft weitere Forschung betrieben werden kann.\\
Obwohl erste Vergleiche mit den experimentellen Daten nur wenige Hinweise darauf geben, dass das von uns genutzte Model dazu in der Lage ist, die von Bastian Trepka hergestellten Gele zu beschreiben - was das ursprüngliche Ziel dieser Arbeit war - halten wir diese Arbeit dennoch insgesamt für einen Erfolg und denken, dass unsere Ergebnisse äußerst vielversprechend sind.\\
Es war uns möglich das Paper von Santos, Campanella und Carignano \cite{kugelgel} so vollständig zu reproduzieren, wie wir dies anfangs geplant hatten. Gleichzeitig konnten wir ihre Ergebnisse für Kugeln auf deutlich kompliziertere Systeme bestehend aus klebenden Stäbchen erweitern.\\
Weiter konnten wir einen eindeutigen Zusammenhang zwischen der Volumendichte dieser Stäbchen und ihrem Aspekteverhältnis zu ihrer Eigenschaft, perkolierende, poröse Netzwerke zu bilden feststellen. Obwohl dieses Ergebnis wenig überraschend und eigentlich wohlbekannt ist, wurde es dem Stand unserer Kenntniss nach bisher nicht für Stäbchen mit Kihara-Lennard-Jones Wechselwirkung beobachtet.\cite{Stabpercol} Unsere Analyse der topologischen Struktur dieser Netzwerke zeigt interessante Ergebnisse und könnte nützliche Einblicke in vergleichbare Systeme geben. Gleichzeitig könnte die Methode genutzt werden, um Bilder von echten Gelen untereinander oder mit den Resultaten von Simulationen  automatische zu vergleichen.\\
Zuletzt zeigt die rheologische Analyse, dass wir in der Tat in der Lage waren, einige Systeme von klebenden Stäbchen zu simulieren, welche die charakteristischen Eigenschaften von Gelen aufweisen.\\
Damit wurde der Nachweis erbracht, dass unsere Werkzeuge und Methoden funktionieren und benutzt werden können, um derartige Gelstrukturen auf Basis von Stäbchen zu generieren. Ausgehend von diesen Resultaten stehen viele Türen für spannende zukünftige Projekte weit offen.
					  
					  	\cleardoublepage
					\appendix
					\section{Appendix}
					\subsection{Shortest distance between spherocylinders \label{App:distspher}}
\begin{figure}[H]
	\centering
	\includegraphics[width=0.5\linewidth]{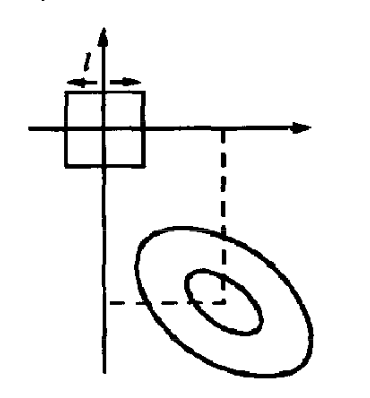}
	\caption{Illustration of an ellipse with $d(l_1(\lambda_1),l_2(\lambda_2)^2=c^2$ and the area covered by $\left(-l/2,l/2\right)\times \left(-l/2,l/2\right)$ \cite{VEGA} }
	\label{fig:ellipse}
	\end{figure}

\begin{figure}[H]
	\centering
	\includegraphics[width=0.5\linewidth]{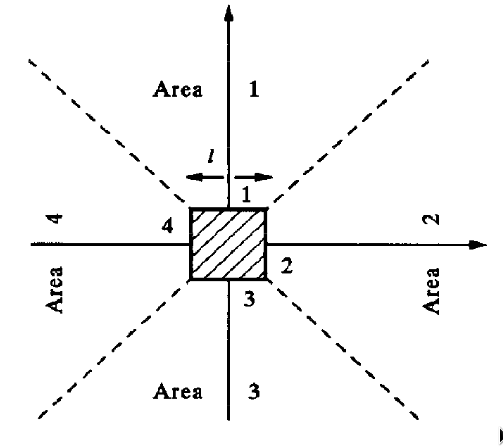}
	\caption{Division of $\lambda_1,\lambda_2$-plane into four areas to determine, which side of the square $\left(-l/2,l/2\right)\times \left(-l/2,l/2\right)$ will be touch first by growing ellipses.\cite{VEGA} }
	\label{fig:area}
	\end{figure}

					In order to calculate the forces between spherocylinders interacting with a Kihara-potential we need to know the shortest distance between the surface of two spherocylinders. To calculate this distance the program of Ullrich Siems used the algorithm described in \cite{VEGA}, which will be explained below. \\
					Due to the fact that the shortest line between to objects is always perpendicular on their surfaces and that the shortest distance between every point of the line segment a spherocylinder to its surface is $\frac{\sigma}{2}$ it is sufficient to calculate the distance between the line segments of spherocylinders and subtract $\sigma$ from the result to get the actual distance between the surfaces.\\
					First of all we parametrize the line segments of two spherocylinders as \begin{align*}
						l_1(\lambda_1)=r_1+\lambda_1{e}_1 \hspace{1cm} \lambda_1\in \left[-l/2,l/2\right] \\
						l_2(\lambda_2)=r_2+\lambda_2{e}_2 \hspace{1cm} \lambda_2\in \left[-l/2,l/2\right],
					\end{align*}
					where $r_{1/2}$ are the center of mass positions of the spherocylinders and ${e}_{1/2}$ are their orientations\\
					We will first assume that the spherocylinders are not parallel to each other, i.e. ${e}_1\cdot{e}_2 \neq 1$. In this case the function $d(l_1(\lambda_1),l_2(\lambda_2))^2$ is a convex function, whose unique minimum can be found by solving the equation \begin{align}
						\nabla d(l_1(\lambda_1),l_2(\lambda_2))^2=0 \\
						\Leftrightarrow \begin{matrix}					
						\lambda_1^0=\frac{r_{12}\cdot {e}_1-{e}_1\cdot{e}_2r_{12}\cdot e_2}{1-({e}_1\cdot{e}_2)^2}\\
							\lambda_2^0= \frac{-r_{12}\cdot {e}_2+{e}_1\cdot{e}_2r_{12}\cdot e_2}{1-({e}_1\cdot{e}_2)^2}\end{matrix},
					\end{align}
					 where $r_{12}=r_1-r_2=-r_{21}$. If $(\lambda_1,\lambda_2)$ is in $\left(-l/2,l/2\right)\times \left(-l/2,l/2\right)$ we can use these two parameters to calculate the shortest distance. Otherwise $(\lambda_1,\lambda_2)$ has to be on the boundary of  $\left(-l/2,l/2\right)\times \left(-l/2,l/2\right)$, i.e $\lambda_1=\pm l/2$ or $\lambda_2=\pm l/2$.\\
					 To figure out which $\lambda_i$ obeys this condition we have a closer look, at the equation \\ $d(l_1(\lambda_1),l_2(\lambda_2)^2=c^2$ and realize that, if we go to coordinates $u=\lambda_1+\lambda_2$ and $v=\lambda_1-\lambda_2$, this is an elliptic equation for ellipses with half axis parallel to $\lambda_1=\lambda_2$ and $\lambda_1=-\lambda_2$. The center of these ellipses can  be determined by searching the minimum of $d(l_1(\lambda_1),l_2(\lambda_2)^2$, which means we already calculated the solution. An illustration of such an ellipse can be found in \ref{fig:ellipse}. \\
						 If we continuously increase $c^2$ these ellipses will touch the boundary of \\$\left(-l/2,l/2\right)\times \left(-l/2,l/2\right)$, which is a square in the $\lambda_1,\lambda_2$-plane, at some point. It turns out that the $c$ at which this first touching happens is the actual shortest distance between our line segments. From the intersection point we can read of $\lambda_1$ and $\lambda_2$ corresponding to it. \\
						 It makes life a lot easier that the side of the square, that is the boundary of \\$\left(-l/2,l/2\right)\times \left(-l/2,l/2\right)$, which is touched first by a given ellipse can be solely determined by looking at the center of the ellipse, i.e. $(\lambda_1^0,\lambda_2^0)$. Figure \ref{fig:area} shows the $\lambda_1,\lambda_2$-plane divided into four areas and $\left(-l/2,l/2\right)\times \left(-l/2,l/2\right)$. If the center of one ellipse like this is within one of these areas, the point where it first touches $\left(-l/2,l/2\right)\times \left(-l/2,l/2\right)$ will be an element of the side of the square touching the respective area. If the center is for example within area 1, we can conclude that $\lambda_1=l/2$, if it were within area 2 we could tell that $\lambda_2=l/2$, etc.\\
						 Now that one of the parameter is chosen to be $\pm l/2$ the remaining one can simply be found by minimizing $d^2$ with respect to the remaining parameter. The result is necessarily in $\left[-l/2,l/2\right]$. \\
						 With the resulting parameters we can determine the shortest distance, as well as the vector along this shortest distance, which is uses as attack point for forces depending on the shortest distance.\\
						 
		If our spherocylinders are now parallel to each other, we have to differentiate between two cases. The first case covers the scenario, in which it is possible to cause an overlap of the two spherocylinders by moving them along a direction perpendicular to their orientation. In this case $\lambda_1$ and $\lambda_2$ can be calculated as the positive/negative projection of the orientation $e_1=e_2$ onto $r_{12}$ over $2$:\begin{align*}
		\lambda_{1/2} = \frac{\pm r_{12}\cdot e_{1/2}}{2}.
	\end{align*} 
	This result can be inserted into $d(l_1(\lambda_1),l_2(\lambda_2)$ to get the shortest distance. This procedure correspond to choosing the centers of the parts of the spherocylinders that would overlap and calculating the shortest distance by connecting these centers. This connection vector is also used as attacking point for the forces calculated with the shortest distance.\\
	If on the other hand it is not possible to produce such an overlap between the two spherocylinders the shortest distance of the line segments is giving by the distance between the  nearest endpoints. One can simple compare all three possibilities of pairing endpoints and chose the shortest distance in this case.

					{	 \subsection{Radius of gyration}
						 As a further indicator for percolating networks we wanted to use the so called \textbf{radius of gyration}$R_g$. It is a measure for the compactness of a system, i.e. the smaller $R_g$ the compacter the system and the bigger $R_g$ the further spread throughout the box are the colloids. The square of the radius of gyration is defined as the mean square distance of the whole system to its center of mass. Due to lack of statistics the results concerning the radius of gyration were completely inconclusive. However since it might be of interest in further analysis of networks of spherocylinders we wanted to include the necessary basics to apply it on such systems here. \\
						 For a collection of $N$ point particles it is given by \begin{align}
						 	R_g^2=\frac{1}{N}\sum\limits_{i=1}^N |\vec{r}_i-\vec{r}_s|^2,
						 \end{align} where $\vec{r}_i$ is the position of the $i$th particle and $\vec{r}_s$ is the center of mass of the system. If the particles have volume, i.e. the mass of the system is given by a mass density $\rho(\vec{r})$ then this translates to \begin{align}
						 R_g^2=\frac{1}{M}\int\limits_{\mathbb{R^d}}\rho(\vec{r})  |\vec{r}_i-\vec{r}_s|^2 d\vec{r}.
						\end{align}
						For identical point symmetric particles, of homogeneous mass density the radius of gyration can be written in terms of the radius of gyration $R_{g,p}$ of a single particle $i$  with respect to its own center of mass $r_{is}$ and the mean square distance of the centers of mass of the particles to the center of mass of the system:
						\begin{align}
							R^2_g=R_{g,p}+\frac{1}{N}\sum\limits_{i=1}^N |\vec{r}_{is}-\vec{r}_s|^2. \label{eq:rogsimple}
						\end{align}  
						To apply these to the systems we simulated only $R_g^p$ needs to be known. For a sphere of radius $r$ it is given by \begin{align}
							R_{g,sphere}^2=\frac{3}{5}r^2. \label{eq:rogsphere}
						\end{align}
						For spherocylinders with cylinder hight $l$ and radius $r$ the radius of gyration $R_{g,sc}$ can be derived from the radius of gyration the cylinder and the two half spheres. The result is \begin{align}
							R_{g,sc}^2=\frac{1}{l+\frac{4}{3}r}\left[\left(\frac{r^2}{4}+\frac{l^2}{12} \right)l+\frac{4}{3}r\left(\left(\frac{3}{5}-\frac{9}{64}\right)r^2+\left(\frac{l}{2}+\frac{3}{8}r\right)^2   \right)\right] \label{eq:rogsphero}.
						\end{align}
						A short proof of equation \eqref{eq:rogsimple} and the derivation of \eqref{eq:rogsphero} can be found in the appendix in section \ref{App:rogproof} and \ref{App:rogsphero}.
					\subsubsection{Radius of gyration for homogeneous identical, point symmetric particles\label{App:rogproof}}
					In this section we proof equation \eqref{eq:rogsimple} for the radius of gyration of a system of $N$ identical, homogeneous, point symmetric particles, who with respect to their own center of mass have a radius of gyration of $R^p_g$. \\
					To do so we first introduce the variable $R^i_g(\vec{r_0})$, whose square is defined as \begin{align}
						\left(R_g^i(\vec{r}_0)\right)^2=\frac{1}{V_i}\int\limits_{V_i}
						|\vec{r}-\vec{r}_{is}|^2 d\vec{r}, 
					\end{align}
					where $V_i$ is the volume of the $i$th particle. $R_g^i(\vec{r}_0)$ can be interpreted as radius of gyration of the $i$th particle with respect to $\vec{r}_0$ rather than with respect to the center of mass of the system.\\
					We can now show that \begin{align}		
					\left(R_g^i(r_s)\right)^2=\left(R^i_g(r_{is})\right)^2+|\vec{r}_s-\vec{r}_0|^2,\label{eq:rogsteiner}
					\end{align} where $\vec{r}_{is}$ is the center of mass of the $i$th particle. Without loss of generality we chose a coordinate system, in which $r_{is}$ is zero. We can then calculate \begin{align}
						V_i R_g^i(\vec{r}_s)=\int\limits_{V_i}|\vec{r}-\vec{r_s}|^2d\vec{r}=\int\limits_{V_i} \vec{r}^2 d\vec{r}+\int\limits_{V_i} \vec{r_s^2}d\vec{r}+\int\limits_{V_i}\vec{r}\cdot \vec{r}_s d\vec{r}
					\end{align} 
					The first term here is $V_iR_g^i(\vec{r_{is}})$, while the second equals $V_i|r_s-r_{is}|^2$. The last one is zero, because $V_i$ is point symmetric, hence proofing equation \eqref{eq:rogsteiner}.\\
					With this we can now verify equation \eqref{eq:rogsimple}.
					\begin{align}
						R_g^2=\frac{1}{V}\int\limits_V |\vec{r}-r_s|^2d\vec{r}=\frac{1}{V}\sum\limits_{i=1}^N \int\limits_{V_i}|\vec{r}-r_s|^2d\vec{r}=\frac{1}{N}\sum\limits_{i=1}^N  \left(R_g^i(\vec{r_s})\right)^2=\\ \frac{1}{V}\sum\limits_{i=1}^N V_i \left(R_g^i(\vec{r_{is}})\right)^2+|\vec{r}_{is}-\vec{r}_s|^2=\left(R_g^p\right)^2+\frac{1}{N}\sum\limits_{i=1}^N|\vec{r}_{is}-\vec{r}_s|^2.
					\end{align}
					
					\subsubsection{Radius of gyration of a spherocylinder\label{App:rogsphero}}
					The radius of gyration $R_{g,sc}$ of a spherocylinder with radius $r$  and cylinder height $l$  is given in equation \eqref{eq:rogsphero}. To derive it we assume that the spherocylinder's center of mass is situated at the origin of our coordinate system. Then the cylinder part of the spherocylinder also has its center of mass at the origin, while the centers of mass of the two half-spheres have a distance to the origin of $d=l/2+\frac{3}{8}r^2$. \\
					The radius of gyration can now be calculated via \begin{align}
						R_{g,sc}^2=\frac{1}{V_{sc}}\left( R_{g,c}^2 V_c+2R_{g,hs}^2(d)V_{hs} \right),\label{eq:rogdersc}
					\end{align} where $R_{c}$ is the radius of gyration of the cylinder, that makes up the spherocylinder and $R_{g,hs}(d)$ is the mean square distance of the two half-sphere from the center of mass of the spherocylinder. With equation \eqref{eq:rogsteiner} we can express this mean square distance in terms of the radius of gyration of half-spheres $R_{g,hs}$ and the distance $d$: \begin{align}
						R^2_{g,hs}(d)=R_{g,hs}^2+d^2.
					\end{align}
					The radius of gyration of a half-sphere is $R_{g,hs}^2=\left(\frac{3}{5}-\frac{9}{64}\right)r^2$, while the radius of gyration of the cylinder is $R_{g,c}^2=\frac{r^2}{4}+\frac{l^2}{^2}$. Inserting all this back into equation \eqref{eq:rogdersc} finally gives us, after some simple algebra, equation \eqref{eq:rogsphero}. \newpage }

				    {	
				    	\begin{figure}[H]
				    		\centering
				    		$\Phi=0.26\%$\hspace{2.7cm}$\Phi=0.31\%$\hspace{2.5cm}$\Phi=0.41\%$\\
				    		\rotatebox[]{90}{\hspace{5cm}$p=10$}	\input{Jeff1/links5000stickystick510.tex}\hspace{-1.25cm}
				    		\input{Jeff1/6000stickystick510.tex}\hspace{-1.25cm}
				    		\input{Jeff1/8000stickystick510.tex}\vspace{-3.8cm}\\
				    		\rotatebox[]{90}{\hspace{5cm}$p=20$}    	\input{Jeff1/links5000stickystick520.tex}\hspace{-1.25cm}
				    		\input{Jeff1/6000stickystick520.tex}\hspace{-1.25cm}
				    		\input{Jeff1/8000stickystick520.tex}\vspace{-3.8cm}\\
				    		\rotatebox[]{90}{\hspace{5cm}$p=30$}	\input{Jeff1/links5000stickystick530.tex}\hspace{-1.25cm}
				    		\input{Jeff1/6000stickystick530.tex}\hspace{-1.25cm}
				    		\input{Jeff1/8000stickystick530.tex}\vspace{-3.8cm}\\
				    		\rotatebox[]{90}{\hspace{5cm}$p=40$}
				    		\input{Jeff1/links5000stickystick540.tex}\hspace{-1.25cm}
				    		\input{Jeff1/6000stickystick540.tex}\hspace{-1.25cm}
				    		\input{Jeff1/8000stickystick540.tex}\vspace{-3.8cm}\\
				    		\rotatebox[]{90}{\hspace{5cm}$p=50$} \input{Jeff1/untenlinks5000stickystick550.tex}\hspace{-1.25cm}
				    		\input{Jeff1/unt6000stickystick550.tex}\hspace{-1.25cm}
				    		\input{Jeff1/unt8000stickystick550.tex}\vspace{-3.8cm}\\
				    		\input{cb.tex}
				    		\label{fig:cosinuskurve}
				    		\end{figure}
				    		\begin{figure}[H]
				    			\centering
				    			$\Phi=0.52\%$\hspace{2.7cm}$\Phi=0.63\%$\hspace{2.5cm}$\Phi=0.73\%$\\
				    			\rotatebox[]{90}{\hspace{5cm}$p=10$}	\input{Jeff1/links10000stickystick510.tex}\hspace{-1.25cm}
				    			\input{Jeff1/12000stickystick510.tex}\hspace{-1.25cm}
				    			\input{Jeff1/14000stickystick510.tex}\vspace{-3.8cm}\\
				    			\rotatebox[]{90}{\hspace{5cm}$p=20$}    	\input{Jeff1/links10000stickystick520.tex}\hspace{-1.25cm}
				    			\input{Jeff1/12000stickystick520.tex}\hspace{-1.25cm}
				    			\input{Jeff1/14000stickystick520.tex}\vspace{-3.8cm}\\
				    			\rotatebox[]{90}{\hspace{5cm}$p=30$}	\input{Jeff1/links10000stickystick530.tex}\hspace{-1.25cm}
				    			\input{Jeff1/12000stickystick530.tex}\hspace{-1.25cm}
				    			\input{Jeff1/14000stickystick530.tex}\vspace{-3.8cm}\\
				    			\rotatebox[]{90}{\hspace{5cm}$p=40$}
				    			\input{Jeff1/links10000stickystick540.tex}\hspace{-1.25cm}
				    			\input{Jeff1/12000stickystick540.tex}\hspace{-1.25cm}
				    			\input{Jeff1/14000stickystick540.tex}\vspace{-3.8cm}\\
				    			\rotatebox[]{90}{\hspace{5cm}$p=50$} \input{Jeff1/untenlinks10000stickystick550.tex}\hspace{-1.25cm}
				    			\input{Jeff1/unt12000stickystick550.tex}\hspace{-1.25cm}
				    			\input{Jeff1/unt14000stickystick550.tex}\vspace{-3.8cm}\\
				    			\input{cb.tex}
				    			\end{figure}
				    			\begin{figure}[H]
				    				\centering
				    				$\Phi=0.73\%$\hspace{2.7cm}$\Phi=0.74\%$\hspace{2.5cm}$\Phi=0.94\%$\\
				    				\rotatebox[]{90}{\hspace{5cm}$p=10$}	\input{Jeff1/links14000stickystick510.tex}\hspace{-1.25cm}
				    				\input{Jeff1/16000stickystick510.tex}\hspace{-1.25cm}
				    				\input{Jeff1/18000stickystick510.tex}\vspace{-3.8cm}\\
				    				\rotatebox[]{90}{\hspace{5cm}$p=20$}    	\input{Jeff1/links14000stickystick520.tex}\hspace{-1.25cm}
				    				\input{Jeff1/16000stickystick520.tex}\hspace{-1.25cm}
				    				\input{Jeff1/18000stickystick520.tex}\vspace{-3.8cm}\\
				    				\rotatebox[]{90}{\hspace{5cm}$p=30$}	\input{Jeff1/links14000stickystick530.tex}\hspace{-1.25cm}
				    				\input{Jeff1/16000stickystick530.tex}\hspace{-1.25cm}
				    				\input{Jeff1/18000stickystick530.tex}\vspace{-3.8cm}\\
				    				\rotatebox[]{90}{\hspace{5cm}$p=40$}
				    				\input{Jeff1/links14000stickystick540.tex}\hspace{-1.25cm}
				    				\input{Jeff1/16000stickystick540.tex}\hspace{-1.25cm}
				    				\input{Jeff1/18000stickystick540.tex}\vspace{-3.8cm}\\
				    				\rotatebox[]{90}{\hspace{5cm}$p=50$} \input{Jeff1/untenlinks14000stickystick550.tex}\hspace{-1.25cm}
				    				\input{Jeff1/unt16000stickystick550.tex}\hspace{-1.25cm}
				    				\input{Jeff1/unt18000stickystick550.tex}\vspace{-3.8cm}\\
				    				\input{cb.tex}
				    				\end{figure}
				    				\begin{figure}[H]
				    					\centering
				    					$\Phi=1.05\%$\hspace{2.7cm}$\Phi=1.57\%$\hspace{2.5cm}$\Phi=2.09\%$\\
				    					\rotatebox[]{90}{\hspace{5cm}$p=10$}	\input{Jeff1/links20000stickystick510.tex}\hspace{-1.25cm}
				    					\input{Jeff1/30000stickystick510.tex}\hspace{-1.25cm}
				    					\input{Jeff1/40000stickystick510.tex}\vspace{-3.8cm}\\
				    					\rotatebox[]{90}{\hspace{5cm}$p=20$}    	\input{Jeff1/links20000stickystick520.tex}\hspace{-1.25cm}
				    					\input{Jeff1/30000stickystick520.tex}\hspace{-1.25cm}
				    					\input{Jeff1/40000stickystick520.tex}\vspace{-3.8cm}\\
				    					\rotatebox[]{90}{\hspace{5cm}$p=30$}	\input{Jeff1/links20000stickystick530.tex}\hspace{-1.25cm}
				    					\input{Jeff1/30000stickystick530.tex}\hspace{-1.25cm}
				    					\input{Jeff1/40000stickystick530.tex}\vspace{-3.8cm}\\
				    					\rotatebox[]{90}{\hspace{5cm}$p=40$}
				    					\input{Jeff1/links20000stickystick540.tex}\hspace{-1.25cm}
				    					\input{Jeff1/30000stickystick540.tex}\hspace{-1.25cm}
				    					\input{Jeff1/40000stickystick540.tex}\vspace{-3.8cm}\\
				    					\rotatebox[]{90}{\hspace{5cm}$p=50$} \input{Jeff1/untenlinks20000stickystick550.tex}\hspace{-1.25cm}
				    					\input{Jeff1/unt30000stickystick550.tex}\hspace{-1.25cm}
				    					\input{Jeff1/unt40000stickystick550.tex}\vspace{-3.8cm}\\
				    					\input{cb.tex}
				    					\label{fig:nsinuskurveend}
				    					\end{figure}}  
					\listoffigures\addcontentsline{toc}{section}{ \hspace{0.5cm} List of Figures}
					
					\listoftables\addcontentsline{toc}{section}{\hspace{0.6cm} List of tables }
					

					\bibliography{bibfile}
					\bibliographystyle{ieeetr}
				
						\addcontentsline{toc}{section}{ \hspace{0.5cm} References}
	%
	%
					 
					 \end{document}